\documentclass[%
 aip,
 amsmath,amssymb,
 reprint,%
aip,
]{revtex4-1}
\usepackage{siunitx}

\usepackage{graphicx}
\usepackage{dcolumn}
\usepackage{bm}
\usepackage{color}

\usepackage[utf8]{inputenc}
\usepackage[T1]{fontenc}
\usepackage{mathptmx}
\usepackage{etoolbox}

\makeatletter
\def\@email#1#2{%
 \endgroup
 \patchcmd{\titleblock@produce}
  {\frontmatter@RRAPformat}
  {\frontmatter@RRAPformat{\produce@RRAP{*#1\href{mailto:#2}{#2}}}\frontmatter@RRAPformat}
  {}{}
}%
\makeatother
\begin{document}

\preprint{AIP/123-QED}

\title[Terahertz time-domain spectroscopy of materials under high pressure in a diamond anvil cell]{Terahertz time-domain spectroscopy of materials under high pressure in a diamond anvil cell}
\author{T. Suter}
\thanks{T. Suter and Z. Macdermid contributed equally.}
\affiliation{%
 Institute for Quantum Electronics, ETH Zürich, 8093 Zürich, Switzerland
}%

\author{Z. Macdermid}%
\thanks{T. Suter and Z. Macdermid contributed equally.}
\affiliation{%
 Institute for Quantum Electronics, ETH Zürich, 8093 Zürich, Switzerland
}%

\author{Z. Chen}
\affiliation{%
 Department of Materials, ETH Zürich, 8093 Zürich, Switzerland
}%

\author{S.L. Johnson}
\affiliation{%
 Institute for Quantum Electronics, ETH Zürich, 8093 Zürich, Switzerland
}%
\affiliation{SwissFEL, Paul Scherrer Institute, 5232 Villigen PSI, Switzerland} 

\author{E. Abreu} 
\email{elsabreu@pyhs.ethz.ch}
\affiliation{%
 Institute for Quantum Electronics, ETH Zürich, 8093 Zürich, Switzerland
}%

\date{\today}
\begin{abstract}
We present the combination of a broadband terahertz time-domain spectroscopy system (\mbox{0.1 - \qty{8}{\tera\hertz}}), a diamond anvil cell capable of generating high pressure conditions of up to \qty{10}{\giga\pascal} and a cryostat reaching temperatures as low as \qty{10}{\kelvin}. This combination allows us to perform equilibrium and time-resolved THz spectroscopy measurements of a sample while continuously tuning its temperature and pressure conditions. In this work, the procedures and characterizations necessary to carry out such experiments in a tabletop setup are presented. Due to the large modifications of the terahertz beam as it goes through the diamond anvil cell (DAC), standard terahertz time-domain spectroscopy analysis procedures are no longer applicable. New methods to extract the pressure dependent material parameters are presented, both for samples homogeneously filling the DAC sample chamber as well as for bulk samples embedded in pressure media. Different pressure media are tested and evaluated using these new methods, and the obtained material parameters are compared to literature values. Time resolved measurements under pressure are demonstrated using an optical pump - THz probe scheme.
\end{abstract}

\maketitle

\section{Introduction}\label{sec:1:Introduction}
The THz range is situated between the microwave and infrared frequency ranges, between 0.1 THz (3000 µm wavelength) and 10 THz (30 µm wavelength). As it is positioned between frequencies accessible through commonly used electronic and optics technologies, it was historically denominated as the THz gap. The emergence of high power, femtosecond pulsed laser systems in the 1980s has led to this gap being gradually filled from the optics side in tabletop systems \cite{lee-2012,koch-2023,jepsen-2010}, revolutionizing the capabilities of linear, and more recently also nonlinear, THz spectroscopy.
Significant progress was achieved with the development terahertz time-domain spectroscopy (THz TDS). With this technique, the electric field $E(t)$ of a THz pulse is detected in the time-domain \cite{smith-1988, van-exter-1989}, contrary to conventional infrared or optical techniques where only the electric field intensity $I\propto|E|^2$ is measured. THz TDS therefore enables the extraction of the full spectral response of the sample, including amplitude and phase. Comparing a sample measurement to a reference measurement (typically air or a substrate, in transmission, or a mirror, in reflection) allows the complex dielectric properties of the material to be calculated directly, without the need for the Kramers-Kronig approximations that are required when only the transmitted or reflected spectral intensity is detected.

As the energy of THz photons is in the low meV range, they are non-ionizing and sensitive to all infrared active material responses that happen in this energy range. These include free carrier dynamics, phonon and magnon resonances, superconducting and charge density wave gaps, which are relevant in condensed matter studies, as well as vibrational and rotational modes of molecules, of relevance for biochemistry, among other subjects. THz TDS has been extensively used in the past few years, most notably in physics, material science, chemistry and biology applications~\cite{jepsen-2010,lee-2012,tonouchi-2007}. 
Going beyond the characterization of the equilibrium state of the system, time-resolved techniques can be used in which the material is brought out of equilibrium by a pump pulse. Its response to photoexcitation is tracked by means of a probe pulse, delayed in time with respect to the pump. Pump-probe techniques are powerful in enabling the study of ultrafast phenomena in the material, including the timescales for excitation, coupling, and recovery of the excitations that exist in equilibrium, the dynamics of phase transitions, and the exploration of transient phases that become accessible only via non-equilibrium pathways. The ability to control both the pump and probe pulse characteristics and the sample environment determines which properties of the material will be excited and which can be probed. In particular, using THz pulses as a probe can be used to follow the transient response of the free electron distribution or of low frequency modes of the system.

\begin{figure}[!htbp]
    \centering
    \includegraphics[width=0.7\linewidth]{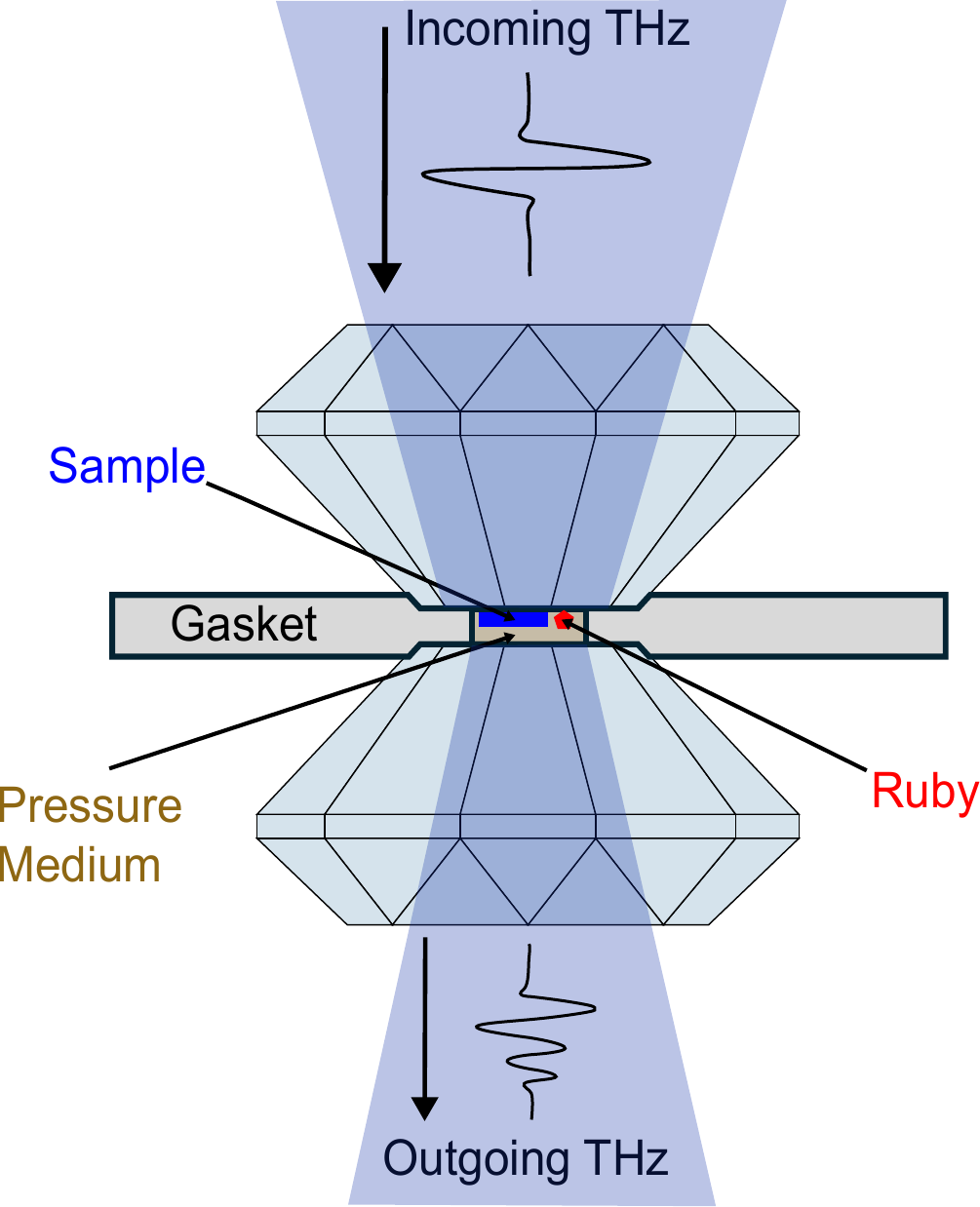}
    \caption{Schematic of a diamond anvil cell showing the two diamonds and the gasket. The sample chamber includes the sample, a pressure medium and a ruby crystal used as a pressure gauge. The THz spot size being larger than the gasket hole, the THz beam is truncated by the gasket. The THz pulse is further modulated by the sample and the pressure medium.}
    \label{fig:DAC_Sketch}
\end{figure}

Experiments in condensed matter physics typically require control of the sample environment. In the context of this work, we focus on temperature and pressure. Varying the temperature, e.g. by using a heater or a cryostat, is a standard procedure in many laboratories. Variation of pressure is far less common, as the available techniques are more challenging. Limited research has been conducted using a diamond anvil cell (DAC) in a THz TDS system\cite{cantaluppi-2018, xu-2021, wang-2023} with the majority of existing studies being performed at synchrotron facilities \cite{guidi-2004,kimura-2010, kimura-2013, voute-2016} or using Fourier-transform infrared spectrometer setups\cite{wenzel-2023, varma-2023}.
A DAC setup allows for isotropic high pressure well above 1 GPa to be applied on a sample, while preserving optical access \cite{weir-1959, jamieson-1959, jayaraman-1983}. A schematic of the DAC is shown in \mbox{Figure \ref{fig:DAC_Sketch}}. In this system, two opposing diamond anvils are pressed against each other. A gasket is placed between the diamonds, consisting of a thin metal plate with a hole in the center, which acts as the sample chamber. The sample is placed in the hole, along with a pressure transmitting medium. In this way, a uniaxial force on the diamonds is converted by the pressure medium into a (quasi) hydrostatic pressure on the sample.

\begin{figure}[!htbp]
    \centering
    \includegraphics[width=\linewidth]{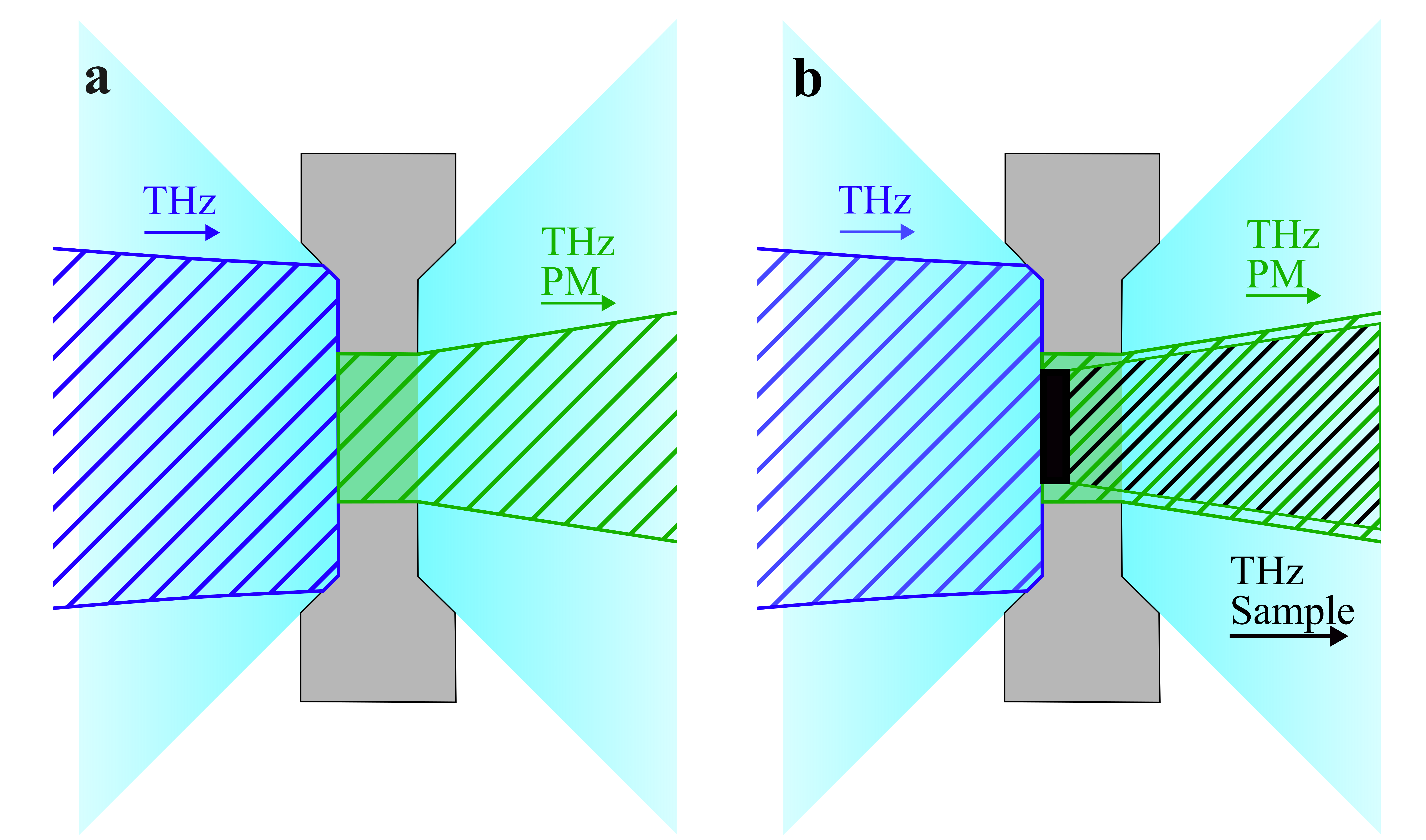}
    \caption{Schematic of a diamond anvil cell (diamond in blue, gasket in gray) with (a) only pressure medium (PM) and (b) pressure medium and a sample in the sample chamber. The incoming THz beam is depicted in blue, whereas the beam depicted in green is modulated by the pressure medium alone and the beam depicted in black is modulated by both the pressure medium and the sample. Not shown are diffraction effects, which are expected to occur at the edges of the sample chamber and of the sample itself.}
    \label{fig:DAC_THz_Sketch}
\end{figure}

In this work, we combine the THz TDS and DAC techniques and propose a methodology to study pressure-driven phenomena in the THz range, also as a function of temperature. The main challenge of these experiments arises from the small dimensions of the gasket hole, which are on the order of the THz wavelength. The focused THz spot size therefore exceeds the gasket hole dimensions. Two effects follow as a consequence: 1) the gasket acts as an aperture, clipping the THz beam and introducing diffraction effects, and 2) the THz beam does not propagate only through the sample but also through the pressure medium, as illustrated in \mbox{Figure \ref{fig:DAC_THz_Sketch}}. Standard THz TDS material parameter extraction procedures that rely on the comparison between a sample and a reference measurement can therefore no longer be used, and alternative methods must be evaluated. As the pressure medium inevitably contributes to the spectroscopic measurements, a suitable material needs to be identified which can preserve the bandwidth of the incident THz pulses. The THz response of this pressure medium should ideally display a negligible pressure and temperature dependence.

\begin{figure*}
\includegraphics[width=0.7\linewidth]{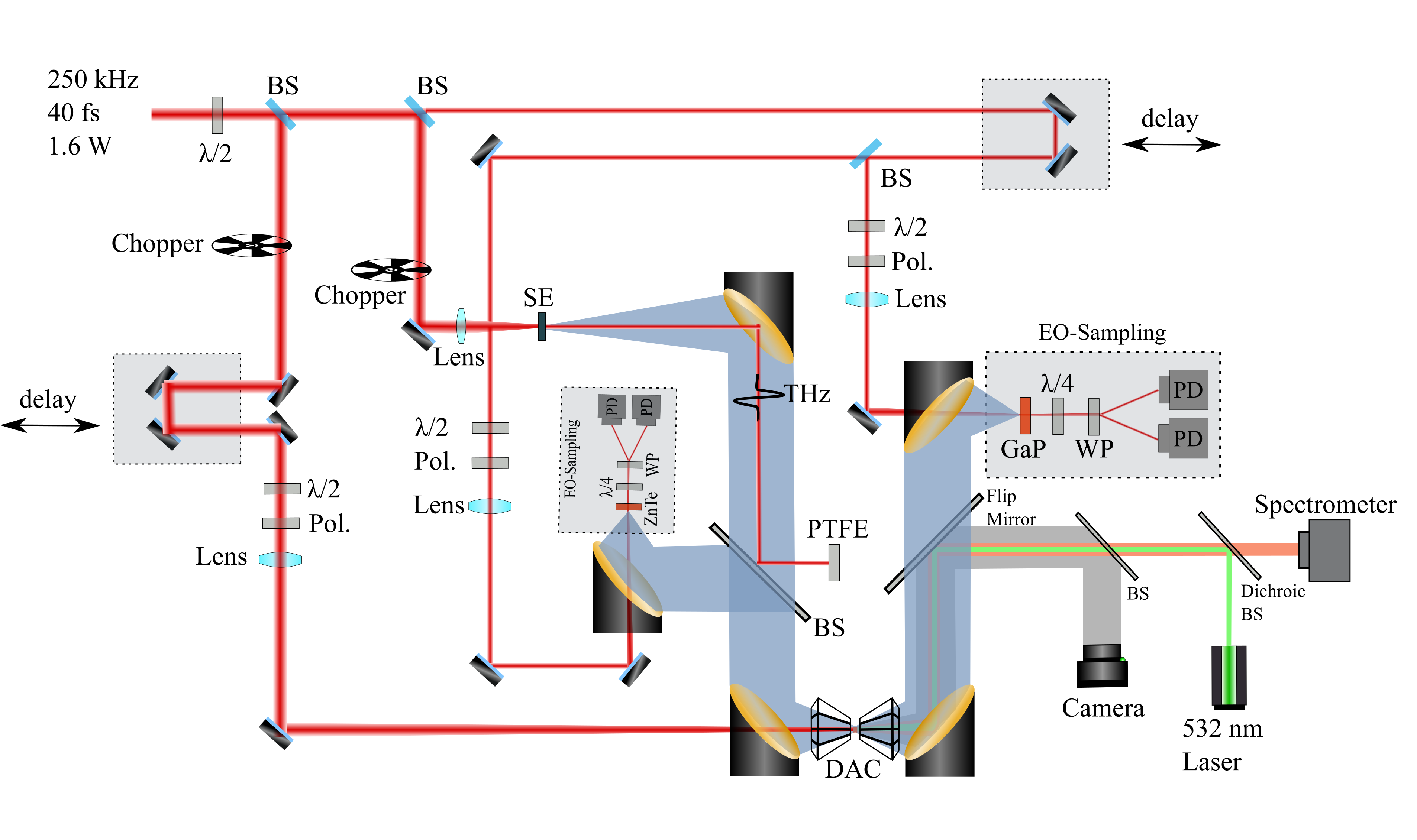}
\caption{\label{fig:TDSSchematic}Schematic of the THz TDS setup used in this work. Optics are abbreviated as BS: Beampslitter, WP: Waveplate, Pol: Polarizer, PD: Photodiode, SE: Spintronic Emitter. The entire THz path from generation to detection is enclosed in an environment which can be purged with nitrogen to avoid absorption of the THz pulse by water vapor in air.}
\end{figure*}

We first describe our experimental setup that combines THz TDS and pressure, with the option to perform \qty{800}{\nano\metre} pump - THz probe measurements. Also included is a characterization of the gasket hole size and how it affects the THz measurements, as well as a description of a typical THz TDS trace.
We then describe a few alternative ways to analyze our THz data under pressure and extract the complex material parameters that characterize the materials present in the sample chamber.
A characterization of different pressure media in the THz range is then presented, as a function of pressure and temperature, and their suitability for the type of experiments we propose is discussed and compared.
Finally, we apply our technique and analysis to a sample of high resistivity silicon, in equilibrium and following photoexcitation by an optical pump pulse. We finish by analyzing and discussing our results and our approach.

\section{Setup and techniques}\label{Sec:2:SetupAndTechniques}

In our setup, high pressure is applied using a CryoDAC Nitro from Almax Easylab \cite{almax-easylab-2023}. This DAC is made of a high strength beryllium copper alloy and consists of two main parts: the cylinder and the piston. The cylinder holds the lower diamond while the piston houses the upper diamond. The position of the lower diamond and the tilt of the upper diamond can be adjusted using two sets of screws. The assembled DAC has a diameter of \qty{38}{\milli\metre} and a height of \qty{29}{\milli\metre}, its small size allowing it to be inserted into an optical cryostat. In the temperature dependent measurements presented in this work, the DAC is used in a Janis ST-100 cryostat \cite{lakeshore-no-date} fitted with windows optimized for THz transmission (topas 6013 polymer).
Optical access through the DAC, onto and out of the diamond surfaces, happens through conically shaped holes cut at a 52 degree angle. The force on the anvils is applied through a gas membrane, which can be pressurized either with helium or nitrogen and allows for in-situ control of the DAC pressure, even when it is used inside the cryostat. The cell is equipped with synthetic type IIas Diacell design diamonds of \qty{0.8}{\milli\metre} anvil diameter --- which define the sample chamber dimensions and the maximum pressure of the \mbox{DAC ---,} \qty{1.55}{\milli\metre} height and \qty{2.55}{\milli\metre} girdle diameter. Depending on the application, we can use \qty{200}{\micro\metre} thick gaskets made of copper beryllium or \qty{250}{\micro\metre} thick gaskets made of stainless steel, which we pre-indent to a thickness of 80 - \qty{120}{\micro\metre} in the region between the diamonds. The indentation thickness is a limiting factor to the maximum achievable pressure before gasket hole expansion sets in, with thinner gaskets allowing for higher pressures. If the gasket indentation thickness is too small, however, the risk of diamond damage increases \cite{dunstan-1989}. In ideal conditions, the gasket hole contracts slightly upon initial compression and remains stable until the pressure limit is reached, at which point gasket hole expansion sets in. Note that if the gasket hole is overfilled with pressure medium, the gasket hole expands immediately \cite{spain-1989}.

The THz TDS setup relies on a Coherent RegA 9040 laser system, providing \qty{1.6}{\watt} of laser power at a repetition rate of \qty{250} {\kilo\hertz} and a pulse duration of \qty{40} {\femto\second}. 
\mbox{Figure \ref{fig:TDSSchematic}} shows a full schematic of the setup, which is suitable for measuring THz pulses in both transmission and reflection with the added capability of performing \qty{800}{\nano\metre} pump - THz probe measurements. The main part of the \qty{800}{\nano\metre} beam is used to generate THz radiation in a trilayer spintronic emitter\cite{seifert-2016}, \mbox{W(2 nm)}/\mbox{$\textrm{Co}_{\textrm{40}}\textrm{Fe}_{\textrm{40}}\textrm{B}_{\textrm{20}}$(1.8 nm)}/\mbox{Pt(2 nm)}, on a \mbox{500 µm} thick sapphire substrate. The THz pulses are detected using electro-optic sampling in \qty{300}{\micro\metre} thick GaP in transmission and in \qty{1}{\milli\metre} thick ZnTe in reflection. This combination of generation and detection provides bandwidths of \mbox{0.1 -- \qty{8}{\tera\hertz}} in transmission and \mbox{0.1 -- \qty{3}{\tera\hertz}} in reflection. The diameter of the THz beam at the DAC position was measured to be \qty{400}{\micro\metre} FWHM using a knife-edge technique. All measurements presented in this work are performed in a purged nitrogen environment to avoid absorption of the THz pulse by water vapor in air. 
A camera is used to image the inside of the DAC, to monitor the position and dimension of the gasket hole and to have a reference for the sample orientation. A Laserglow LCS-0532-TSC DPSS laser system (\qty{80}{\milli\watt} output power) \cite{laserglow-technologies-no-date} is used to excite the ruby crystals inside the DAC. The ruby fluorescence is then detected by an OceanOptics Maya Pro 2000 spectrometer \cite{ocean-optics-no-date}. Optical access for the camera and the ruby pressure gauge is realized through a mirror that can be moved in and out of the THz beam. The pressure in the DAC is calculated from the ruby fluorescence using the international practical pressure scale ruby gauge \cite{shen-2020}.

An important consideration with THz TDS is the inclusion of reflections in the time-domain measurements. Such reflections can originate from any material that the THz beam transmits through, such as the generation and detection crystals, but also the sample itself. For measurements in reflection, it is important to keep in mind that the part of the THz beam which is reflected at the first air-diamond interface arrives at the detection before the reflection from the diamond-gasket interface inside the DAC, which includes the contribution from the sample reflection. Finally, care must be taken in choosing the thickness and refractive index of the THz generation and detection crystals, so as to avoid any temporal overlap of their internal reflections with the THz pulse that is reflected off of the diamond-gasket interface.
Another key parameter of a THz TDS measurement with a DAC is the spot size of the focused THz beam. Small misalignment of the off-axis parabolic mirrors can lead to severe astigmatism in the THz focus. In that case, the transmission through the gasket aperture is reduced and additional reflections originating from the outer diamond surfaces can appear. Before every high-pressure experiment, it is therefore important to measure the THz spot size along the beam propagation axis near the focus and to check that the transmission through the DAC with an air-filled gasket is consistent with previous measurements.

\begin{figure}[!hbtp]
    \centering
    \includegraphics[width=\linewidth]{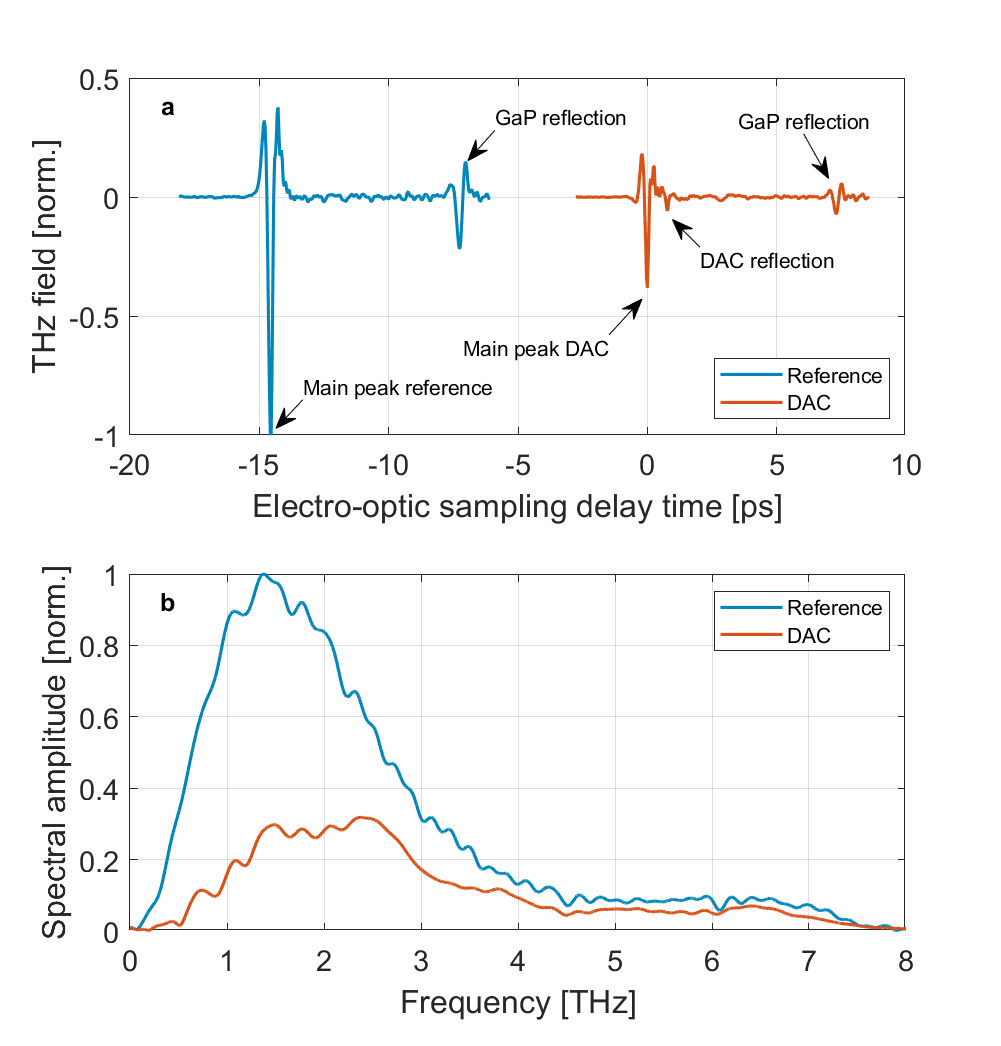}
    \caption{(a) THz electric field transmitted through air without a DAC (blue) and through the air-filled DAC with a gasket hole size of \qty{435}{\micro\metre} in the time-domain (red). The reflection from the GaP crystal \qty{7.3}{\pico\second} after the main peak was excluded from the Fourier transform shown in (b). In both a and b, the traces are normalized to the peak value of the reference THz measurement in air without the DAC (blue traces).}
    \label{fig:EmptyDAC}
\end{figure}

A measurement of the transmission through the DAC with an air-filled gasket is shown in \mbox{Fig. \ref{fig:EmptyDAC}}, in both the time-domain (\mbox{Fig. \ref{fig:EmptyDAC}a}) and the spectral-domain (\mbox{Fig. \ref{fig:EmptyDAC}b}).
The reflections originating from the GaP detection crystal, \qty{7.3}{\pico\second} after the main THz peak, were not included in the range used to calculate the Fourier transform (approximately between \qty{-3}{\pico\second} and \qty{5}{\pico\second} relative to the main THz peak). There is, however, a small reflection at \mbox{0.77 ps} originating from the sample chamber, labeled ``DAC reflection'' in \mbox{Fig. \ref{fig:EmptyDAC}a}, which is too close to the main pulse to be excluded from the Fourier transform. Reflections originating from the outer diamond interfaces can generally be neglected due to the relatively large dimension ($L_{\mathrm{D}} = $\mbox{1.55 mm}) and refractive index ($n_{\mathrm{D}} = 2.4$) \cite{kubarev-2009} of the  diamonds, which mean that their first internal reflection is expected to be delayed by $\Delta t = 2L_{\mathrm{D}}n_{\mathrm{D}}/c\approx$ \qty{25}{\pico\second} with respect to the main pulse going through the DAC. Compared to a reference in air, the main pulse going through the DAC is expected to be delayed by $\Delta t =2L_{\mathrm{D}}(n_{\mathrm{D}} - n_{\mathrm{Air}})/c \approx$ \mbox{14.5 ps}, which matches the delay observed between the blue and red traces in \mbox{Figure \ref{fig:EmptyDAC}a}.

\begin{figure}[!htbp]
    \centering
    \includegraphics[width=\linewidth]{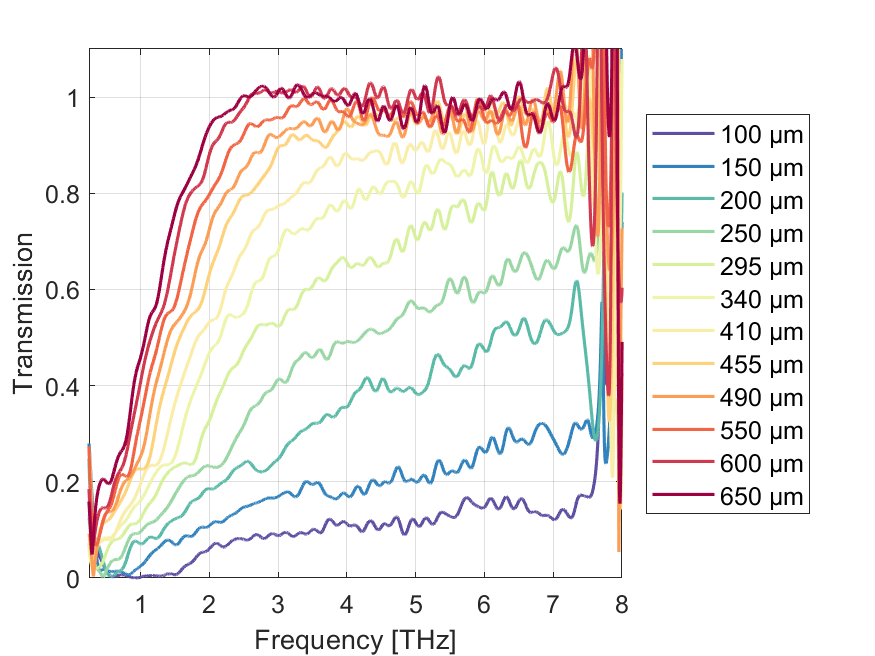}
    \caption{THz transmission through \qty{100}{\micro\metre} thick molybdenum apertures with diameters ranging from \qty{100}{\micro\metre} to \qty{650}{\micro\metre}. The transmission partially exceeds unity for diameters of 535, 600 and \qty{650}{\micro\metre}, which is attributed to diffraction effects. The transmission is calculated using the spectrum of free space THz propagation through air as a reference.}
    \label{fig:aperture}
\end{figure}

The size of the diamond anvils is a defining parameter of a DAC. Smaller anvil diameters are necessary to reach higher pressure conditions, but in turn they set a limit to the focal spot size of the THz beam\cite{saleh-2007} that can propagate through the DAC. Assuming Gaussian beams, the $1/e^2$ diameter of the focused THz beam, $2w_0$, depends on the wavelength $\lambda$ of the THz radiation as
\begin{equation}
    2w_0 = \frac{4\lambda f}{\pi D},
    \label{eq:GaussianSpotSize}
\end{equation}
where $f$ is the focal length and $D$ is the initial THz beam diameter.
The choice of the diamond anvil size therefore sets a limit for the achievable transmission through the DAC. It also affects the transmitted spectrum, since lower frequencies are focused less tightly and are therefore less transmitted than higher frequencies through the small aperture. The metals typically used as gasket materials have plasma frequencies on the order of several eV and therefore reflect THz radiation, so that the gasket hole effectively acts as an aperture in THz transmission measurements. The inverse relation between the maximum achievable pressure and the diamond anvil area, combined with the large THz focal spot size, sets a limit on the maximum pressure at which THz TDS can be used in transmission.
In order to determine the gasket hole size that is better suited for the THz spectrum in our TDS system, we have studied the THz transmission through a series of apertures in easily machined molybdenum disks, with diameters ranging from \qty{100}{\micro\metre} to \qty{650}{\micro\metre}. To match the conditions inside the DAC as closely as possible, the disks were chosen to have the same \qty{100}{\micro\metre} thickness as the gaskets. The resulting THz transmission was calculated using free space propagation as a reference and is shown in \mbox{Figure \ref{fig:aperture}}.

As expected, the overall transmission decreases with decreasing aperture diameter. The decrease in the low frequencies is larger compared to the decrease at higher frequencies, confirming the frequency dependence of the THz spot size expected from \mbox{Eq. \ref{eq:GaussianSpotSize}}.
It is worth noting that the THz beams in our setup are not perfectly Gaussian, so that we are not able to focus all frequencies to the theoretical minimum. Referring back to \mbox{Eq. \ref{eq:GaussianSpotSize}}, the spot size could be reduced by 1) using an off-axis parabolic mirror with a shorter focal length --- which would, however, reduce the available space for the DAC and the cryostat; or by 2) expanding the beam size before focusing onto the sample position --- which would require additional off-axis parabolic mirrors, thus complicating the setup alignment. The numerical aperture of the DAC must also be considered as a limiting factor in focusing the THz beam.

Given the data in \mbox{Fig. \ref{fig:aperture}}, we set \qty{200}{\micro\metre} as a minimum gasket hole diameter to be used in combination with THz TDS. The gasket hole diameter, which defines the aperture, is typically set to be no larger than half the diamond anvil diameter \cite{dunstan-1989, spain-1989}, therefore a diamond anvil diameter of at least \qty{400}{\micro\metre} allows for a gasket hole diameter of \qty{200}{\micro\metre}. 
According to the data collected by O'Bannon et al. \cite{obannon-2018}, this choice would result in a maximum achievable pressure of \mbox{$\approx$ \qty{40}{\giga\pascal}}. As a final word in this section, it is important to note that the transmission data shown in \mbox{Fig. \ref{fig:aperture}} is not corrected for losses due to reflections at the diamond interfaces or due to propagation through the pressure medium.

\section{Material parameter extraction}\label{Sec:6:MaterialParameterExtraction}
In this section, we present novel parameter extraction techniques for THz TDS measurements in a DAC, starting with the case of a homogeneously filled sample chamber and then extending it, through a mixed medium approach, to the situation where a small bulk sample is suspended in another material, typically a pressure medium.
For a homogeneously filled sample chamber, we propose three different methods to extract material parameters from THz TDS measurements in a DAC, which differ in the choice of the reference measurement that is used. The presented parameter extraction techniques are all based on measurements in transmission. For THz TDS measurements in reflection, a reference is typically acquired by placing a highly reflective material at the sample position. However, deviations in the spatial position of the sample and reference plane as small as a few micrometers already lead to a non-negligible phase difference between the sample and the reference measurement\cite{jepsen-2010}, making parameter extraction for reflection measurements quite challenging.
While a reflection geometry is sometimes unavoidable, to study material phases that strongly reflect or absorb THz radiation \cite{verseils-2023}, in this work we limit our discussion of the analysis methods to measurements performed in transmission. 

In the course of this section, we refer to the Fresnel coefficients that describe the propagation of the THz electric field $\tilde{E}(\omega)$ from \mbox{medium I} to \mbox{medium J} as $t_{\textrm{I,J}}$ and $r_{\textrm{I,J}}$ for transmission and reflection, respectively. We express the propagation through a medium of thickness $L$ by the function $\tilde{P}(\tilde{n},L) = e^{-i\omega\tilde{n}L/c}$, where $\omega$ is the frequency, $\tilde{n}$ denotes the complex index of refraction and $c$ is the speed of light in vacuum. For the sake of readability, we do not explicitly write the frequency dependence of the refractive index, the propagation term or the Fresnel coefficients. As the THz pulse propagates through a material, the real part of the refractive index generally contributes to a phase shift of the THz electric field, whereas the imaginary part leads to absorption.

In standard THz TDS measurements on bulk samples surrounded by air, the parameter extraction relies on the transfer function $\tilde{H}(\omega)$, which consists of the ratio between the THz electric field propagating through the sample, $\tilde{E}_{\textrm{sam}}(\omega)$, and the one propagating through a reference, $\tilde{E}_{\textrm{ref}}(\omega)$. Formally, $\tilde{H}(\omega)$ is defined as
\begin{align}
\begin{split}
    \tilde{H}(\omega, \tilde{n}_S, L_{\mathrm{S}}) &= \frac{\tilde{E}_{\textrm{sam}}(\omega)}{\tilde{E}_{\textrm{ref}}(\omega)} \\
    &= \frac{t_\textrm{A,S}\tilde{P}(\tilde{n}_S, L_{\mathrm{S}})t_\textrm{S,A}}{\tilde{P}(\tilde{n}_A, L_{\mathrm{S}})} \sum^{M}_{j = 0}(r_\textrm{S,A}\tilde{P}(\tilde{n}_S, L_{\mathrm{S}}))^{2j} \\
    &= \frac{4\tilde{n}_S}{(\tilde{n}_S+1)^2}e^{-i\frac{\omega L_{\mathrm{S}}(\tilde{n}_S-1)}{c}}\sum^{M}_{j = 0}\left(\frac{\tilde{n}_S-1}{\tilde{n}_S+1}e^{-i\frac{\omega L_{\mathrm{S}} \tilde{n}_S}{c}} \right)^{2j},
    \end{split}
\label{eq:regularTDS}
\end{align}
where the subscripts A and S denote air and sample, respectively, $L_{\mathrm{S}}$ is the sample thickness and the parameter M describes the number of reflections of the THz pulse inside the sample \cite{jepsen-2010}. In the last expression, the $t_{\textrm{I,J}}$ terms are expanded assuming normal incidence conditions. Analytically inverting this formula is challenging due to the periodic nature of $\tilde{H}(\omega)$, encoded in the complex exponential terms. Using a model description of the refractive index in the material under study, Eq. \ref{eq:regularTDS} can however be used to find the material parameters in a fitting procedure.

The transfer function in Eq. \ref{eq:regularTDS} relies on the assumption that the only change between the sample and reference measurements, $\tilde{E}_{\textrm{sam}}(\omega)$ and $\tilde{E}_{\textrm{ref}}(\omega)$, is the removal of the sample. This approach is no longer possible when using a DAC, since the diamonds and the aperture formed by the gasket hole strongly affect the THz beam propagation and since the sample cannot be removed once the DAC is under pressure. We therefore propose three different methods, using three different references, with which we expand the standard THz parameter analysis techniques to accommodate the challenges posed by the DAC. We first consider the case of a sample chamber of thickness $L_{\mathrm{S}}$ which is homogeneously filled with the sample material, before expanding the analysis to bulk samples surrounded by a pressure medium.
In each of the proposed methods, we construct a transfer function $\tilde{H}_{\textrm{sim}}(\omega)$. We obtain the simulated spectrum as $\tilde{E}_{\textrm{sim}}(\omega) = \tilde{H}_{\textrm{sim}}(\omega) \cdot \tilde{E}_{\textrm{ref}}(\omega)$, where $\tilde{E}_{\textrm{ref}}(\omega)$ is a reference measurement. A different reference is used for each method --- air, the empty gasket or the sample itself --- and the corresponding transfer functions $\tilde{H}_{\textrm{sim}}(\omega)$ are constructed accordingly.

\textbf{Air referenced method.} In this first method, labeled a.r., the transmission through the DAC is modeled according to \mbox{Eq. \ref{eq:2:M1:DACSimpleModelReflections}}, where the spectral response of each optical element in the DAC is explicitly accounted for. The spectral amplitude $\tilde{E}_{\textrm{sim, a.r.}}(\omega)$ transmitted through the DAC can then be expressed as
\begin{align}
\begin{split}
        \tilde{H}_{\textrm{sim, a.r.}}(\omega) =& \frac{\tilde{E}_{\textrm{sim, a.r.}}(\omega)}{\tilde{E}_0(\omega)} = \frac{\tilde{E}_{\textrm{sim, a.r.}}(\omega)}{\tilde{E}_\textrm{ref, a.r.}(\omega) \cdot \mathrm{exp}(i\omega L_{\mathrm{Air}}/c)} \\
        =& t_\mathrm{A,D}  \tilde{P}(\tilde{n}_\mathrm{D}, L_\mathrm{D})  t_\mathrm{D,S}  \tilde{T}_\mathrm{G}(\omega)  \tilde{P}(\tilde{n}_\mathrm{S}, L_{\mathrm{S}}) \\
        &\cdot \sum^{M}_{m=0} (r_\mathrm{S,D}\tilde{P}(\tilde{n}_\mathrm{S}, L_{\mathrm{S}}))^{2m} \cdot t_\mathrm{S,D}  \tilde{P}(\tilde{n}_\mathrm{D}, L_\mathrm{D})  t_\mathrm{D,A},
\end{split}
\label{eq:2:M1:DACSimpleModelReflections}
\end{align}
where $\tilde{T}_\mathrm{G}(\omega)$ denotes the frequency dependent transfer function through the aperture formed by the gasket hole, which is determined from the transmission measurements discussed in \mbox{Section \ref{Sec:2:SetupAndTechniques}} (\mbox{Fig. \ref{fig:aperture}}). The gasket hole size during the pressure dependent measurement is estimated to a precision of $\pm$\qty{5}{\micro\metre} using a camera (Fig. \ref{fig:TDSSchematic}). For aperture sizes in between the ones measured in \mbox{Fig. \ref{fig:aperture}}, the transfer function $\tilde{T}_\mathrm{G}(\omega)$ is obtained by linear interpolation.
It is furthermore assumed that the diffraction effects captured in the aperture measurement are comparable to those occurring in the actual sample measurement.
The initial electric field $\tilde{E}_{0}(\omega)$ can be retrieved from a reference measurement $\tilde{E}_\textrm{ref, a.r.}(\omega)$, measured in air. This reference measurement has propagated through the distance $L_{\mathrm{Air}} = L_\mathrm{D} + L_{\mathrm{S}} + L_\mathrm{D}$ in air instead of through the DAC. The electric field incident on the DAC, $\tilde{E}_{0}(\omega)$, can therefore be reconstructed as $\tilde{E}_{0}(\omega) = \tilde{E}_\textrm{ref, a.r.}(\omega) \cdot \mathrm{exp}(i\omega L_{\mathrm{Air}}/c)$, where we correct for the additional phase term acquired over the distance $L_{\mathrm{Air}}$. This is especially important when comparing the simulation to the measurement in the time-domain via the inverse Fourier transform. Omitting this correction will lead to a discrepancy in the timing of the simulation compared to the measurement.

\begin{figure}[!hbtp]
    \centering
    \includegraphics[width=\linewidth]{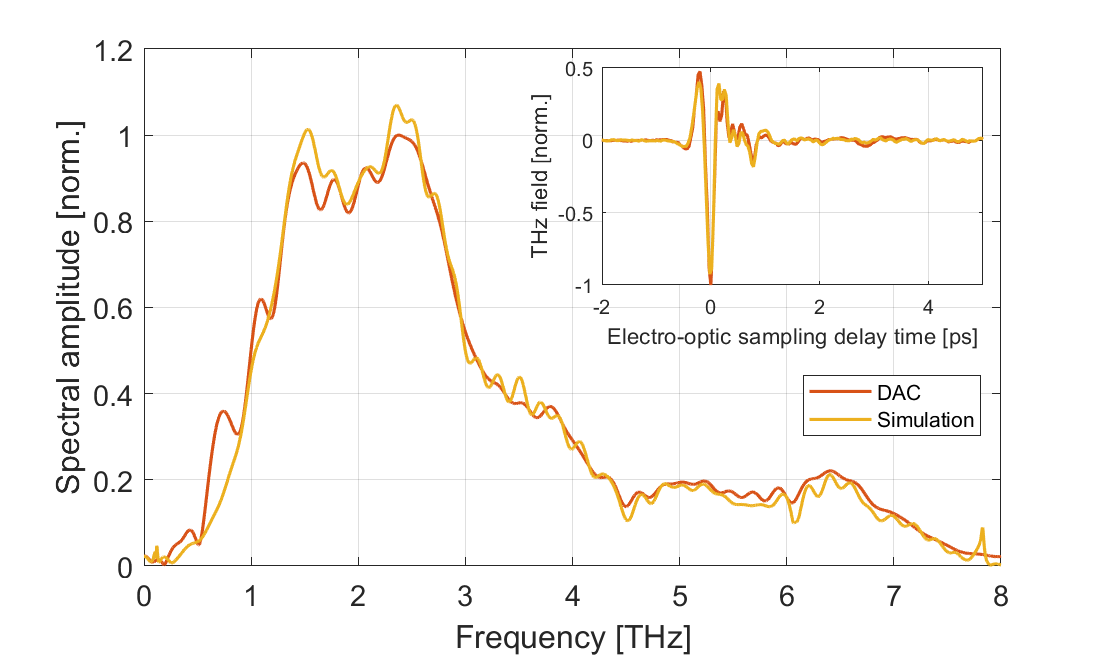}
    \caption{Spectrum of the THz electric field transmitted through the DAC with an air-filled gasket (red) and a simulation following the air-referenced method (yellow). The corresponding time-domain traces are shown in the inset. Both traces in each panel were normalized by the peak value of the respective DAC measurement (red traces).}
    \label{fig:AirReferenced_EmptyDAC}
\end{figure}

An initial evaluation of the air-referenced approach can be performed using the measurements shown in \mbox{Fig. \ref{fig:EmptyDAC}}, which consist of a reference measurement in air and a measurement of the THz transmission through the DAC with an air-filled gasket. A comparison of the simulation (yellow) with the measurement (red) is shown in \mbox{Fig. \ref{fig:AirReferenced_EmptyDAC}}.
To match the timing of the reflection in the simulation, the distance between the diamonds was set to \mbox{117 µm}, whereas the actual gasket thickness was \mbox{96 µm}. This discrepancy can be explained by a small difference in gasket rotation and diamond rotation with respect to the conditions under which the gasket was indented. If these two rotations do not exactly reproduce the conditions under which the gasket was indented, the diamond anvil cannot be inserted completely into the gasket indent by hand. This discrepancy was visible, but not quantifiable, under a microscope. When a force is applied, the diamonds quickly move into full contact with the gasket surface. This discrepancy does, however, limit the use of empty gaskets at ambient conditions as a reference to extract pressure dependent material parameters, since the distance between the diamonds is then significantly larger than in the measurements under pressure.

\textbf{Gasket referenced method.} In this second method, labeled g.r., an empty gasket is used as a reference as described in \mbox{Eq. \ref{eq:2:M2:TransferFunctionDAC}}. Ideally, the same gasket is used as a reference and for the experiment. We assume that the propagation through the diamonds, $\tilde{P}(\tilde{n}_\mathrm{D}, L_\mathrm{D})$, the Fresnel coefficients into the first diamond and out of the second diamond, $t_\mathrm{A,D}$ and $t_\mathrm{D,A}$, as well as the transfer function of the gasket (including diffraction effects), $\tilde{T}_\mathrm{G}(\omega)$, are identical for the reference and sample measurement. These terms therefore cancel out in the transfer function $\tilde{H}_{\textrm{sim, g.r.}}(\omega) = \tilde{E}_{\textrm{sim, g.r.}}(\omega) / \tilde{E}_\textrm{ref, g.r.}(\omega)$, which can then be expressed as
\begin{align}
\begin{split}
     \tilde{H}_{\textrm{sim, g.r.}}(\omega) =& \frac{\tilde{E}_{\textrm{sim, g.r.}}(\omega)}{\tilde{E}_\textrm{ref, g.r.}(\omega)} \\
     =& \frac{t_\mathrm{D,S} \tilde{P}(\tilde{n}_\mathrm{S}, L_{\mathrm{S}}) t_\mathrm{S,D} }{t_\mathrm{D,A} \tilde{P}(\tilde{n}_{A}, L_{\mathrm{G}}) t_\mathrm{A,D} } \tilde{P}(\tilde{n}_{A}, L_{\mathrm{G}} - L_{\mathrm{S}}) \\
     & \cdot \frac{\sum_{m=0}^{M}\left(r_\mathrm{D,S}r_\mathrm{S,D}\tilde{P}(\tilde{n_S},L_{\mathrm{S}})^{2}\right)^m}{\sum_{m=0}^{M}\left(r_\mathrm{D,A}r_\mathrm{A,D}\tilde{P}(\tilde{n_A},L_{\mathrm{G}})^{2}\right)^{m}}.
\end{split}
\label{eq:2:M2:TransferFunctionDAC}
\end{align}
Here, the difference between the gasket thickness $L_{\mathrm{G}}$ in the reference measurement and the sample thickness $L_{\mathrm{S}}$ in the sample measurement requires the additional phase term $P(\tilde{n}_{A}, L_{\mathrm{G}} - L_{\mathrm{S}})$ to be included in \mbox{Eq. \ref{eq:2:M2:TransferFunctionDAC}} (see the discussion surrounding \mbox{Fig. \ref{fig:AirReferenced_EmptyDAC}}, above). This additional phase term is acquired in air outside of the DAC, as it arises from a difference in the total DAC length.

\textbf{Self referenced method.} In this third method, labeled s.r., we use information about the refractive index $\tilde{n}_{\mathrm{S}_i}$ of the sample at one value of pressure $p_i$ to calculate its refractive index $\tilde{n}_{\mathrm{S}_j}$ at $p_j$, where $p_j>p_i$. The change in refractive index between two measurements is evaluated according to \mbox{Eq. \ref{eq:2:M3:TDSRelativeChange}}. This method relies on knowing the dielectric properties of the sample at ambient conditions and on performing a measurement close to ambient pressure as a part of the pressure dependent series. Diffraction effects from the edges of the gasket hole are assumed to be similar between two measurements and are therefore not considered in this method.
As in the gasket referenced measurement above, $\tilde{P}(\tilde{n}_\mathrm{D}, L_\mathrm{D})$, $t_\mathrm{A,D}$ and $t_\mathrm{D,A}$ cancel out, and the transfer function relating two measurements at two different pressure values can be written as

\begin{align}
    \begin{split}
                \tilde{H}_{\textrm{sim, s.r.}}(\omega) =& \frac{\tilde{E}_{\textrm{sim, s.r.}}(\omega)}{\tilde{E}_\textrm{ref, s.r.}(\omega)} =
                \frac{\tilde{E}_{\mathrm{p}_j}(\omega)}{\tilde{E}_{\mathrm{p}_i}(\omega)} \\
                =& \frac{  t_{\mathrm{D},\mathrm{S}_j}  \tilde{P}(\tilde{n}_{\mathrm{S}_j}, L_{\mathrm{S}})}{t_{\mathrm{D},\mathrm{S}_i}  \tilde{P}(\tilde{n}_{\mathrm{S}_i}, L_{\mathrm{S}})} \cdot \frac{ \sum^{M}_{m=0} (r_{\mathrm{S}_j\mathrm{D}}\tilde{P}(\tilde{n}_{\mathrm{S}_j}, L_{\mathrm{S}}) )^{2m}}{   \sum^{M}_{m=0} (r_{\mathrm{S}_i\mathrm{D}}\tilde{P}(\tilde{n}_{\mathrm{S}_i}, L_{\mathrm{S}}))^{2m}}
                \cdot \frac{ t_{\mathrm{S}_j\mathrm{D}}}{t_{\mathrm{S}_i\mathrm{D}}} \\
                =& \left(\frac{\tilde{n}_\mathrm{D} +\tilde{n}_{\mathrm{S}_i}}{\tilde{n}_\mathrm{D} + \tilde{n}_{\mathrm{S}_j}}\right)^2 \frac{\tilde{n}_{\mathrm{S}_j}}{\tilde{n}_{\mathrm{S}_i}}e^{-i\frac{\omega L_\mathrm{S}}{c}(\tilde{n}_{\mathrm{S}_j} - \tilde{n}_{\mathrm{S}_i})} \\
                & \cdot \frac{\sum^{M}_{m=0} (r_{\mathrm{S}_j,\mathrm{D}}\tilde{P}(\tilde{n}_{\mathrm{S}_j}, L_{\mathrm{S}}) )^{2m}}{\sum^{M}_{m=0} (r_{\mathrm{S}_i,\mathrm{D}}\tilde{P}(\tilde{n}_{\mathrm{S}_i}, L_{\mathrm{S}}))^{2m}}.    
    \end{split}
    \label{eq:2:M3:TDSRelativeChange}
\end{align}

In cases where the ratio of the Fabry-Pérot reflections is close to 1, \mbox{Eq. \ref{eq:2:M3:TDSRelativeChange}} simplifies to 
\begin{equation}
     \frac{\tilde{E}_{\mathrm{p}_j}(\omega)}{\tilde{E}_{\mathrm{p}_i}(\omega)} \approx \left(\frac{\tilde{n}_\mathrm{D} +\tilde{n}_{\mathrm{S}_i}}{\tilde{n}_\mathrm{D} + \tilde{n}_{\mathrm{S}_j}}\right)^2 \frac{\tilde{n}_{\mathrm{S}_j}}{\tilde{n}_{\mathrm{S}_i}}e^{-i\frac{\omega L_\mathrm{S}}{c}(\tilde{n}_{\mathrm{S}_j} - \tilde{n}_{\mathrm{S}_i})}.
    \label{eq:2:M3:TDSRelativeChange_NoReflections}
\end{equation}
The transfer function $\tilde{H}_{\textrm{sim, s.r.}}(\omega)$ can then be expressed as
\begin{align}
\begin{split}
        \tilde{H}_{\textrm{sim, s.r.}}(\omega) &= \frac{\tilde{E}_{p_j}(\omega)}{\tilde{E}_{p_i}(\omega)} = T(\omega)e^{i\phi(\omega)} \\
        &\approx \left(\frac{\tilde{n}_\mathrm{D} +\tilde{n}_{\mathrm{S}_i}}{\tilde{n}_\mathrm{D} + \tilde{n}_{\mathrm{S}_j}}\right)^2 \frac{\tilde{n}_{\mathrm{S}_j}}{\tilde{n}_{\mathrm{S}_i}}e^{(\alpha_{\mathrm{S}_j} - \alpha_{\mathrm{S}_i})L_\mathrm{S}/2}e^{-i\frac{\omega L_\mathrm{S}}{c}(n_{\mathrm{S}_j} - n_{\mathrm{S}_i})},
\end{split}
\label{eq:2:M3:DAC_TDS_SimpleModel_Setup2}
\end{align}
where in the last step the complex refractive index $\tilde{n}$ was rewritten as  $\tilde{n} = n' + in'' = n' + i \frac{c\alpha}{2\omega}$.

For all three methods introduced in this section, the complex dielectric function can be modeled using Lorentzian oscillators as
\begin{equation}
    \epsilon(\omega) = \epsilon_\infty + \sum_{j}^{N}\frac{\omega_{p,j}^2}{\omega_{0,j}^2 - \omega^2 - i\omega\gamma_j},
    \label{eq:Lorentz}
\end{equation}
and converted to a complex refractive index $\tilde{n}$ for use in the methods described above. Parameters $\omega_{p,j}$, $\omega_{0,j}$ and $\gamma_j$ correspond to the amplitude, frequency and damping of mode $j$, respectively, and $\epsilon_\infty$ accounts for higher frequency contributions to $\epsilon(\omega)$.
No Drude term was included in the model in Eq. \ref{eq:Lorentz} due to the low spectral content transmitted through the DAC for frequencies up to \qty{0.75}{\tera\hertz}, which is a consequence of the aperture formed by the gasket. The addition of a Drude term would, however, be necessary if studying metals. The parameters of the model are obtained via fitting, using literature values as starting points, where available. The fitting procedure in our case was carried out using the Nelder-Mead \cite{lagarias-1998} and particle swarm \cite{kennedy-1995, wang-2017} optimization algorithms in order to minimize the discrepancy between the simulation and the experimental result, which can be evaluated either in the spectral-domain or in the time-domain following
\begin{align}
    \Delta \Tilde{E}(\omega) &= \sum_{\omega}|\Tilde{E}_{\mathrm{sam}}(\omega) -\Tilde{E}_{\mathrm{sim}}(\omega) |^2, \label{eq:5:fittingSD} \\
    \Delta E(t) &= \sum_{t}|E_{\mathrm{sam}}(t) - E_{\mathrm{sim}}(t)|^2. \label{eq:5:fittingTD} 
\end{align}
According to Parseval's theorem, fitting in the spectral or time-domain is equivalent due to the unitarity of the Fourier transform \cite{kaplan-1991}. 
\\
\textbf{Extension to bulk samples.}
The case, labeled mix, where a small bulk sample is suspended in another material, typically a pressure medium, can be handled by modifying either of these three analysis methods, here we opt for the air referenced method (Eq. \ref{eq:2:M1:DACSimpleModelReflections}). We consider the case where a bulk sample is in direct contact with one of the diamond surfaces, as illustrated in \mbox{Figure \ref{fig:DAC_THz_Sketch}}. In a simple mixed medium approach, we divide the initial THz beam $\tilde{E}_0(\omega)$ into two fractions, $\zeta$ and $(1-\zeta)$, as
\begin{equation}
     \tilde{E}_0(\omega)= \zeta\cdot\tilde{E}_0(\omega) + (1 - \zeta)\cdot\tilde{E}_0(\omega), \quad \textrm{with} \quad 0 \leq \zeta \leq 1.
\end{equation}
The first part, $\zeta\cdot\tilde{E}_0(\omega)$, is then simulated to pass through the part of the sample chamber occupied by the sample, with transfer function $\tilde{H}_\textrm{sim, S+PM}(\omega)$. The remaining fraction, \mbox{$(1 - \zeta)\cdot\tilde{E}_0(\omega)$}, passes only through the part of the sample chamber filled with pressure medium, and is modeled by the transfer function $\tilde{H}_\textrm{sim, PM}(\omega)$. 
The detected THz spectrum can then be written as the superposition of both simulated beams as
\begin{align}
\begin{split}
        &\tilde{E}_\textrm{sim, mix}(\omega)\\
        &= \zeta\tilde{E}_0(\omega)\tilde{H}_\textrm{sim, S+PM}(\omega) + \left(1-\zeta \right)\tilde{E}_0(\omega)\tilde{H}_\textrm{sim, PM}(\omega)
\end{split}
\end{align}
with 
\begin{align}
\begin{split}
        \tilde{H}_\textrm{sim, S+PM}(\omega) =& t_\mathrm{A,D}  \tilde{P}(\tilde{n}_\mathrm{D}, L_\mathrm{D})  t_\mathrm{D,S}  T_\mathrm{G}(\omega)  \tilde{P}(\tilde{n}_\mathrm{S}, L_{\mathrm{S}}) \\
        &\cdot \sum^{M}_{m=0} (r_\mathrm{S,D}r_\mathrm{S,PM}\tilde{P}_S(\tilde{n}_\mathrm{S}, L_{\mathrm{S}})^2)^{m}\cdot t_\mathrm{S,PM} \\
        &\cdot \tilde{P}(\tilde{n}_\mathrm{S}, L_\mathrm{G} - L_\mathrm{S}) \cdot \sum^{M}_{m=0} (r_\mathrm{PM,S}r_\mathrm{PM,D}\\
        &\cdot \tilde{P}(\tilde{n}_\mathrm{S}, L_\mathrm{G} - L_\mathrm{S})^2)^{m} \cdot t_\mathrm{PM,D}  \tilde{P}(\tilde{n}_\mathrm{D}, L_\mathrm{D})  t_\mathrm{D,A},
\end{split}
\label{eq:Sample_H_SandPM}
\end{align}
simulating the transmission through the sample and the pressure medium, with $L_\mathrm{S}$ as the sample thickness and $L_\mathrm{G}$ as the gasket thickness, and
\begin{align}
\begin{split}
        \tilde{H}_\textrm{sim, PM}(\omega) =& t_\mathrm{A,D}  \tilde{P}(\tilde{n}_\mathrm{D}, L_\mathrm{D})  t_\mathrm{D,PM}  T_\mathrm{G}(\omega)  \tilde{P}(\tilde{n}_\mathrm{PM}, L_{\mathrm{G}}) \\
        &\cdot \sum^{M}_{m=0} (r_\mathrm{PM,D}\tilde{P}(\tilde{n}_\mathrm{PM}, L_{\mathrm{G}}))^{2m} \cdot t_\mathrm{PM,D}  \tilde{P}(\tilde{n}_\mathrm{D}, L_\mathrm{D})  t_\mathrm{D,A},
\end{split}
\label{eq:Sample_H_PMOnly}
\end{align}
simulating the transmission through the DAC considering only the pressure medium.
In our simple approximation, the parameter $\zeta$ is frequency independent and expected to equal the surface fraction occupied by the sample in the sample chamber. In general, however, $\zeta$ is expected to be frequency dependent, and to be influenced by various experimental parameters such as the THz spot size, the sample chamber dimensions, and the sample geometry and position.

\section{Pressure media}\label{Sec:5:PressureMedia}
In this section we present the investigation of different commonly used pressure media so as to test their suitability for THz TDS measurements. The chosen pressure media include solids (CsI, KBr, PTFE) and liquids (4:1 methanol-ethanol, Daphne 7575, silicone oil), where ``silicone oil'' refers to dimethyl polysiloxane \cite{silicone_oil}. All pressure media were examined at room temperature, for PTFE and Daphne 7575 additional measurements were performed where the temperature was lowered while a high pressure was maintained. Gaseous pressure media were not included for two main reasons: 1) they require dedicated equipment to load, and 2) they typically lead to significant gasket hole contraction when building up pressure which, as mentioned before, strongly affects the THz transmission and hinders the analysis of THz measurements. 
For the solid pressure media, a TDS measurement can be performed at ambient conditions, when the pressure medium is loaded in the DAC but no force is yet applied on the diamonds. With liquid pressure media, however, the sample chamber needs to be sealed to prevent leaking and evaporation. As a consequence, measurements with liquid pressure media always require the application of some finite pressure beyond the ambient value. 

\begin{figure}
    \centering
    \includegraphics[width=\linewidth]{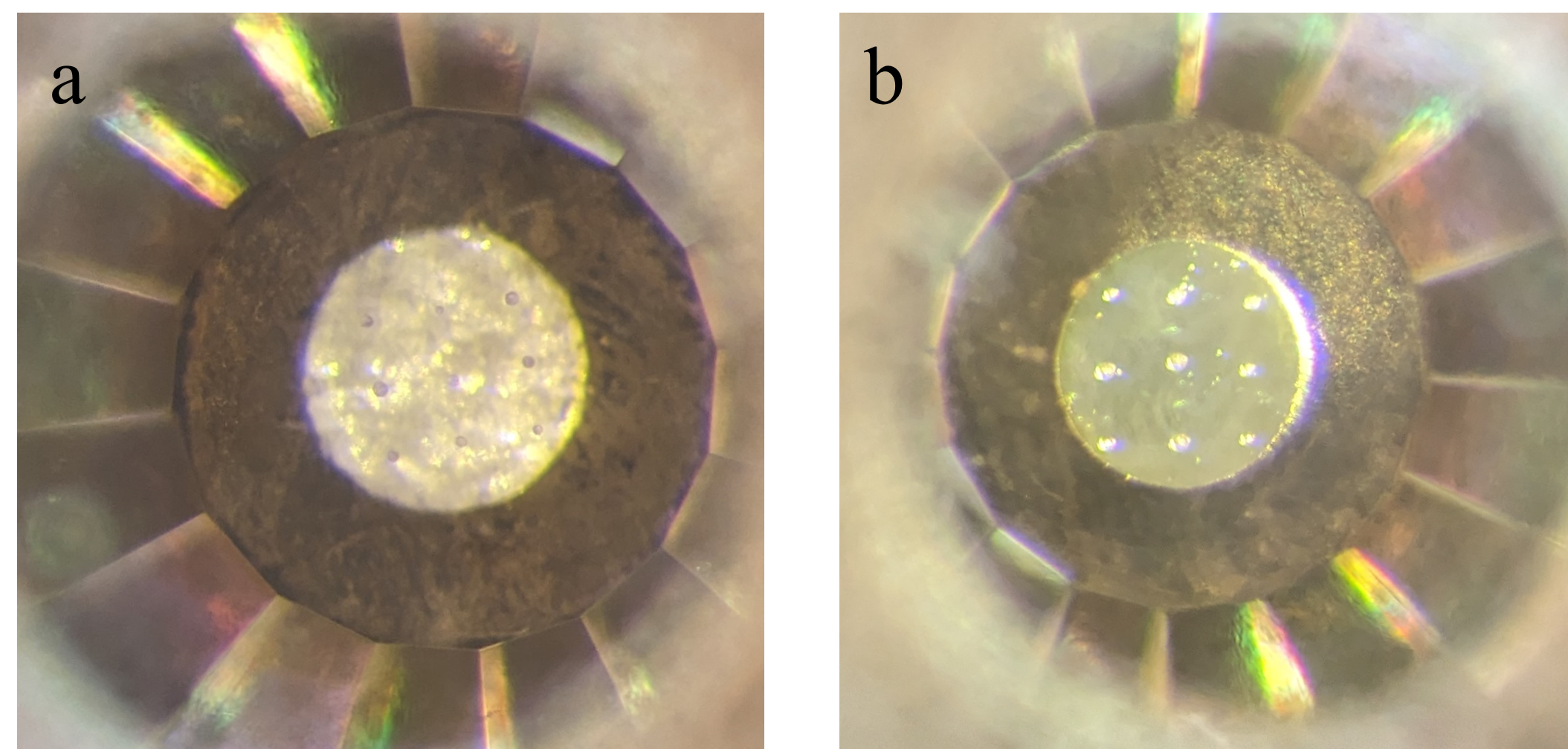}
    \caption{Square pattern of nine ruby crystals in (a) CsI, and (b) Daphne 7575. These microscope images are taken at ambient pressure, before the DAC is inserted into the THz TDS setup.}
    \label{fig:rubies}
\end{figure}

\begin{figure}[!hbtp]
    \centering
    \includegraphics[width=\linewidth]{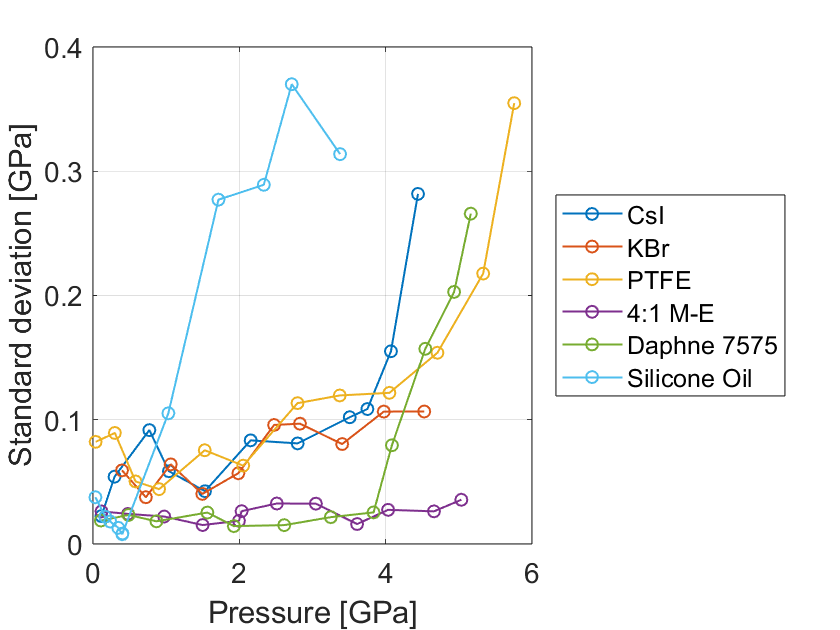}
    \caption{Pressure dependence of the standard deviation for all pressure media covered in this work.}
    \label{fig:Hydrostaticity}
\end{figure}

A procedure to quantify the hydrostaticity of a pressure medium was introduced by Bell and Mao \cite{bell-1981}. In this procedure, multiple pressure gauges are placed in the sample volume. These sensors can then be individually excited and measured. To quantify the hydrostaticity, the standard deviation $\sigma$ of the pressure $P_i$ from $N$ pressure markers is then evaluated as
\begin{equation}
    \sigma = \sqrt{\frac{1}{N-1}\sum_{i=1}^{N} (P_i - \bar{P})^2},   
    \label{eq:hydrostaticity:std}
\end{equation}
where $\bar{P}$ denotes the mean pressure. Nine rubies were used as pressure markers and placed in a grid as shown in \mbox{Fig. \ref{fig:rubies}}. The standard deviation $\sigma$ as a function of pressure is shown in \mbox{Fig. \ref{fig:Hydrostaticity}} for all pressure media under investigation.

\begin{figure}[!hbtp]
    \centering
    \includegraphics[width=1\linewidth]{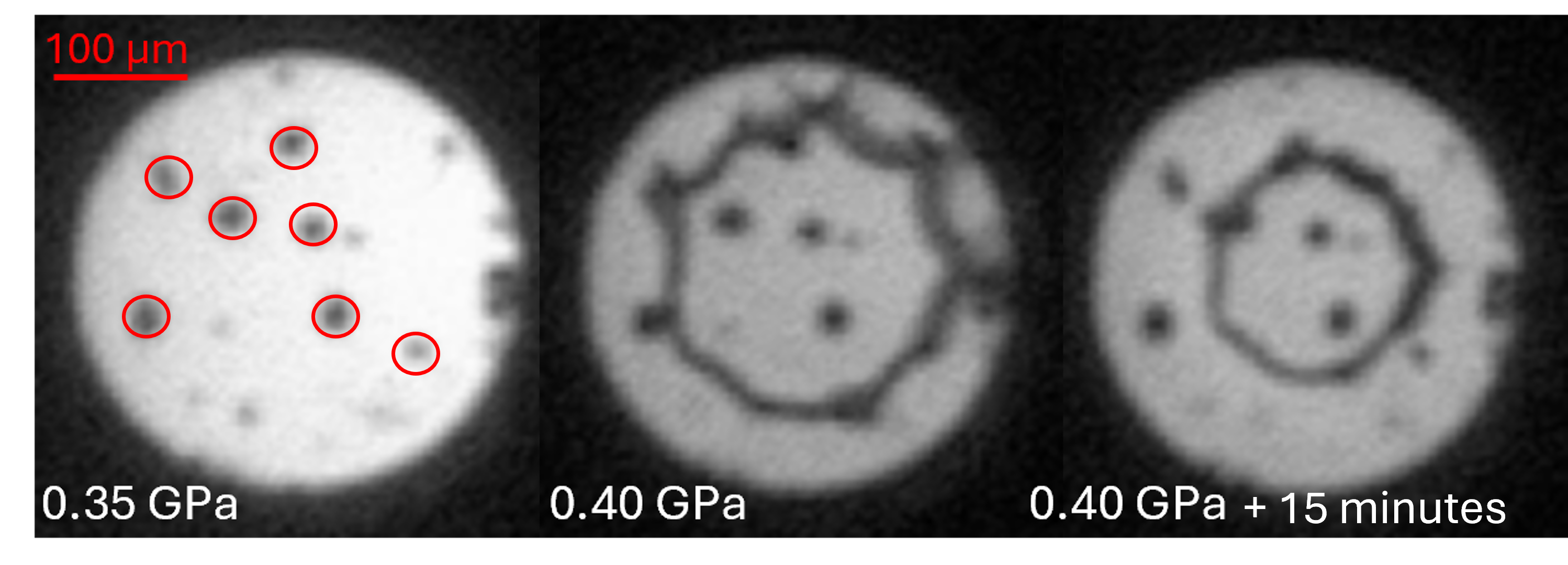}
    \caption{Camera images of the sample chamber filled with silicone oil under pressure. The pattern of seven ruby crystals is highlighted by red circles in the leftmost image. A domain forms between \qty{0.35}{\giga\pascal} (left) and \qty{0.4}{\giga\pascal} (center), which evolves with time and stabilizes in shape after about 15 minutes (right).} 
    \label{fig:Siliconoil}
\end{figure}

\begin{figure*}[!hbpt]
\includegraphics[width = \textwidth]{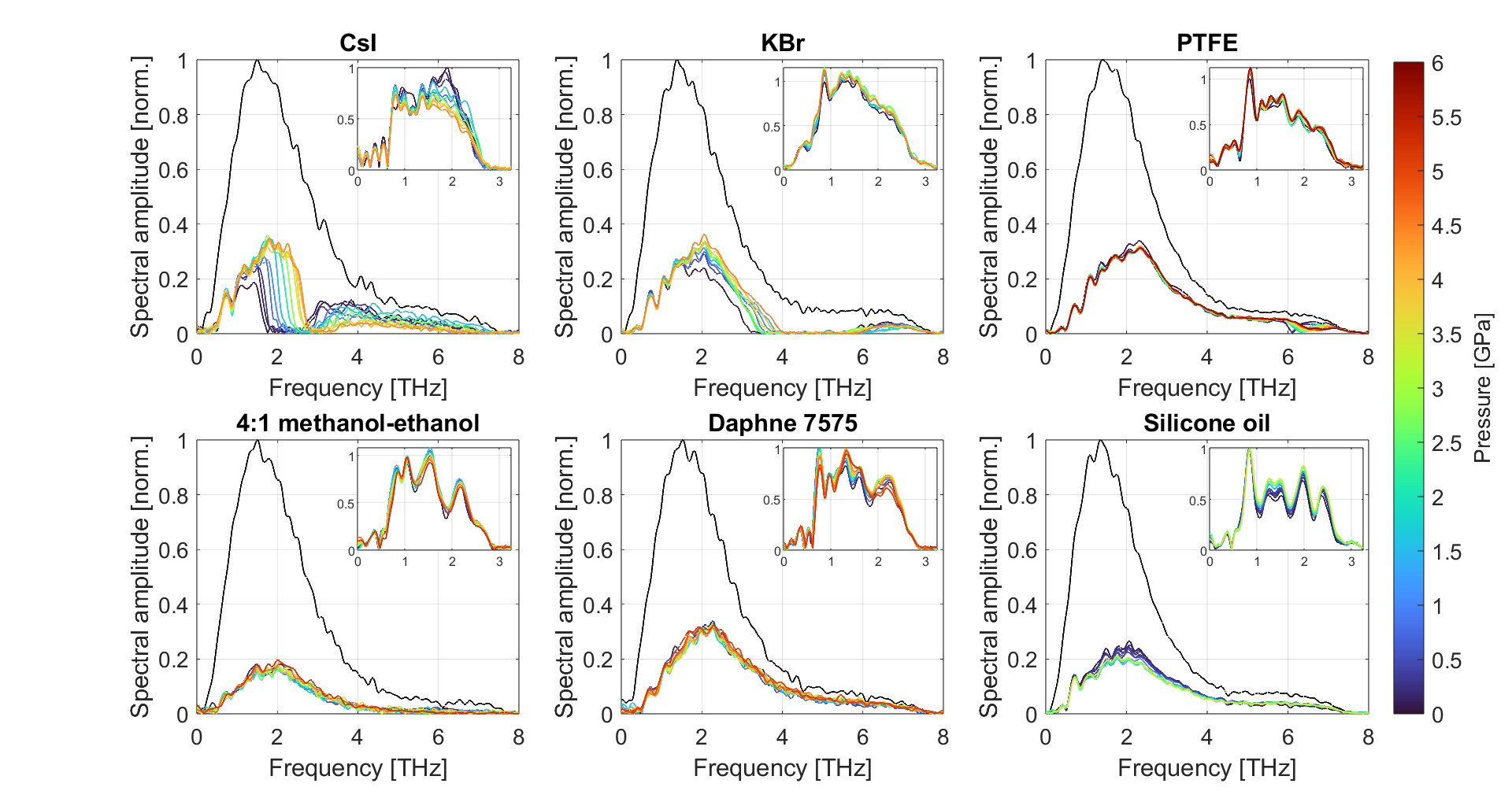}
\caption{Spectra of the THz electric field transmitted through the different pressure media with increasing values of pressure. The pressure value corresponding to a given trace is given by the color of the line, using the color bar as a reference. The spectra were normalized to the peak of their respective reference measurements in air (black traces). THz spectra measured in reflection are shown in the inset, normalized to the peak of the measurement at the lowest available pressure for each pressure medium (0.1 -- \qty{0.4}{\giga\pascal}).}
\label{fig:pressuremediatransmission}
\end{figure*}
A low value of $\sigma$ corresponds to hydrostatic pressure being applied throughout the sample volume. A low $\sigma$ attesting not only to uniform pressure but indeed to hydrostatic pressure is related to the fact that the optical response of ruby crystals to stress is anisotropic\cite{syassen-2008}, and that the different ruby crystals are oriented randomly when placed within the pressure medium.
Aside from silicone oil, all of the pressure media included in this study provide reasonable hydrostatic conditions up to \mbox{$\approx$ \qty{4}{\giga\pascal}}, after which CsI, PTFE and Daphne 7575 show a significant deviation from hydrostaticity. The hydrostaticity of KBr remains relatively unchanged as a function of pressure. Overall, the liquid pressure media performed better, as expected, with the 4:1 methanol-ethanol mixture as the only pressure medium to remain truly hydrostatic over the whole pressure range covered in this work. For silicone oil, the value of $\sigma$ increases rapidly already below 1 GPa. 
Our hydrostaticity measurements for CsI, KBr and PTFE are in good agreement with the study by Celeste et al. \cite{celeste-2019}. To the best of our knowledge, an evaluation of the hydrostaticity of Daphne 7575 relying on the standard deviation of rubies is not available in the literature. Our measurement is, however, in agreement with the results published by Stasko et al.\cite{stasko-2020} which were conducted using three manganin manometers in a piston-cylinder type cell. Their findings showed that the standard deviation of the pressure gauges for Daphne 7575 remains small up to its solidification at 3.9-\qty{4}{\giga\pascal}, after which a steep increase in the standard deviation can be observed. The excellent hydrostatic conditions observed for 4:1 methanol-ethanol mixture are in agreement with a previous study by Klotz et al. \cite{klotz-2009}, where hydrostatic conditions were sustained up to the solidification point at \qty{10.5}{\giga\pascal}. Our silicone oil data show similarities to those reported by Klotz et al. \cite{klotz-2009} but the response we observe is always dominated by the formation of domains. As shown in \mbox{Fig. \ref{fig:Siliconoil}}, a dark ring forms between 0.35 and \qty{0.4}{\giga\pascal}, and stabilizes over a 15 -- 20 min timescale. The increase in $\sigma$, up to almost \qty{0.4}{\giga\pascal}, is mostly due to the difference in pressure between the inside and the outside of the ring, with the inner domain being at a lower pressure than the outer domain.
The domain formation likely results from pressure induced demixing of the silicone oil, which consists of a mixture of molecular chains of varying lengths. This is also likely the origin of the variability in solidification pressures reported in the literature.

The THz transmission through the DAC for increasing values of pressure is shown in \mbox{Fig. \ref{fig:pressuremediatransmission}} for all investigated pressure media. The data are normalized by reference measurements performed in air without the DAC, allowing for a direct comparison of the spectral amplitude between the pressure media. THz reflection data are shown in the insets of \mbox{Fig. \ref{fig:pressuremediatransmission}}, normalized to the peak of the measurement at the lowest available pressure for each pressure medium (0.1 -- \qty{0.4}{\giga\pascal}). We will focus our discussion of the data, as well as the analysis of Section \ref{Sec:9:Analysis}, on the results of the transmission measurements. The reflection results exhibit the same qualitative behavior but the geometry renders parameter extraction more challenging, as discussed in Section \ref{Sec:6:MaterialParameterExtraction}.

In PTFE,  little to no pressure dependence up to the weak phonon absorption at \qty{6.1}{\tera\hertz} can be observed, whereas for CsI and KBr $\sim$\qty{2}{\tera\hertz} wide gaps in the transmitted spectra are visible, which a exhibit a blue shift as pressure increases. These gaps can be attributed to phonon modes in the material, for which the frequency hardens under pressure \cite{lowndes-1974, postmus-1968}. In 4:1 methanol-ethanol, Daphne 7575 and silicone oil, a featureless transmission was observed, with no significant pressure dependence. Crossing the solidification point of Daphne 7575 at \qty{3.9}{\giga\pascal}\cite{stasko-2020}, despite affecting the hydrostaticity in the sample chamber (Fig. \ref{fig:Hydrostaticity}) did not result in a significant change in the transmitted THz spectrum. 
However, 4:1 methanol-ethanol, Daphne 7575 and silicone oil exhibit significant differences in transmitted amplitude and overall bandwidth. The  peak amplitude (averaged over all pressure values) of the transmitted spectrum for Daphne 7575 is larger by a factor of 1.87 compared to 4:1 methanol-ethanol and by a factor of 1.43 compared to silicone oil. The bandwidth transmitted through 4:1 methanol-ethanol is narrower than through the other two liquid pressure media, with less spectral weight in the higher frequency region of the spectrum: the transmitted spectrum of 4:1 methanol-ethanol decreases to 10\% of its peak amplitude at \qty{4.48}{\tera\hertz}, while the same decrease is observed only at \qty{6.7}{\tera\hertz} for Daphne 7575 and at \qty{7}{\tera\hertz} for silicone oil.
Although no previous measurements of the refractive index of 4:1 methanol-ethanol in the THz domain are available, measurements of pure ethanol and methanol exhibit strong absorption in the THz domain\cite{yomogida-2010A, yomogida-2010B, zhang-2020}.

\begin{figure}[!hbtp]
    \centering
    \includegraphics[width=\linewidth]{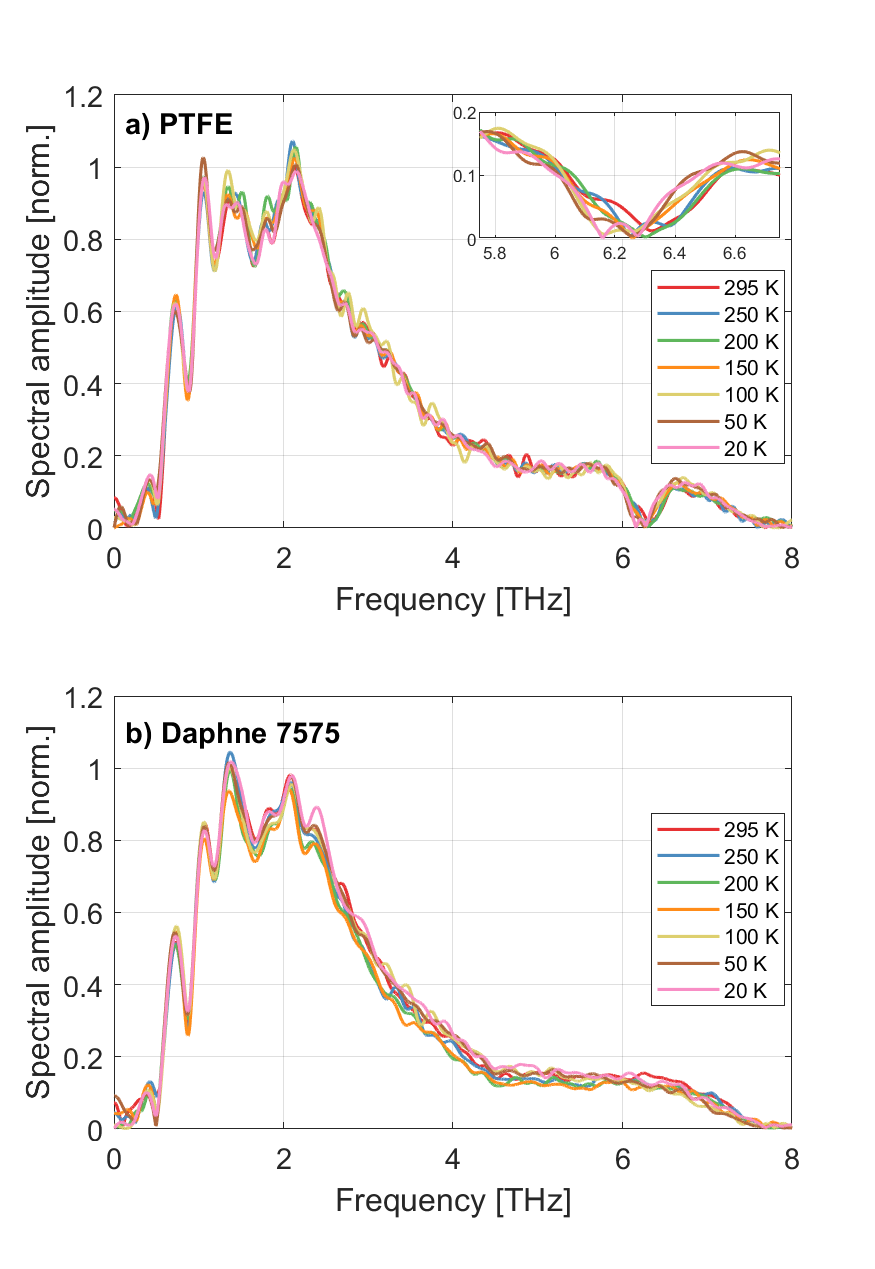}
    \caption{Temperature dependence, between \qty{295}{\kelvin} and \qty{20}{\kelvin}, of the spectra of the THz electric field transmitted  through (a) PTFE and (b) Daphne 7575 under pressure (\qty{1.63}{\giga\pascal} for PTFE and \qty{1.55}{\giga\pascal} for Daphne 7575, at \qty{295}{\kelvin}). The inset in (a) highlights the temperature dependent softening of the phonon mode in PTFE. In both a and b, the spectra have been normalized to the peak value of the room temperature spectum (red traces).} 
    \label{fig:PTFE_and_daphne:TDependence}
\end{figure}

Among all the materials studied in this work, we consider Daphne 7575 and PTFE as the most suitable for THz TDS studies under pressure. Indeed, in the examined pressure range they guarantee reasonable hydrostatic conditions and exhibit a good THz transmission with no pressure dependence except from the phonon mode in PTFE.
We therefore examine also the temperature dependence of the hydrostaticity and THz transmission under pressure for these two selected pressure media. The pressure was first increased to 1.55 -- 1.63 GPa at room temperature, after which the DAC temperature was lowered down to 20 K while maintaining the high pressure. As the DAC cools down, the gas membrane and the gas tubes inside the cryostat cool down as well. The subsequent decrease in membrane gas temperature leads to a decrease in membrane gas pressure, which we opted not to compensate for. Consequently, the pressure in the DAC decreases slightly with temperature, by -\qty{0.26}{\giga\pascal} and -\qty{0.23}{\giga\pascal} at 20 K for Daphne 7575 and PTFE, respectively. The advantage of the choice not to compensate for the decrease in pressure is that the initial DAC pressure is recovered when the DAC is heated back to room temperature. The temperature dependent THz transmission for PTFE and for Daphne 7575 is shown in \mbox{Fig. \ref{fig:PTFE_and_daphne:TDependence}}. No significant change in THz transmission was observed in the whole temperature range for both pressure media, again with the exception of the phonon mode in PTFE. The hydrostaticity of the pressure media remains also essentially unchanged as temperature is varied, i.e. $\sigma$ stays at the value shown in \mbox{Fig. \ref{fig:Hydrostaticity}} for $\approx$\qty{1.6}{\giga\pascal}. Crossing the solidification temperature for Daphne 7575, estimated at \qty{215}{\kelvin} for  \qty{1.5}{\giga\pascal} following Stasko et al. \cite{stasko-2020}, does not lead to a significant change in THz transmission or hydrostaticity. This is in contrast to the clear increase in $\sigma$ which we see in \mbox{Fig. \ref{fig:Hydrostaticity}} when crossing the \qty{3.9}{\giga\pascal} solidification pressure \cite{stasko-2020}.

\section{THz TDS on bulk silicon}\label{sec:bulk silicon}
\begin{figure}[!hbtp]
    \centering
    \includegraphics[width=\linewidth]{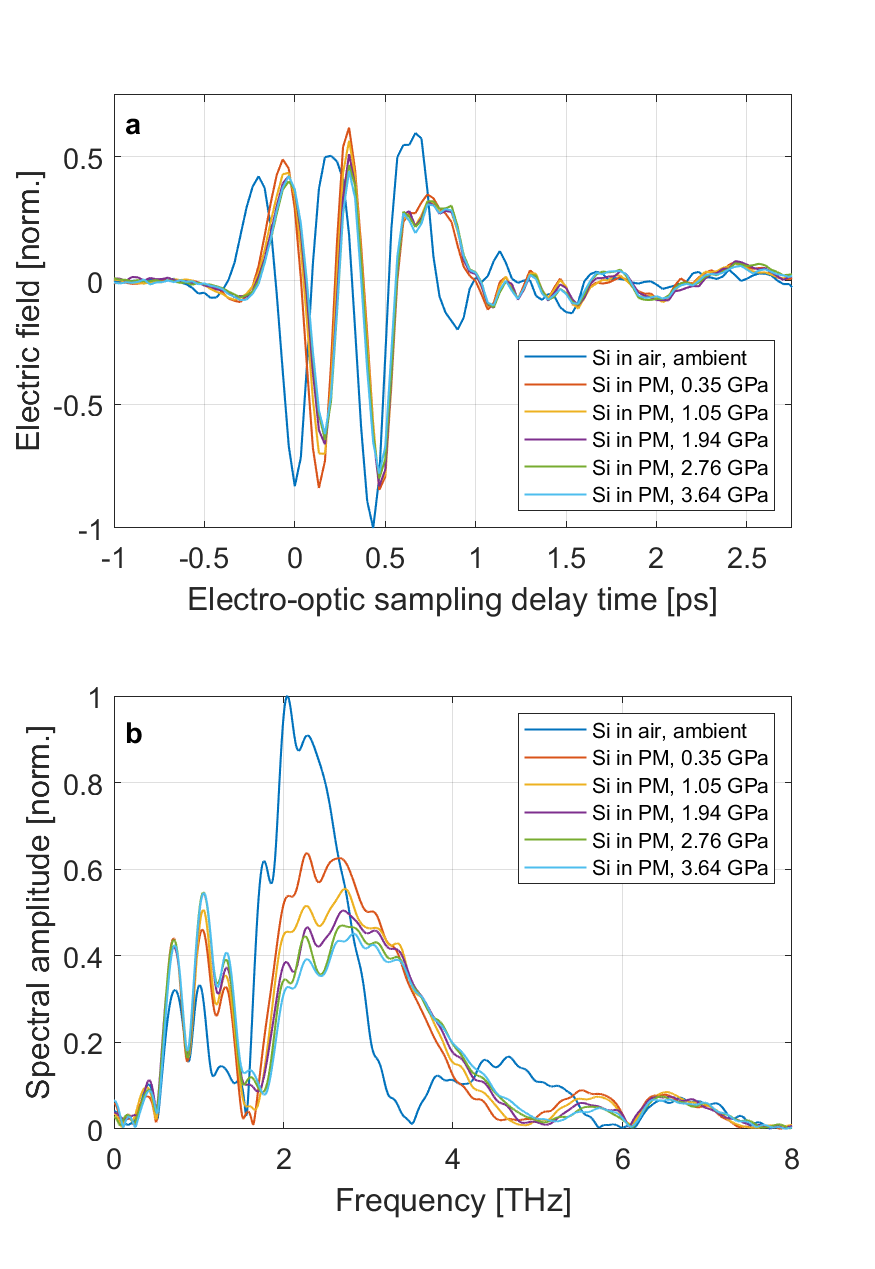}
    \caption{THz electric field transmitted through silicon in the DAC in (a) the time domain and (b) the spectral domain, at ambient conditions surrounded by air (blue) and for increasing values of pressure with Daphne 7575 as the pressure medium (PM). All plots are normalized by the peak value of the THz electric field with the sample in the air-filled gasket at ambient conditions (blue trace).}
    \label{fig:Si_SDTD}
\end{figure}

We measured the pressure dependent THz response of a bulk high resistivity silicon sample using Daphne 7575 as a pressure medium. The sample was a \qty{56}{\micro\metre} thick silicon slab, which covered 39\% of the gasket hole area at ambient conditions. Silicon is an ideal test sample given that it is very well characterized in the THz domain, exhibiting a constant refractive index and negligible absorption \cite{franta-2017}. With a bulk modulus of $K =$\qty{98.7}{\giga\pascal}\cite{martienssen-2005} we expect a decrease in volume of 3.7\% at a pressure of \qty{3.64}{\giga\pascal}. Assuming isotropic compression, this corresponds to a decrease in thickness of 1.25\%.
The time-domain THz transmission data obtained for each pressure value are shown in \mbox{Fig. \ref{fig:Si_SDTD}a}. The corresponding spectra, obtained by Fourier transforming the data in \mbox{Fig. \ref{fig:Si_SDTD}a}, is shown in \mbox{Fig. \ref{fig:Si_SDTD}b}. A measurement taken at ambient conditions before the pressure medium was loaded (i.e. with the DAC sample chamber containing only the silicon slab, surrounded by air) is also added in blue, in \mbox{Fig. \ref{fig:Si_SDTD}}.
Compared to the measurements of the air-filled gasket shown in \mbox{Figure \ref{fig:EmptyDAC}a}, where a distinct main THz peak is visible, two peaks are visible in all the measurements in \mbox{Fig. \ref{fig:Si_SDTD}a}, labeled ``1'' and ``2''. When Daphne 7575 is added as a pressure medium, peak 1 (at $\approx$ \qty{0}{\pico\second}) is delayed more than peak 2 (at $\approx$ \qty{0.5}{\pico\second}) when compared to the measurement in air (blue trace). While no strong spectral features are expected from air, Daphne 7575 or silicon, a strong modulation with sharp dips can be observed in \mbox{Fig. \ref{fig:Si_SDTD}b}.
These findings indicate that peak 1 in the time-domain can be attributed to the part of the beam which passes only through the pressure medium and not through the sample, as illustrated in \mbox{Fig. \ref{fig:DAC_THz_Sketch}b} and discussed in \mbox{Section \ref{Sec:6:MaterialParameterExtraction}}, while peak 2 includes the sample response. For a refractive index of Daphne 7575 larger than that of air, the optical path difference between the two parts of the beam is shortened in the Daphne 7575 data and hence the delay time between the two peaks in \mbox{Fig. \ref{fig:Si_SDTD}a} is reduced, leading also to changes in the THz spectra of \mbox{Fig. \ref{fig:Si_SDTD}b}. As the transmission of Daphne 7575 does not show a pressure dependence in the covered pressure region, spectral changes must be attributed to pressure induced changes in the material parameters of silicon or in the geometry of the sample chamber.

\section{Time-resolved measurements of silicon}\label{Sec:8:DynamicMeasurements}
The sample described in Section \ref{sec:bulk silicon} was further used in time-resolved measurements at high pressure.
The silicon slab was pumped by an \qty{800}{\nano\metre} pulse and probed by the THz pulse. The spot size of the pump beam at the DAC position was set to \qty{940}{\micro\metre} FWHM, far exceeding the gasket hole dimensions. The fluence corresponding to this spot size would be \qty{70}{\micro\joule/\centi\metre^2} in free space. Using Fresnel equations to determine the transmission into the first diamond and into the silicon sample ($\tilde{n} = 3.6756 + i \: 0.005977$ at \qty{800}{\nano\metre}\cite{franta-2017}), we estimate the fluence absorbed by the sample inside the DAC to be \qty{55}{\micro\joule/\centi\metre^2}.

\begin{figure}[!hbtp]
    \centering
    \includegraphics[width=\linewidth]{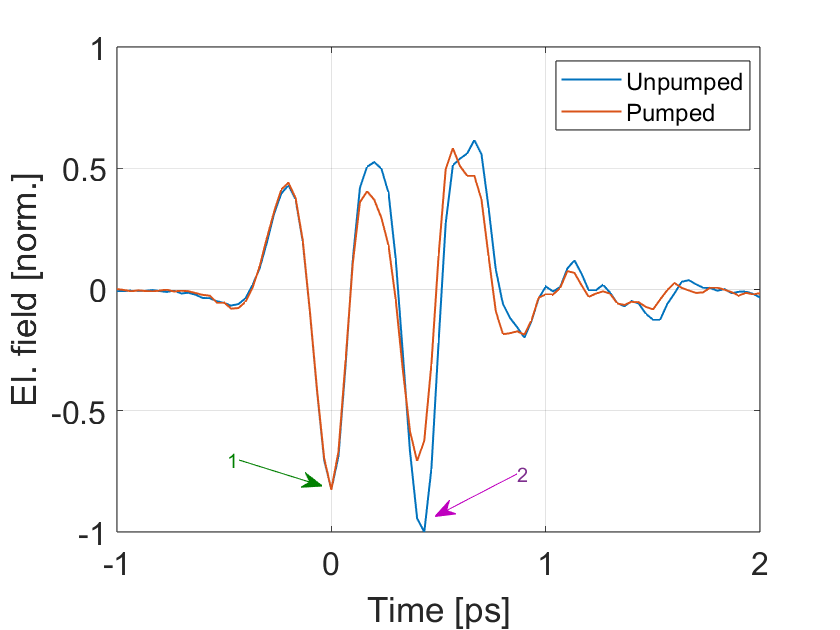}
    \caption{\qty{800}{\nano\metre} pump -- THz probe measurement of the silicon sample in the DAC at ambient conditions surrounded by air (without a pressure medium). The THz electric field transmitted through the sample is shown in the equilibrium case (blue) as well as \qty{13}{\pico\second} after photoexcitation by the \qty{800}{\nano\metre} pump (red). The plots are normalized by the peak value of the unpumped electric field trace (blue).} 
    \label{fig:on_off_ambient}
\end{figure}

First, we investigate the situation where the silicon sample is placed in the DAC without a pressure medium (i.e. surrounded by air in the sample chamber), at ambient pressure. In \mbox{Fig. \ref{fig:on_off_ambient}}, the transmitted THz field measured through the unpumped sample (\qty{800}{\nano\metre} pump pulse blocked) is compared with the THz transmitted through the sample \qty{13}{\pico\second} after the arrival of the the pump pulse.
While a clear modulation is visible in peak 2 of the THz trace, peak 1 remains unchanged. This supports the assumption that only peak 2 contains the response of the sample, as discussed in \mbox{Sections \ref{Sec:6:MaterialParameterExtraction} and \ref{sec:bulk silicon}}.

\begin{figure}[!hbtp]
    \centering
    \includegraphics[width=\linewidth]{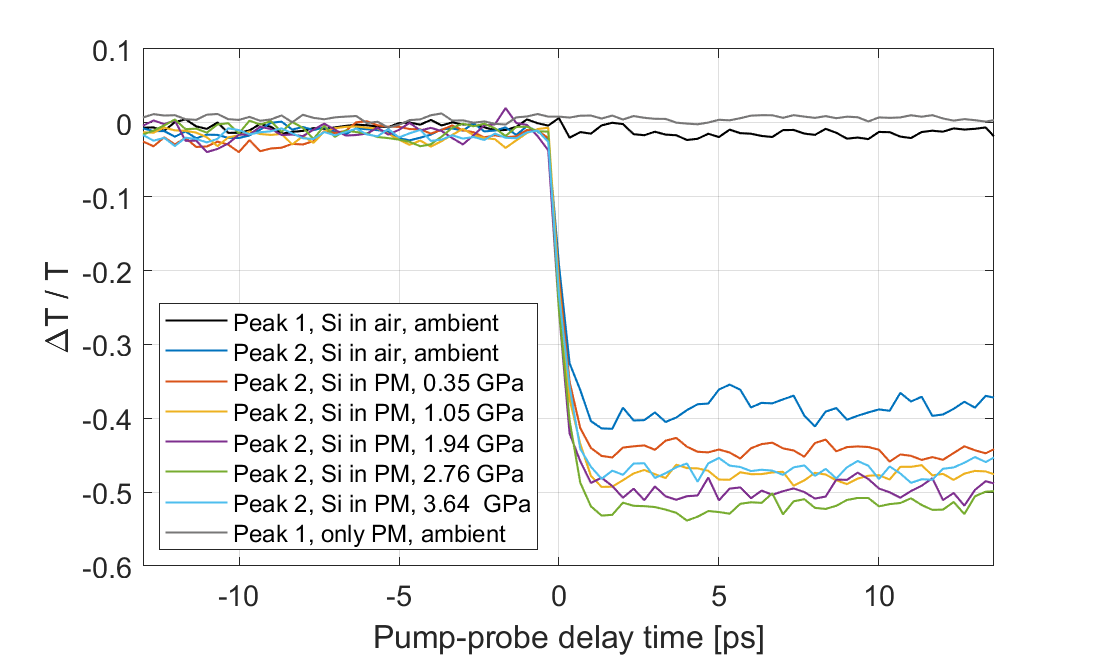}
    \caption{\qty{800}{\nano\metre} pump -- THz probe measurement of the silicon sample (relative change in transmitted THz electric field) at the electro-optic sampling delay corresponding to peak 2 or peak 1, as labeled: at ambient conditions surrounded by air (peak 1 in black, peak 2 in blue), and for increasing values of pressure with Daphne 7575 as the pressure medium (PM). A measurement of Daphne 7575 alone \mbox{(peak 1)} at ambient conditions without the sample is shown in gray.}
    \label{fig:dynamics}
\end{figure}

The pressure medium, Daphne 7575, is then added and pressure is applied. At each pressure value, unpumped and pumped THz traces are measured, as described for \mbox{Fig. \ref{fig:on_off_ambient}}. Additionally, the THz field value at the electro-optic sampling delay corresponding to peak 1 and peak 2 are measured as a function of pump-probe delay and shown in \mbox{Figure \ref{fig:dynamics}}, both in air (no pressure medium) and in Daphne 7575, as labeled. The same measurement is repeated with only Daphne 7575 in the sample chamber, at ambient conditions, at the e-o sampling delay corresponding to peak 1 (the only peak, in that case).
The pump-probe trace on the single peak of the configuration with only Daphne 7575 was acquired as a baseline and indeed shows no photoinduced modification of the THz response of the DAC assembly in the absence of the sample. 
The measurement on peak 1 of the silicon in air configuration shows a very small photoinduced signal, over 10 times smaller than the one detected on peak 2 in all other cases. The magnitude of the decrease in the THz signal at the e-o sampling delay corresponding to peak 2 is comparable to previous \qty{800}{\nano\metre} pump -- THz probe measurements of silicon at ambient conditions with comparable pump fluences \cite{li-2012}. We further show that this photoinduced decrease of the THz transmission varies with applied pressure. 

We finally remark that, given the high repetition rate of the laser used in this study, care must be taken that the sample returns to equilibrium in the \qty{4}{\micro\second} period between consecutive pump pulses. We verified that this requirement was met in this study, using an absorbed fluence of \qty{55}{\micro\joule/\centi\metre^2}. Using a larger piece (about \mbox{2 mm x 2 mm}) of the same \qty{56}{\micro\metre} thick silicon slab and performing \qty{800}{\nano\metre} pump -- THz probe measurements outside the DAC, in free space, we observed that a long lived photoinduced effect persists between pump pulses for an absorbed fluence of \qty{14}{\micro\joule/\centi\metre^2}. The maximum allowed absorbed fluence for the sample to be able to recover in \qty{4}{\micro\second} is therefore between 7 and \qty{14}{\micro\joule/\centi\metre^2} in free space. The difference in relaxation time of the sample in the two configurations is likely related to the improved heat diffusion when the sample is inside the DAC, in contact with the pressure media and the diamond, compared to when it is in air.

\section{Analysis and discussion}\label{Sec:9:Analysis}

In this section we present the extraction of material parameter based on the data shown in Sections \ref{Sec:5:PressureMedia}, \ref{sec:bulk silicon} and \ref{Sec:8:DynamicMeasurements}, using the techniques described in \mbox{Section \ref{Sec:6:MaterialParameterExtraction}}.
The thickness of the sample chamber, denoted $L_\mathrm{S}$ in \mbox{Section \ref{Sec:6:MaterialParameterExtraction}}, is measured for the first and last pressure value in each series, and linearly interpolated for pressure values in between.
We consider first a homogeneously filled sample chamber, as in the case of the pressure media measurements, and then move on to mixtures, which describe the case of a bulk sample in a pressure medium environment.

\textbf{Homogeneously filled sample chamber.}
The measurements performed to identify suitable pressure media can be analyzed using the parameter extraction methods for homogeneously filled sample chambers shown in \mbox{Section \ref{Sec:6:MaterialParameterExtraction}}, where the complex dielectric function, $\epsilon(\omega)$, and hence the complex refractive index of the material, $\tilde{n} = n' + in''$, is obtained via fitting of Eq. \ref{eq:Lorentz}.

For both CsI and KBr, a phonon mode completely absorbs the spectral amplitude in a $\sim$\qty{2}{\tera\hertz} wide range around \qty{5}{\tera\hertz} and \qty{3}{\tera\hertz}, respectively, as can be seen in \mbox{Figure \ref{fig:pressuremediatransmission}}. This has two important consequences: 1) the fitting procedure is affected more by noise in the region around the phonon frequency; and 2) the self-referenced method cannot be used for these materials as the phonon hardening would necessitate a reconstruction at a higher pressure from a frequency region where there was no amplitude at a lower pressure.

\begin{figure}[!htbp]
    \centering
    \includegraphics[width=\linewidth]{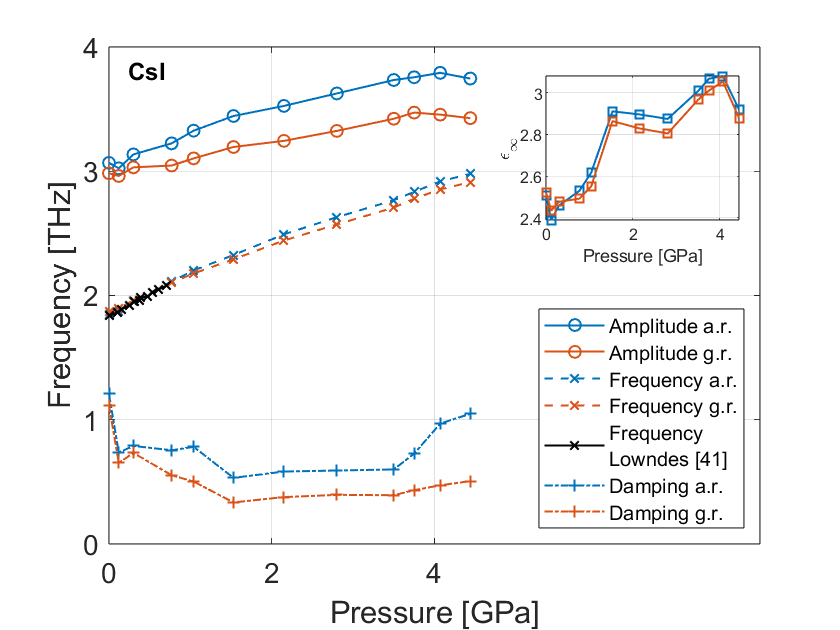}
    \caption{Fitting parameters of the dielectric function with one Lorentzian oscillator in CsI for the air referenced (blue lines, a.r.) and gasket referenced (red lines, g.r.). The phonon frequency as reported by Lowndes \cite{lowndes-1974} is added as a comparison, in black. The high-frequency response $\epsilon_\infty$ is shown in the inset.}
    \label{fig:FittingParametersCsI}
\end{figure}

\begin{figure}[!htbp]
    \centering
    \includegraphics[width=\linewidth]{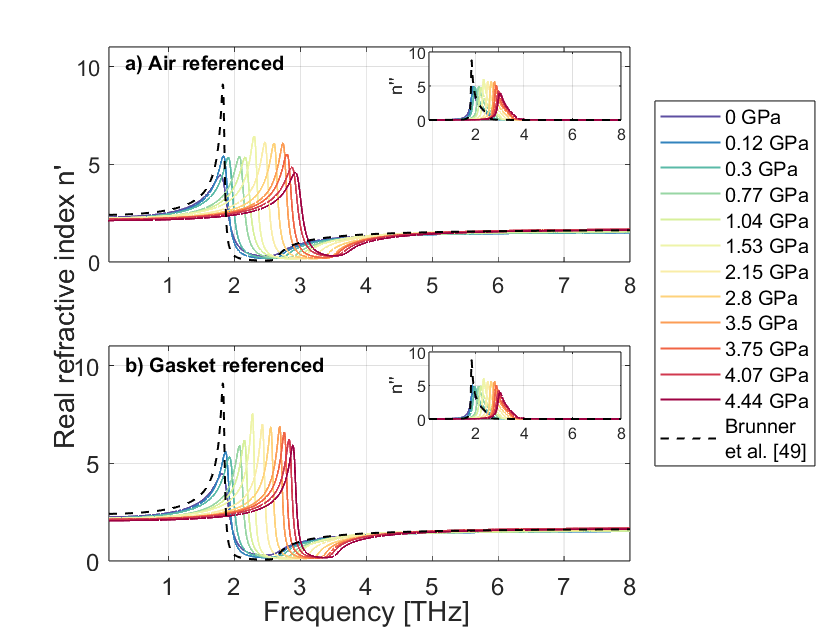}
    \caption{Extracted refractive index of CsI for (a) the air-referenced and (b) the gasket-referenced approach. The imaginary part of the refractive index is shown in the insets. A measurement at ambient conditions by Brunner et al. \cite{brunner-2009} is added as a dashed black line.}
    \label{fig:RefractiveIndexCsI}
\end{figure}

The fitting parameters of the dielectric function for CsI, assuming one Lorentzian oscillator, are shown in \mbox{Fig. \ref{fig:FittingParametersCsI}}, and the corresponding refractive index is shown in \mbox{Fig. \ref{fig:RefractiveIndexCsI}}. 
Both methods show a clear hardening of the phonon frequency with pressure, in excellent agreement with the measurement by Lowndes et al. \cite{lowndes-1974} which we include in \mbox{Fig. \ref{fig:FittingParametersCsI}} for comparison. In turn, the refractive index at atmospheric pressure matches previous reports from Brunner et al \cite{brunner-2009}, in particular the off resonance value of $\tilde{n}$ and the phonon frequency (\qty{1.84}{\tera\hertz} in Brunner et al \cite{brunner-2009}), as shown by the comparison in \mbox{Fig. \ref{fig:RefractiveIndexCsI}}, while the value of $\tilde{n}$ around the phonon frequency differs by a factor of two.

\begin{figure}[!t]
    \centering
    \includegraphics[width=\linewidth]{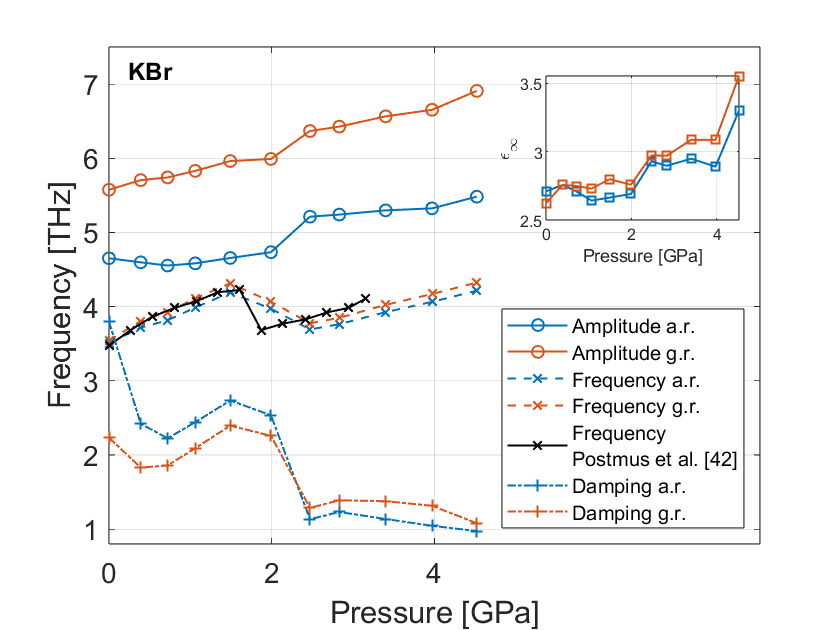}
    \caption{Fitting parameters of the dielectric function with one Lorentzian oscillator in KBr for the air referenced (blue lines, a.r.) and gasket referenced (red lines, g.r.) methods. The phonon frequency as reported by Postmus et al. \cite{postmus-1968} is added as a comparison, in black. The high-frequency response $\epsilon_\infty$ is shown in the inset.}
    \label{fig:FittingParametersKBr}
\end{figure}

\begin{figure}[!t]
    \centering
    \includegraphics[width=\linewidth]{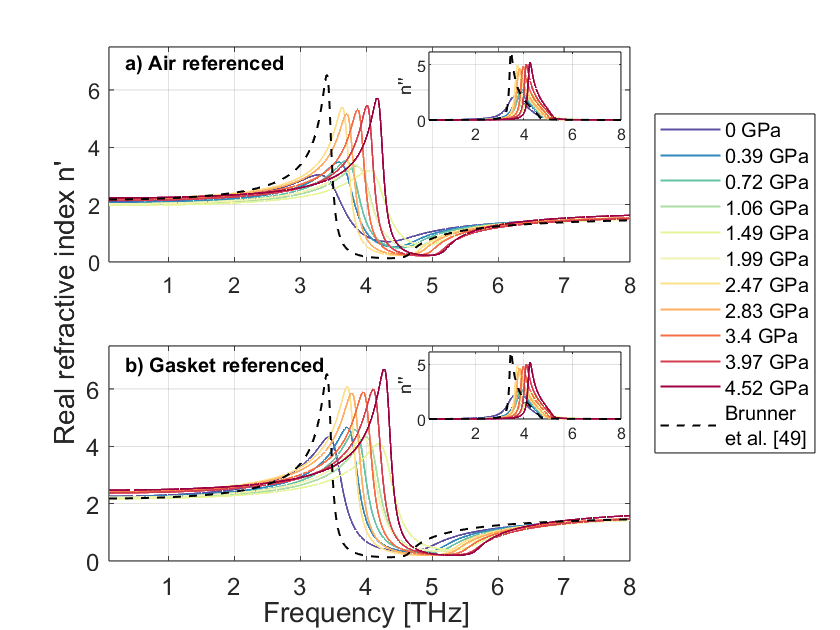}
    \caption{Extracted refractive index of KBr for (a) the air-referenced and (b) the gasket-referenced approach. The imaginary part of the refractive index is shown in the insets. A measurement at ambient conditions by Brunner et al. \cite{brunner-2009} is added as a dashed black line.}
    \label{fig:RefractiveIndexKBr}
\end{figure}

The parameters of the dielectric function for KBr are shown in \mbox{Fig. \ref{fig:FittingParametersKBr}} and the corresponding refractive index is shown in \mbox{Fig. \ref{fig:RefractiveIndexKBr}}, also assuming one Lorentzian oscillator.
The pressure dependent phonon frequency we obtain shows good agreement with previous measurements by Postmus et al.\cite{postmus-1968}.
At moderately low pressures, between \qty{1.8}{\giga\pascal} and \qty{2.5}{\giga\pascal} \cite{postmus-1968,singh-2000,zhao-2015}, KBr undergoes a structural phase transition from Fm3m at low pressure to Pm3m at high pressure \cite{dewaele-2012}. This phase transition is clearly visible in our data as a decrease in the phonon frequency over that pressure range. It is also visible in the camera images: the KBr filled sample chamber is transparent below and above the transition but becomes opaque around \qty{2}{\giga\pascal}, possibly indicating the presence of domain boundaries between structurally distinct domains that coexist.
Regarding the refractive index, our measurements shows good agreement with atmospheric pressure data from Brunner et al \cite{brunner-2009} (\mbox{Fig. \ref{fig:RefractiveIndexKBr}}). As in CsI, the off resonance value of $\tilde{n}$ and the phonon frequency (\qty{3.44}{\tera\hertz} in Brunner et al \cite{brunner-2009}) of KBr match the observations in Brunner et al \cite{brunner-2009} while the value of $\tilde{n}$ around the phonon frequency differs by about a factor of two.

For both CsI and KBr the fitting parameters are very comparable for both parameter extraction methods with the exception of the phonon amplitude in KBr, which is about 20\% larger in the gasket referenced method compared to the air reference one. Uncertainties in determining the phonon amplitude in CsI and KBr, and consequently also in extracting the exact value of $\tilde{n}$ around the phonon frequency, are expected given the strong absorption features in the transmission spectra of \mbox{Fig. \ref{fig:pressuremediatransmission}}, which hinder accurate modeling of the phonon mode.

\begin{figure}[!t]
    \centering
    \includegraphics[width=\linewidth]{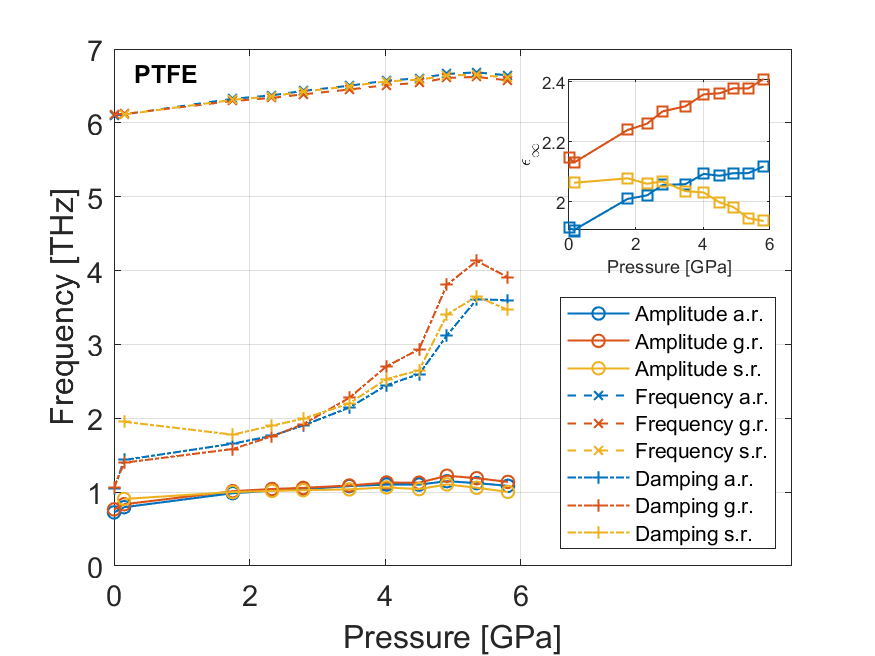}
    \caption{Fitting parameters of the dielectric function with one Lorentzian oscillator in PTFE for the air referenced (blue lines, a.r.), gasket referenced (red lines, g.r.) and self referenced (yellow lines, s.r.) methods. The high-frequency response $\epsilon_\infty$ is shown in the inset.}
    \label{fig:FittingParametersPTFE}
\end{figure}

\begin{figure}[!t]
    \centering
    \includegraphics[width=\linewidth]{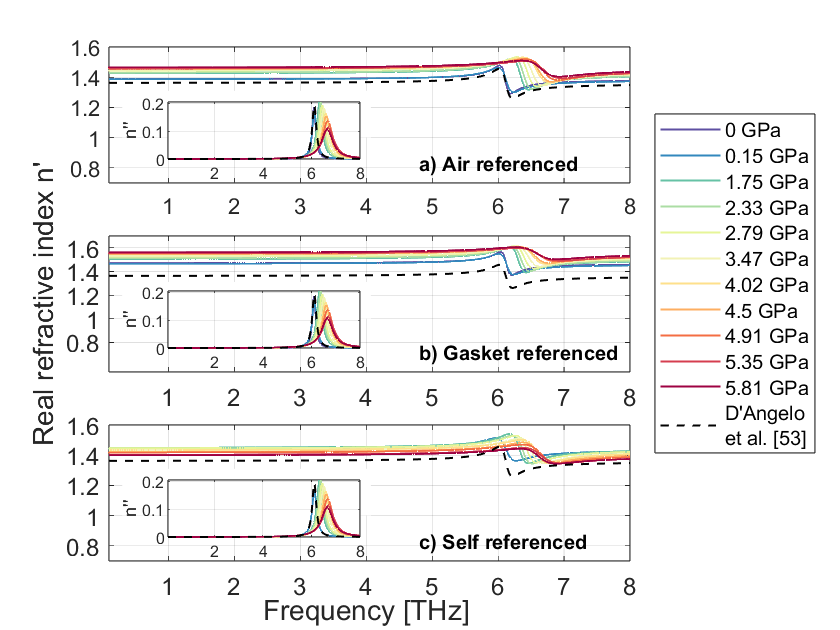}
    \caption{Extracted refractive index of PTFE for all three methods. The imaginary part of the refractive index is shown in the insets. A measurement at ambient conditions by D'Angelo et al. \cite{dangelo-2014} is added as a dashed black line.}
    \label{fig:RefractiveIndex PTFE}
\end{figure}

In contrast to KBr and CsI, PTFE exhibits a weakly absorbing phonon mode at \qty{6.1}{\tera\hertz} \cite{dangelo-2014}, which is noticeable as a drop in the frequency spectrum but still sufficiently weak to enable the use of the self referenced approach. The fitted parameters assuming one Lorentzian oscillator, shown in \mbox{Fig. \ref{fig:FittingParametersPTFE}}, are seen to be mostly insensitive to the chosen extraction method. The main difference between the methods can be found in the values of $\epsilon_\infty$, shown in the inset of \mbox{Fig. \ref{fig:FittingParametersPTFE}}, which increases with pressure for the air referenced and gasket referenced approaches but decreases with increasing pressure for the self-referenced approach.
The extracted refractive index for PTFE as a function of pressure is shown in \mbox{Fig. \ref{fig:RefractiveIndex PTFE}}, with the atmospheric pressure data showing very good agreement with previous reports from D'Angelo et al. \cite{dangelo-2014}, in particular for the air referenced approach.

\begin{figure}[!t]
    \centering
    \includegraphics[width=\linewidth]{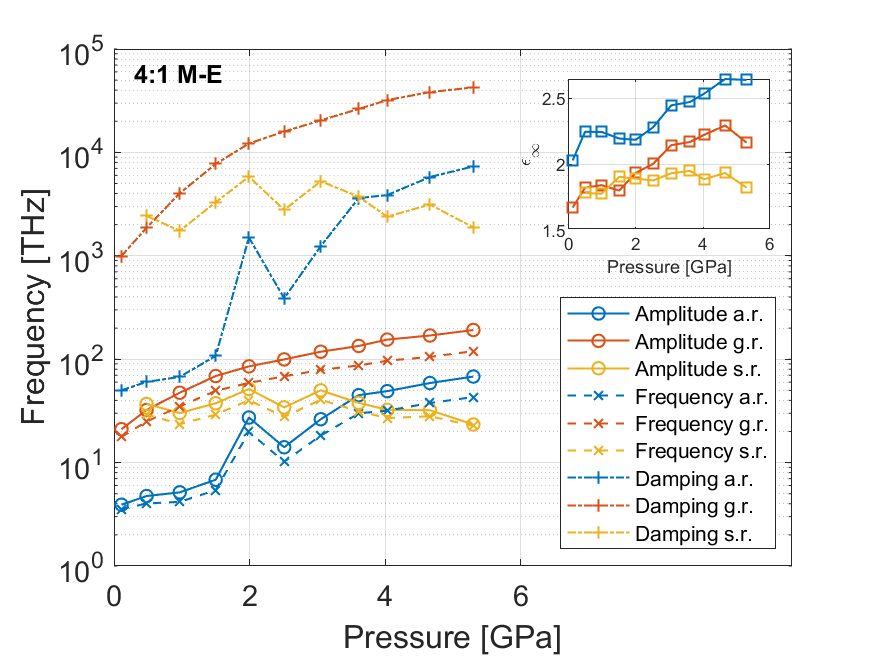}
    \caption{Fitting parameters of the dielectric function with one Lorentzian oscillator in 4:1 methanol-ethanol for the air referenced (blue lines, a.r.), gasket referenced (red lines, g.r.) and self referenced (yellow lines, s.r.) methods. The high-frequency response $\epsilon_\infty$ is shown in the inset.}
    \label{fig:FittingParametersME}
\end{figure}

\begin{figure}[!t]
    \centering
    \includegraphics[width=\linewidth]{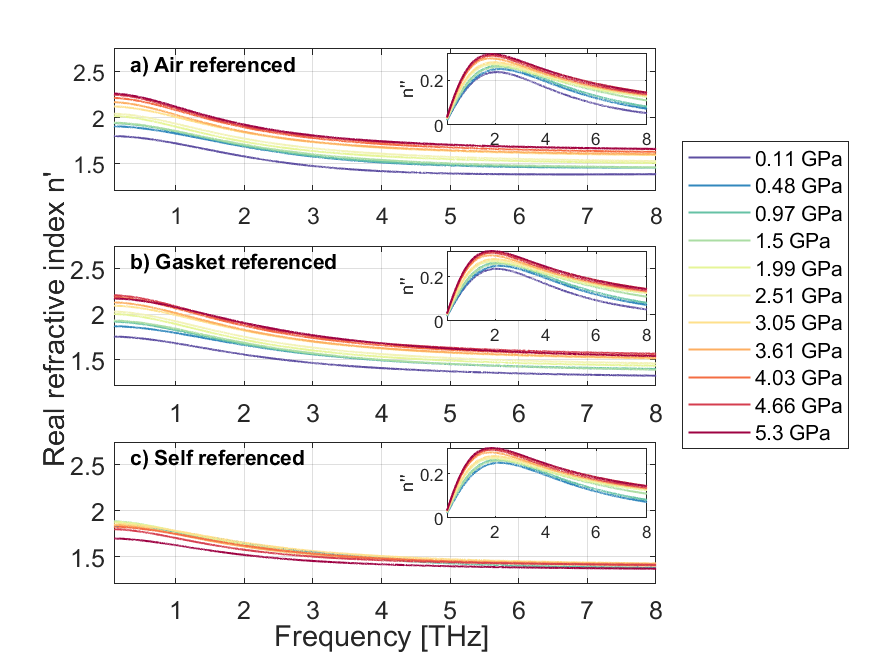}
    \caption{Extracted refractive index of 4:1 methanol-ethanol for all three methods. The imaginary part of the refractive index is shown in the insets.}
    \label{fig:RefIndxME}
\end{figure}

While no literature values are available for the refractive index of 4:1 methanol-ethanol in the THz region, measurements with pure methanol or ethanol show significant absorption in the THz range\cite{yomogida-2010A, yomogida-2010B, zhang-2020}. In line with these previous reports, and even though the 4:1 methanol-ethanol data in \mbox{Figure \ref{fig:pressuremediatransmission}} is relatively featureless, the strong absorption of high frequency spectral components can best be modeled by including one Lorentzian oscillator at high frequencies. This leads to the extracted fitting parameters shown in \mbox{Fig. \ref{fig:FittingParametersME}} and the corresponding refractive index shown in \mbox{Fig. \ref{fig:RefIndxME}}. While the air-referenced and gasket-referenced approaches are somewhat closer in their estimates, the values obtained by the self-referenced approach sometimes differ by over one order of magnitude. The discrepancy between the methods can be partly explained by the absorption being introduced through an oscillator outside the observable spectrum. The refractive indices estimated from the three methods are nevertheless similar within the observable spectrum, due to different fitting parameters compensating each other, with the self-referenced approach providing overall lower values of the real part of $\tilde{n}$.

\begin{figure}[!hbtp]
    \centering
    \includegraphics[width=\linewidth]{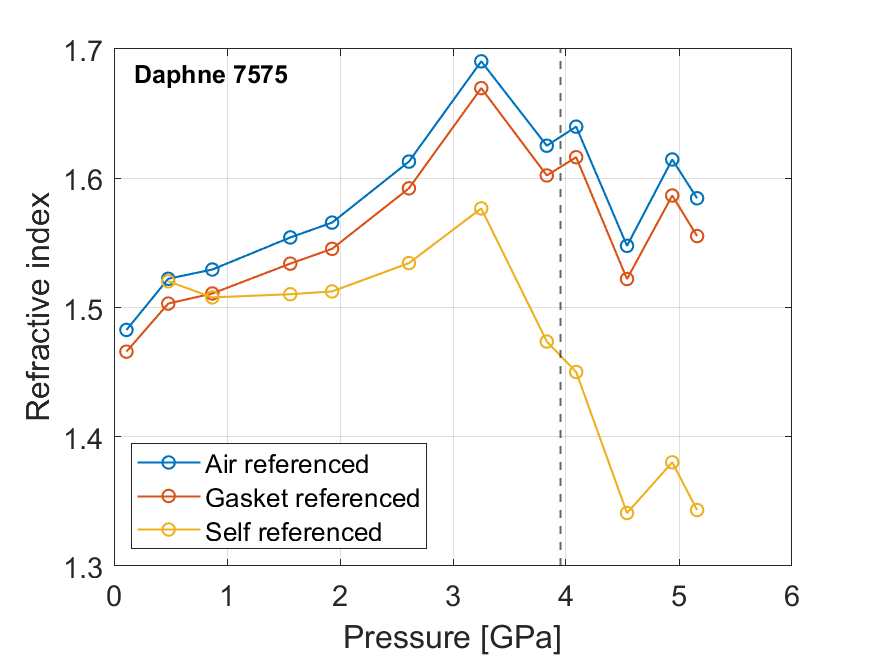}
    \caption{Pressure dependent refractive index (real) of Daphne 7575 for all three methods. The solidification pressure is marked with a dashed line.}
    \label{fig:RefIndexDaphne}
\end{figure}

Due to the featureless transmission of Daphne 7575 over the whole pressure range, the fitting procedure could not be improved significantly when including Lorentzian oscillators. Good agreement with the measured THz spectra could be achieved for all three methods using only $\epsilon_\infty$ as a material parameter. The extracted refractive index for Daphne 7575, which is purely real and equal to $\sqrt{\epsilon_\infty}$, is shown in \mbox{Fig. \ref{fig:RefIndexDaphne}}. The air-referenced and gasket-referenced approaches again show very similar results, with the self-referenced approach deviating at larger pressure values.

\begin{figure}[!hbtp]
    \centering
    \includegraphics[width=\linewidth]{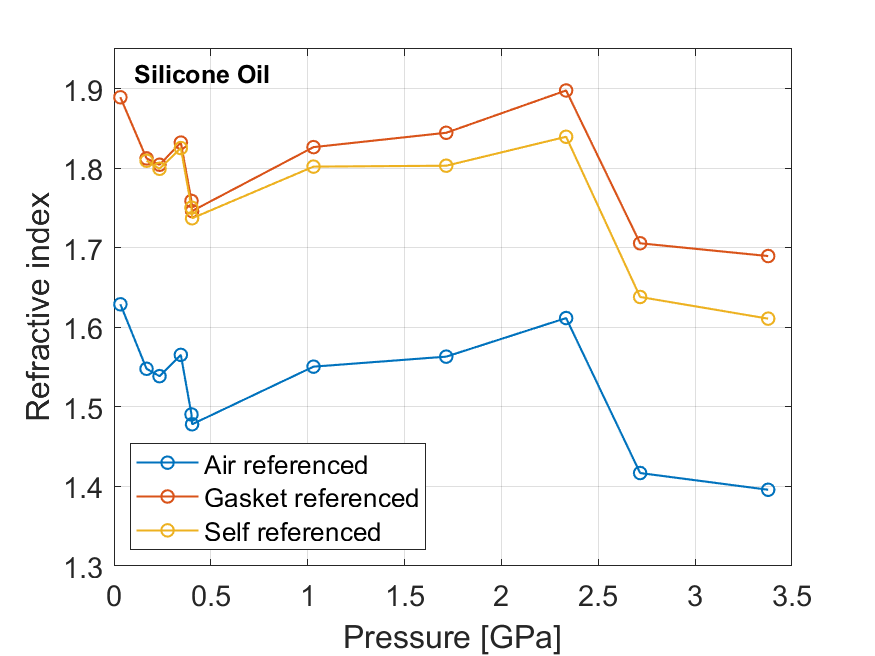}
    \caption{Pressure dependent refractive index (real) of silicone oil for all three methods.}
    \label{fig:RefIndexSiliconeOil}
\end{figure}

 The same frequency-independent approach was taken with silicone oil as with Daphne 7575, for which the extracted refractive index is shown in \mbox{Fig. \ref{fig:RefIndexSiliconeOil}}.

\begin{figure}[!t]
    \centering
    \includegraphics[width=\linewidth]{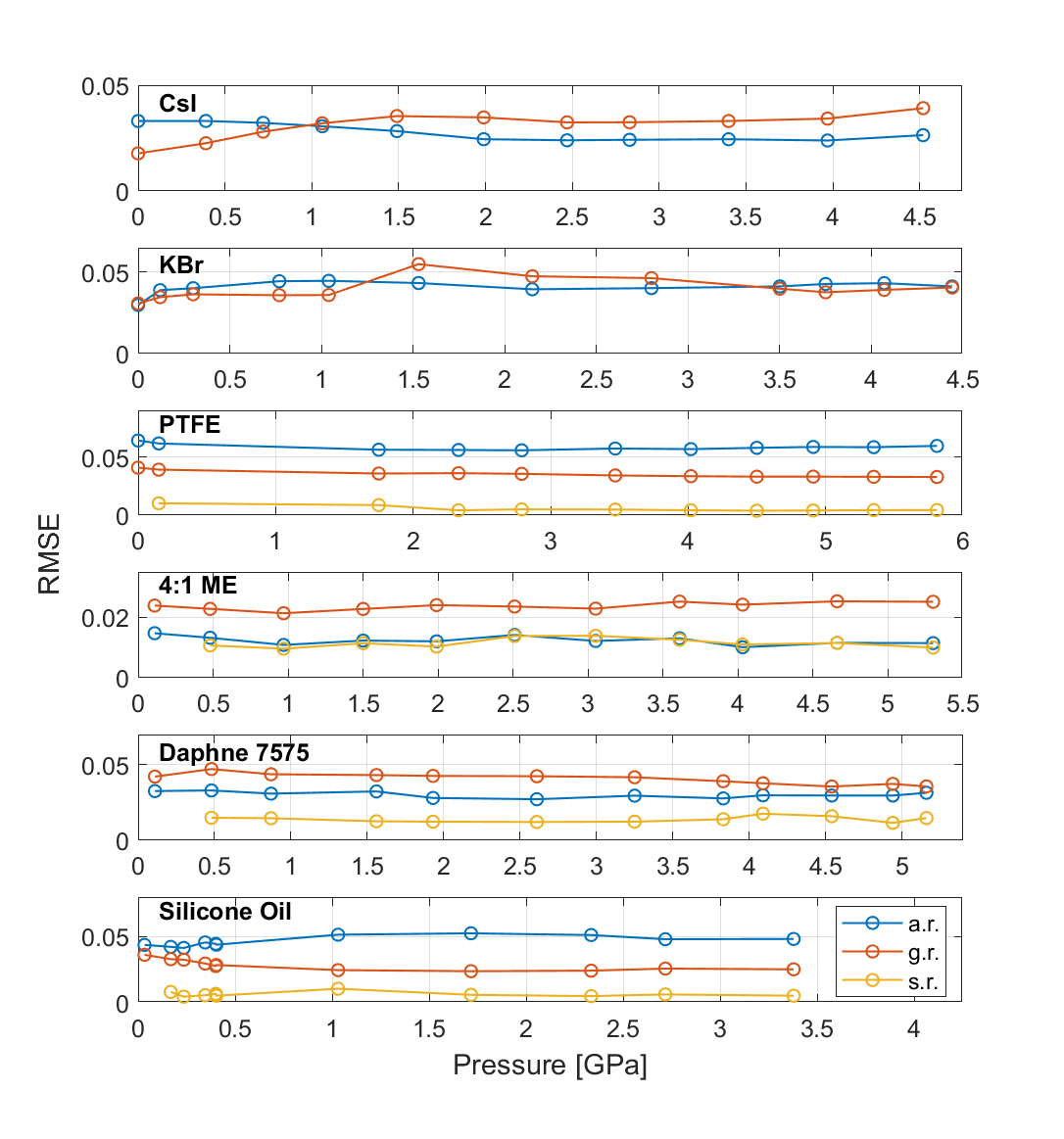}
    \caption{Root mean square error of the air-referenced (blue, a.r.), gasket-referenced (red, g.r.) and self-referenced (yellow, s.r.) parameter extraction methods.}
    \label{fig:RMSE_all_models}
\end{figure}

Previous work on the pressure dependence of material parameters in the THz domain remains very limited. For the materials covered in this work, only the phonon frequency for CsI and KBr had previously been studied as a function of pressure \cite{lowndes-1974,postmus-1968}. Focusing on the general trends exhibited by the pressure dependence of the pressure media, we observe a frequency hardening and amplitude increase of the phonon modes in the three solid materials. The increase in phonon frequency is consistent with the positive Gr\"uneisen parameters exhibited by CsI and KBr, as determined in previous reports \cite{lowndes-1974,postmus-1968}, and by PTFE, as evidenced by the positive thermal expansion coefficient of this material \cite{kirby-1956}.
\\
\textbf{Evaluation of the parameter extraction methods. }
In order to assess the performance of our approach, in particular the relative performance of the three parameter extraction methods, we compare the simulated spectrum to the measured spectrum by computing the root mean square error (RMSE), defined as
\begin{align}
    \textrm{RMSE} = \frac{1}{N} \sqrt{\sum_{i=1}^N|\Tilde{E}_{\mathrm{sam}}(\omega_i) -\Tilde{E}_{\mathrm{sim}}(\omega_i) |^2},
    \label{eq:RSME}
\end{align}
where \mbox{$\omega_1 = 0.5$ THz} and \mbox{$\omega_N = 7.8$ THz}, and using the same notation as in Eq. \ref{eq:5:fittingSD}.
An example of the data used in this calculation of the RSME is shown in \mbox{Fig. \ref{fig:Appdx:PTFE_ModelComp}a} of the appendix. The RMSE obtained  at every pressure value and for every pressure medium is shown \mbox{Fig. \ref{fig:RMSE_all_models}}. 
All three methods yield errors in the range between 0 and 0.05, overall lowest for the case of 4:1 methanol-ethanol. The relative performance of the three methods is pressure independent for all pressure media shown in \mbox{Fig. \ref{fig:RMSE_all_models}}.
The self referenced approach yields the lowest RMSE values in all cases when it can be applied, while the air referenced and gasket referenced methods are overall similar to each other.
More generally, it is important to consider that the different choice of reference results in distinct limitations for each method.
On the one hand, while all pressure dependent measurements in a DAC critically rely on a good THz focus and a stable THz setup, the air referenced approach strongly depends on a good match between the configuration used for the DAC measurements and the one when the aperture transmission (\mbox{Fig. \ref{fig:aperture}}) was measured.
On the other hand, both the gasket referenced and the self referenced methods are very sensitive to strong gasket deformation, since variations in gasket diameter are not taken into account in the parameter extraction procedure, contrary to the air referenced method.
Finally, the self referenced approach relies on a nonzero spectral amplitude throughout the whole relevant spectral range, as mentioned above, as well as on previous knowledge of the material parameters at ambient pressure, to be used as starting values. This method further requires careful evaluation of the fitting performance at every pressure value, since all errors accumulate and propagate to the calculation of material parameters at higher pressure.

\textbf{Bulk silicon.}
We now switch from a DAC homogeneously filled with a pressure medium to the case when the DAC is filled with a bulk sample placed in a pressure medium environment, as in Section \ref{sec:bulk silicon}. While the spectra in \mbox{Fig. \ref{fig:Si_SDTD}b} reveal a monotonic decrease of the DAC transmission with increasing pressure, the time traces in \mbox{Fig. \ref{fig:Si_SDTD}a} provide more detailed information. Peak 1 corresponds to the part of the beam that goes only through the pressure medium around the sample. While the transmission of the pressure medium does not change with pressure, the amplitude of peak 1 decreases in the first \mbox{$\approx$\qty{1}{\giga\pascal}} due to gasket hole contraction (as mentioned in Section \ref{Sec:2:SetupAndTechniques}) and the consequent further clipping of the THz beam. Peak 2 corresponds to the part of the beam that goes through the sample and is essentially insensitive to the gasket hole contraction, so that the monotonic decrease in transmission with increasing pressure is likely related to a pressure induced change in the THz response of silicon.

More formally, we consider the model of a bulk sample placed in the sample chamber along with the pressure medium, as described in \mbox{Section \ref{Sec:6:MaterialParameterExtraction}}, where the contributions of both the bulk sample and the pressure medium (Daphne 7575, in our case) must be considered. 
We calculate the refractive index of Daphne 7575 as the mean of the values obtained from the three extraction methods shown in \mbox{Fig. \ref{fig:RefIndexDaphne}}. Refractive index values at intermediate pressures were obtained by linear interpolation between the measured pressure data. 
The small silicon sample used in the pressure study was obtained from a larger sample, for which a frequency independent, real refractive index of $\tilde{n} = n' = 3.44$ was obtained from THz TDS measurements performed at ambient conditions, outside the DAC.
In our simple model, we assume that the refractive index of silicon remains real and frequency independent also in high-pressure conditions.

\begin{figure}[!t]
    \centering
    \includegraphics[width=\linewidth]{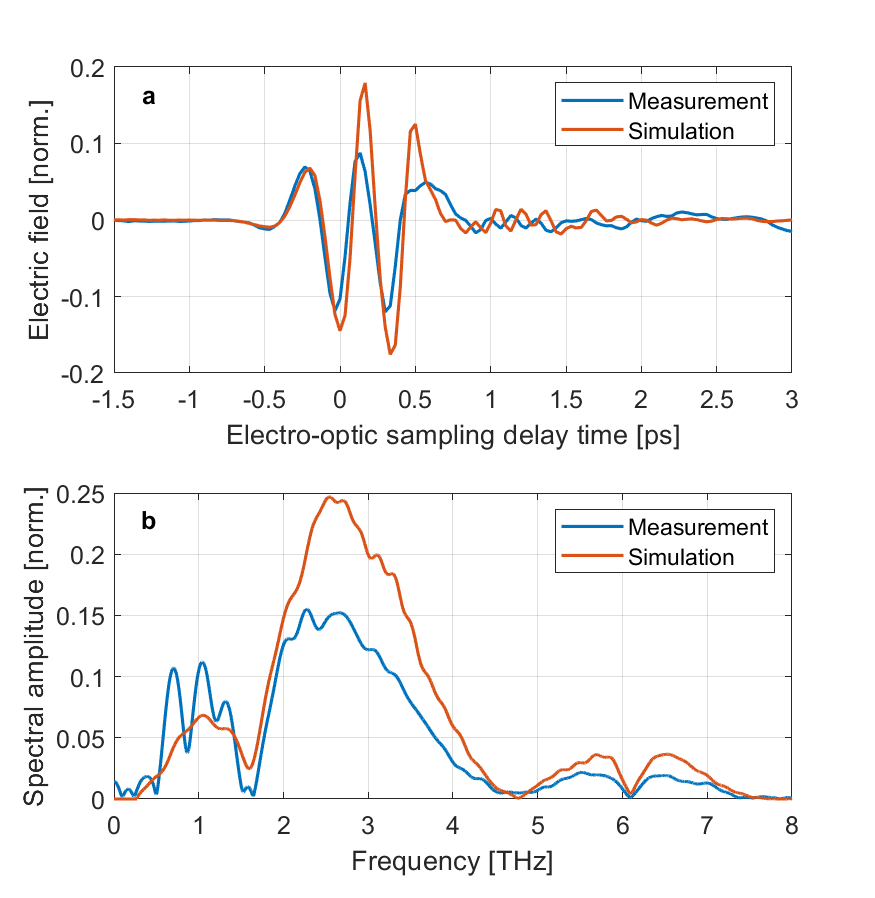}
    \caption{Comparison between the measured and simulated THz electric field transmitted through silicon at \qty{0.35}{\giga\pascal} in (a) the time domain and (b) the frequency domain. In both a and b, the traces are normalized to the peak value of the reference THz measurement in air without the DAC.}
    \label{fig:Si_Fitting}
\end{figure}

\begin{figure}[!hbtp]
    \centering
    \includegraphics[width=\linewidth]{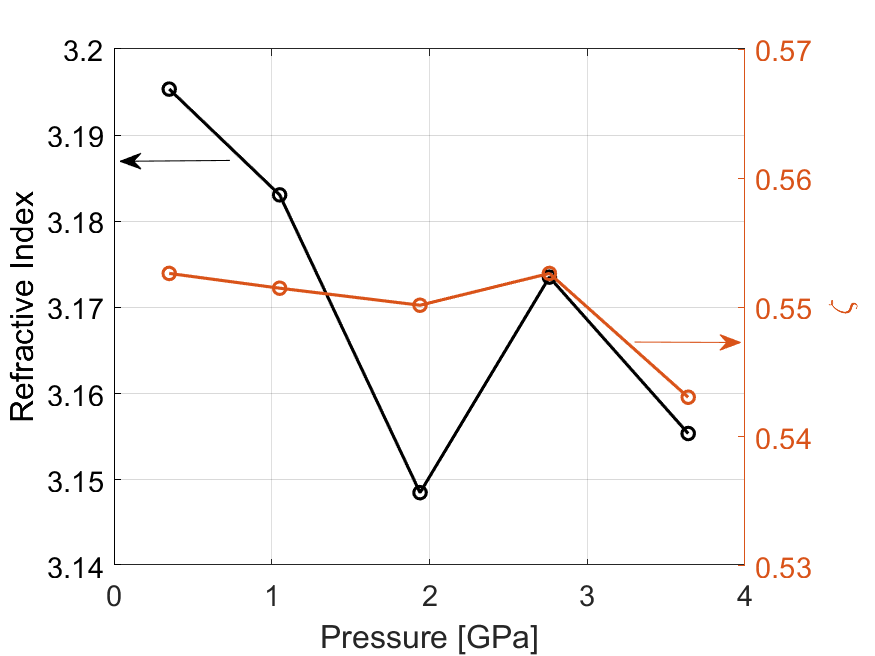}
    \caption{Pressure dependence of the refractive index $n'$ (real) of silicon (black, left y-axis) and the THz field fraction $\zeta$ (red, right y-axis).}
    \label{fig:Si_Parameters}
\end{figure}

The fitting procedure as described in \mbox{Section \ref{Sec:6:MaterialParameterExtraction}}, which extends the air referenced method to include bulk samples, was first carried out on the refractive index and then on the THz field fraction $\zeta$, the two-step fitting approach ensuring the best fit of the position and depth of the features in the spectrum.
A comparison between the measured data and the simulated THz signals for a bulk silicon sample in Daphne 7575 at a pressure of \qty{0.35}{\giga\pascal} is shown in \mbox{Fig. \ref{fig:Si_Fitting}}, for both the time and the frequency domains. The agreement between measurement and simulation is noticeably less accurate than the one obtained for a homogeneously filled sample chamber (cf \mbox{Fig. \ref{fig:Appdx:PTFE_ModelComp}} in the appendix): while the time delay between the two pulses in \mbox{Fig. \ref{fig:Si_Fitting}}a, or equivalently the oscillation in the spectrum in \mbox{Fig. \ref{fig:Si_Fitting}}b, can be fit, a considerable mismatch remains in the amplitudes.
The values obtained for $n'$ and $\zeta$ as a function of pressure are shown in \mbox{Fig. \ref{fig:Si_Parameters}}.
Both remain approximately constant with increasing pressure. Comparing $\zeta \approx 55\%$ with the fraction of the gasket hole area covered by the sample, about 39\% (cf. Section \ref{sec:bulk silicon}), yields a close but imperfect match. It is also noteworthy that $\zeta$ appears to be insensitive to the gasket hole contraction that occurs in the first \mbox{$\approx$\qty{1}{\giga\pascal}}, and which is clearly seen as an amplitude decrease of peak 1 in \mbox{Fig. \ref{fig:Si_SDTD}a}. These observations highlight the limitations of our modeling of the propagation as a mixture of two independent beams.
There are several possible explanations for the discrepancies between our simple model and our measurements, as detailed in the following.
1) Diffraction effects are undoubtedly present during measurements through the DAC, and are further affected by the presence of a bulk sample in the sample chamber, but they are not included in our parameter extraction methods.
2) We define the THz field fraction $\zeta$ as a frequency independent parameter. There are, however, several frequency dependent effects which likely affect $\zeta$, such as the THz beam spot size, the dimensions of the sample chamber, the dimensions and position of the sample in the sample chamber, or diffraction related changes in the propagation of the THz pulse towards the THz detection.
3) The refractive index of silicon was assumed to be real not only at ambient pressure but over the whole pressure range, neglecting any pressure induced increase in absorption which could potentially explain the decrease in THz transmission observed in \mbox{Fig. \ref{fig:Si_Fitting}}.
Regarding the point about the dimensions and position of the sample in 2), a comparison between measurements obtained with different silicon samples shows that $\zeta$, albeit not matching the surface fraction of the sample in the sample chamber, increases with increasing sample area. In a related study we considered the effect of \qty{10}{\micro\metre} thick pieces of aluminum, with different surface areas and shapes, placed at different positions within the sample chamber of the DAC. While the shape and position of these metallic pieces, perfectly opaque to the THz radiation, did not lead to any noticeable change of the transmitted THz field, an increase in sample area was seen to clearly correlate with a decrease in THz transmission, consistent with our results on bulk silicon. 
Having considered all these aspects, it is clear that a description of the THz response of bulk samples under pressure beyond our simple model requires further characterization, which is however beyond the scope of our current work.

Finally, regarding the time resolved measurements on bulk silicon presented in Section \ref{Sec:8:DynamicMeasurements}, we observe a photoinduced decrease of the THz transmission through the material, consistent with previous reports. While a detailed and quantitative analysis of the pressure dependence of the transient response involves better modeling of the material parameters of the sample, as discussed in the previous paragraph, we nevertheless remark that the transient relative change in THz transmission increases with increasing pressure for all but the highest pressure value in \mbox{Fig. \ref{fig:dynamics}}.

\section{Conclusion}\label{Sec:10:Conclusion}

In this work we demonstrate the combination of a THz TDS system with a DAC to study the low frequency response of materials in a high pressure environment and at low temperature, both in equilibrium and following excitation by an \qty{800}{\nano\metre} wavelength pump pulse. We have optimized the DAC parameters to address the challenges posed by integrating it in the THz TDS system, which mainly stem from the fact that the size of the diamond anvils and the gasket hole is comparable to a significant fraction of the wavelengths included in the THz pulse spectrum, and hence to the focused THz spot size.
We characterized the effect of the gasket hole size on the transmitted THz spectrum, settling for a minimum gasket hole diameter of \qty{200}{\micro\metre}, which enables a reasonable transmission over the 0 -- \qty{8}{\tera\hertz} range we are interested in, while allowing for the application of pressures up to \qty{40}{GPa} --- although smaller anvil diameters would be needed to reach that pressure than the ones of the diamonds used here. In addition, the strong dependence of the THz transmission on the gasket hole size means that measurements performed in a DAC are very sensitive to any changes of the gasket hole diameter during the application of pressure.
Given that THz TDS relies on the coherent detection of THz pulses in the time domain, we also consider the effect of reflections from all the interfaces in the THz optical path, discussing how those can be excluded, or instead integrated, into the analysis of the measured data.

We performed a detailed study of the THz transmission and reflection of several materials which are commonly used as pressure media in high pressure experiments, namely solid KBr, CsI and PTFE, and liquid Daphne 7575, 4:1 methanol-ethanol and silicone oil, focusing on the transmitted spectrum, on changes in that spectrum induced by increasing pressure up to \qty{6}{GPa} or by decreasing temperature down to \qty{20}{\kelvin}, as well as on the hydrostaticity of the medium.
Daphne 7575 and PTFE were proven to be ideal pressure media for THz TDS studies. Daphne 7575 exhibits a large transmission in the covered spectral range and excellent hydrostatic conditions up to its solidification pressure, \qty{4}{\giga\pascal} at room temperature. Above \qty{4}{\giga\pascal}, Daphne 7575 becomes less hydrostatic but the THz transmission is unchanged.
The THz transmission of PTFE is mostly pressure independent and limited only by a phonon mode at \qty{6.1}{\tera\hertz} (at ambient pressure). PTFE provides less hydrostatic conditions compared to Daphne 7575 up to \qty{4}{\giga\pascal} due to its solid form at room temperature.
For both Daphne 7575 and PTFE, no major changes in the THz signals were observed when the temperature was decreased to \qty{20}{\kelvin} under a $\approx$\qty{1.6}{\giga\pascal} applied pressure.
A technical advantage of Daphne 7575 over PTFE is the significantly faster loading of a liquid pressure medium compared to a solid one.

Going beyond the case of a homogeneously filled sample chamber, we consider the case where a bulk sample is added to the pressure medium. In order to optimize the measurement of the sample response, the surface fraction covered by the sample should be maximized and a pressure medium with low THz absorption and little to no pressure dependence of the THz transmission should ne used. In our proof of principle experiments, a sample of high resistivity silicon covering 37\% of the gasket hole area is used, with Daphne 7575 as a pressure medium. In addition to the equilibrium THz response as a function of pressure we also performed pressure dependent \qty{800}{\nano\metre} pump -- THz probe measurements, after verifying that no component of the DAC other than the sample contributed to transient changes in the THz transmission following \qty{800}{\nano\metre} photoexcitation.
The THz pulse transmitted through the DAC is temporally split into two, one part going only through pressure medium and one part going through the sample and some pressure medium. This temporal splitting allows us to distinguish the sample response, in both equilibrium and time resolved measurements, but it must be taken into account when analyzing the the data, both in the time and in the frequency domain.

We propose an analysis framework to retrieve complex material parameters from THz TDS measurements under pressure. Even though our THz setup enables measures in both transmission and reflection, we focus our analysis efforts on the transmission configuration. Our approach relies on three methods, each using its own reference --- air referenced, gasket referenced and self referenced --- and each entailing its own assumptions. The three methods are quite successful in modeling the case of a homogeneously filled sample chamber, as happens for the measurements on pressure media. The extracted pressure dependent shifts in phonon frequency for KBr and CsI are in good agreement with previous studies, and we report on the pressure dependence of the refractive indices of PTFE, Daphne 7575, 4:1 methanol-ethanol and silicon oil in the THz range. Regarding mixtures, such as when a bulk silicon sample is added to the pressure medium inside the sample chamber, while our proposed parameter extraction procedure simulates some aspects of the data, further improvements are necessary in order to accurately retrieve the material parameters of silicon or of other bulk material samples.

Our work describes the integration of high pressure and low temperature capabilities in a THz TDS setup, and proposes analysis methods which can be used to recover the pressure dependent complex material parameters of the sample from transmission measurements in the THz range. Analyzing the results of our proof of principle experiments on a bulk silicon sample, we identify two main challenges that should be addressed for this technique to be broadly applicable in the study of bulk material samples: 1) frequency dependent diffraction effects must be quantified and potentially taken into account in the analysis of the data; 2) the data extraction methodology in the case of mixtures must be improved, to go beyond a description of the system based only on relative surface fractions of the components of the mixture. While addressing these challenges goes beyond the scope of the current manuscript, possible directions include efforts to simulate the frequency dependent diffraction and separation of the THz beam through the sample chamber, as well efforts to measure these effects using for example a two-dimensional electro optic sampling technique \cite{wu-1996, usami-2005,blanchard-2022}.

Our work will contribute to extending the study of light-induced dynamics involving THz pulses to materials under pressure. Applying pressure enables a direct and continuous control of the ground state from which a system is excited. This is of interest for ultrafast dynamics experiments in condensed matter in general but in particular for quantum materials, in which rich phase diagrams can be sensitively controlled by external parameters and where some macroscopic properties can only be accessed through the application of external pressure. THz radiation allows us to observe and selectively drive the low energy excitations in such materials, contributing to the understanding of the nature of the different complex phases and of the mechanisms involved in phase transitions.

\begin{acknowledgments}
We would like to thank Nicola Casati for generously providing us a sample of Daphne 7575 and for assisting us in the gasket indentation. We also thank Michele Buzzi, Stefan Klotz, Marine Verseils, Gaston Garbarino, Marco Cammarata and Tomasz Poreba for precious advice and useful discussions. This research was funded by the Swiss National Science Foundation through Ambizione Grant PZ00P2\_179691 and Starting Grant TMSGI2\_211211.

\end{acknowledgments}

\section*{Data Availability Statement}

\begin{center}
\renewcommand\arraystretch{1.2}
\begin{tabular}{| >{\raggedright\arraybackslash}p{0.3\linewidth} | >{\raggedright\arraybackslash}p{0.65\linewidth} |}
\hline
\textbf{AVAILABILITY OF DATA} & \textbf{STATEMENT OF DATA AVAILABILITY}\\  
\hline
Data openly available in a public repository that issues datasets with DOIs
&
The data that support the findings of this study are openly available in the ETH Research Collection at http://doi.org/[doi], reference number [reference number].
\\\hline

\end{tabular}
\end{center}

\appendix

\section{Appendixes}

\begin{figure}[!hbtp]
    \centering
    \includegraphics[width=\linewidth]{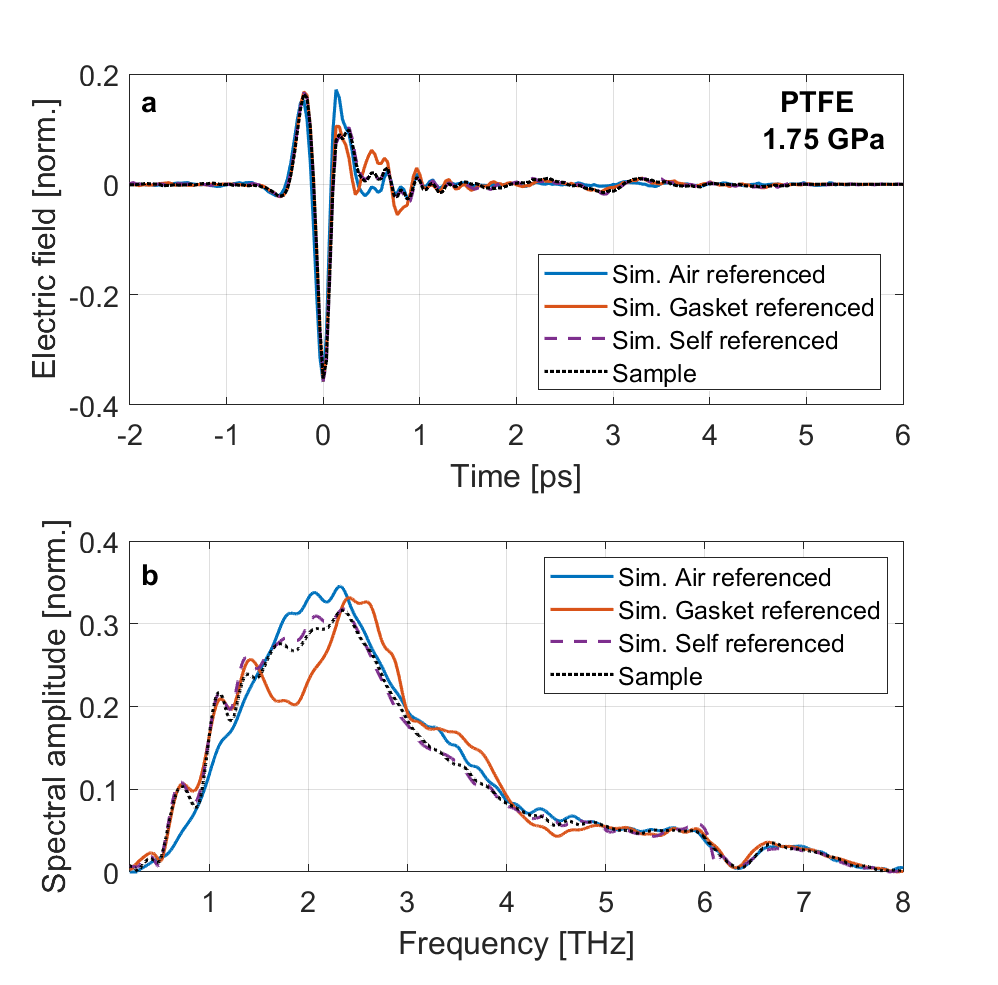}
    \caption{Comparison between measurements and simulations of the THz electric field transmitted through PTFE at \qty{1.75}{\giga\pascal} in (a) the time-domain and (b) the spectral domain, using the three different parameter extraction methods. In both a and b, the traces are normalized to the peak value of the reference THz measurement in air without the DAC.}
    \label{fig:Appdx:PTFE_ModelComp}
\end{figure}

A comparison of the simulated THz signals to the measurements performed in PTFE at a pressure of \qty{1.75}{\giga\pascal} is shown for the time-domain and spectral-domain in \mbox{Fig. \ref{fig:Appdx:PTFE_ModelComp}}.

\nocite{*}
\bibliography{aipsamp}

\providecommand{\noopsort}[1]{}\providecommand{\singleletter}[1]{#1}%
\begin{thebibliography}{57}%
\makeatletter
\providecommand \@ifxundefined [1]{%
 \@ifx{#1\undefined}
}%
\providecommand \@ifnum [1]{%
 \ifnum #1\expandafter \@firstoftwo
 \else \expandafter \@secondoftwo
 \fi
}%
\providecommand \@ifx [1]{%
 \ifx #1\expandafter \@firstoftwo
 \else \expandafter \@secondoftwo
 \fi
}%
\providecommand \natexlab [1]{#1}%
\providecommand \enquote  [1]{``#1''}%
\providecommand \bibnamefont  [1]{#1}%
\providecommand \bibfnamefont [1]{#1}%
\providecommand \citenamefont [1]{#1}%
\providecommand \href@noop [0]{\@secondoftwo}%
\providecommand \href [0]{\begingroup \@sanitize@url \@href}%
\providecommand \@href[1]{\@@startlink{#1}\@@href}%
\providecommand \@@href[1]{\endgroup#1\@@endlink}%
\providecommand \@sanitize@url [0]{\catcode `\\12\catcode `\$12\catcode `\&12\catcode `\#12\catcode `\^12\catcode `\_12\catcode `\%12\relax}%
\providecommand \@@startlink[1]{}%
\providecommand \@@endlink[0]{}%
\providecommand \url  [0]{\begingroup\@sanitize@url \@url }%
\providecommand \@url [1]{\endgroup\@href {#1}{\urlprefix }}%
\providecommand \urlprefix  [0]{URL }%
\providecommand \Eprint [0]{\href }%
\providecommand \doibase [0]{http://dx.doi.org/}%
\providecommand \selectlanguage [0]{\@gobble}%
\providecommand \bibinfo  [0]{\@secondoftwo}%
\providecommand \bibfield  [0]{\@secondoftwo}%
\providecommand \translation [1]{[#1]}%
\providecommand \BibitemOpen [0]{}%
\providecommand \bibitemStop [0]{}%
\providecommand \bibitemNoStop [0]{.\EOS\space}%
\providecommand \EOS [0]{\spacefactor3000\relax}%
\providecommand \BibitemShut  [1]{\csname bibitem#1\endcsname}%
\let\auto@bib@innerbib\@empty
\bibitem [{\citenamefont {Lee}(2012)}]{lee-2012}%
  \BibitemOpen
  \bibfield  {author} {\bibinfo {author} {\bibfnamefont {Y.-S.}\ \bibnamefont {Lee}},\ }\bibfield  {title} {\enquote {\bibinfo {title} {{Principles of Terahertz Science and Technology}},}\ }\href {\doibase 10.1080/00107514.2012.737849} {\bibfield  {journal} {\bibinfo  {journal} {Contemporary Physics}\ }\textbf {\bibinfo {volume} {53}},\ \bibinfo {pages} {526--527} (\bibinfo {year} {2012})}\BibitemShut {NoStop}%
\bibitem [{\citenamefont {Koch}\ \emph {et~al.}(2023)\citenamefont {Koch}, \citenamefont {Mittleman}, \citenamefont {Ornik},\ and\ \citenamefont {Castro-Camus}}]{koch-2023}%
  \BibitemOpen
  \bibfield  {author} {\bibinfo {author} {\bibfnamefont {M.}~\bibnamefont {Koch}}, \bibinfo {author} {\bibfnamefont {D.~M.}\ \bibnamefont {Mittleman}}, \bibinfo {author} {\bibfnamefont {J.}~\bibnamefont {Ornik}}, \ and\ \bibinfo {author} {\bibfnamefont {E.}~\bibnamefont {Castro-Camus}},\ }\bibfield  {title} {\enquote {\bibinfo {title} {{Terahertz time-domain spectroscopy}},}\ }\href {\doibase 10.1038/s43586-023-00232-z} {\bibfield  {journal} {\bibinfo  {journal} {Nature Reviews Methods Primers}\ }\textbf {\bibinfo {volume} {3}} (\bibinfo {year} {2023}),\ 10.1038/s43586-023-00232-z}\BibitemShut {NoStop}%
\bibitem [{\citenamefont {Jepsen}, \citenamefont {Cooke},\ and\ \citenamefont {Koch}(2010)}]{jepsen-2010}%
  \BibitemOpen
  \bibfield  {author} {\bibinfo {author} {\bibfnamefont {P.~U.}\ \bibnamefont {Jepsen}}, \bibinfo {author} {\bibfnamefont {D.~G.}\ \bibnamefont {Cooke}}, \ and\ \bibinfo {author} {\bibfnamefont {M.}~\bibnamefont {Koch}},\ }\bibfield  {title} {\enquote {\bibinfo {title} {{Terahertz spectroscopy and imaging – Modern techniques and applications}},}\ }\href {\doibase 10.1002/lpor.201000011} {\bibfield  {journal} {\bibinfo  {journal} {Laser $\&$ Photonics Reviews}\ }\textbf {\bibinfo {volume} {5}},\ \bibinfo {pages} {124--166} (\bibinfo {year} {2010})}\BibitemShut {NoStop}%
\bibitem [{\citenamefont {Smith}, \citenamefont {Auston},\ and\ \citenamefont {Nuss}(1988)}]{smith-1988}%
  \BibitemOpen
  \bibfield  {author} {\bibinfo {author} {\bibfnamefont {P.~R.}\ \bibnamefont {Smith}}, \bibinfo {author} {\bibfnamefont {D.~H.}\ \bibnamefont {Auston}}, \ and\ \bibinfo {author} {\bibfnamefont {M.~C.}\ \bibnamefont {Nuss}},\ }\bibfield  {title} {\enquote {\bibinfo {title} {{Subpicosecond photoconducting dipole antennas}},}\ }\href {\doibase 10.1109/3.121} {\bibfield  {journal} {\bibinfo  {journal} {IEEE Journal of Quantum Electronics}\ }\textbf {\bibinfo {volume} {24}},\ \bibinfo {pages} {255--260} (\bibinfo {year} {1988})}\BibitemShut {NoStop}%
\bibitem [{\citenamefont {Van~Exter}, \citenamefont {Fattinger},\ and\ \citenamefont {Grischkowsky}(1989)}]{van-exter-1989}%
  \BibitemOpen
  \bibfield  {author} {\bibinfo {author} {\bibfnamefont {M.}~\bibnamefont {Van~Exter}}, \bibinfo {author} {\bibfnamefont {C.}~\bibnamefont {Fattinger}}, \ and\ \bibinfo {author} {\bibfnamefont {D.}~\bibnamefont {Grischkowsky}},\ }\bibfield  {title} {\enquote {\bibinfo {title} {{Terahertz time-domain spectroscopy of water vapor}},}\ }\href {\doibase 10.1364/ol.14.001128} {\bibfield  {journal} {\bibinfo  {journal} {Optics Letters}\ }\textbf {\bibinfo {volume} {14}},\ \bibinfo {pages} {1128} (\bibinfo {year} {1989})}\BibitemShut {NoStop}%
\bibitem [{\citenamefont {Tonouchi}(2007)}]{tonouchi-2007}%
  \BibitemOpen
  \bibfield  {author} {\bibinfo {author} {\bibfnamefont {M.}~\bibnamefont {Tonouchi}},\ }\bibfield  {title} {\enquote {\bibinfo {title} {{Cutting-edge terahertz technology}},}\ }\href {\doibase 10.1038/nphoton.2007.3} {\bibfield  {journal} {\bibinfo  {journal} {Nature Photonics}\ }\textbf {\bibinfo {volume} {1}},\ \bibinfo {pages} {97--105} (\bibinfo {year} {2007})}\BibitemShut {NoStop}%
\bibitem [{\citenamefont {Cantaluppi}\ \emph {et~al.}(2018)\citenamefont {Cantaluppi}, \citenamefont {Buzzi}, \citenamefont {Jotzu}, \citenamefont {Nicoletti}, \citenamefont {Mitrano}, \citenamefont {Pontiroli}, \citenamefont {Riccò}, \citenamefont {Perucchi}, \citenamefont {Di~Pietro},\ and\ \citenamefont {Cavalleri}}]{cantaluppi-2018}%
  \BibitemOpen
  \bibfield  {author} {\bibinfo {author} {\bibfnamefont {A.}~\bibnamefont {Cantaluppi}}, \bibinfo {author} {\bibfnamefont {M.}~\bibnamefont {Buzzi}}, \bibinfo {author} {\bibfnamefont {G.}~\bibnamefont {Jotzu}}, \bibinfo {author} {\bibfnamefont {D.}~\bibnamefont {Nicoletti}}, \bibinfo {author} {\bibfnamefont {M.}~\bibnamefont {Mitrano}}, \bibinfo {author} {\bibfnamefont {D.}~\bibnamefont {Pontiroli}}, \bibinfo {author} {\bibfnamefont {M.}~\bibnamefont {Riccò}}, \bibinfo {author} {\bibfnamefont {A.}~\bibnamefont {Perucchi}}, \bibinfo {author} {\bibfnamefont {P.}~\bibnamefont {Di~Pietro}}, \ and\ \bibinfo {author} {\bibfnamefont {A.}~\bibnamefont {Cavalleri}},\ }\bibfield  {title} {\enquote {\bibinfo {title} {{Pressure tuning of light-induced superconductivity in K3C60}},}\ }\href {\doibase 10.1038/s41567-018-0134-8} {\bibfield  {journal} {\bibinfo  {journal} {Nature Physics}\ }\textbf {\bibinfo {volume} {14}},\ \bibinfo {pages} {837--841} (\bibinfo {year} {2018})}\BibitemShut {NoStop}%
\bibitem [{\citenamefont {Xu}\ \emph {et~al.}(2021)\citenamefont {Xu}, \citenamefont {Huang}, \citenamefont {Liu}, \citenamefont {Zhang}, \citenamefont {Jiang}, \citenamefont {Gou}, \citenamefont {Zeng}, \citenamefont {Wang},\ and\ \citenamefont {Su}}]{xu-2021}%
  \BibitemOpen
  \bibfield  {author} {\bibinfo {author} {\bibfnamefont {S.}~\bibnamefont {Xu}}, \bibinfo {author} {\bibfnamefont {D.}~\bibnamefont {Huang}}, \bibinfo {author} {\bibfnamefont {Z.}~\bibnamefont {Liu}}, \bibinfo {author} {\bibfnamefont {K.}~\bibnamefont {Zhang}}, \bibinfo {author} {\bibfnamefont {H.}~\bibnamefont {Jiang}}, \bibinfo {author} {\bibfnamefont {H.}~\bibnamefont {Gou}}, \bibinfo {author} {\bibfnamefont {Z.}~\bibnamefont {Zeng}}, \bibinfo {author} {\bibfnamefont {T.}~\bibnamefont {Wang}}, \ and\ \bibinfo {author} {\bibfnamefont {F.}~\bibnamefont {Su}},\ }\bibfield  {title} {\enquote {\bibinfo {title} {{Hydrostatic pressure effect of photocarrier dynamics in GaAs probed by time-resolved terahertz spectroscopy}},}\ }\href {\doibase 10.1364/oe.421011} {\bibfield  {journal} {\bibinfo  {journal} {Optics Express}\ }\textbf {\bibinfo {volume} {29}},\ \bibinfo {pages} {14058} (\bibinfo {year} {2021})}\BibitemShut {NoStop}%
\bibitem [{\citenamefont {Wang}\ \emph {et~al.}(2023)\citenamefont {Wang}, \citenamefont {Xu}, \citenamefont {Yang},\ and\ \citenamefont {Su}}]{wang-2023}%
  \BibitemOpen
  \bibfield  {author} {\bibinfo {author} {\bibfnamefont {Y.}~\bibnamefont {Wang}}, \bibinfo {author} {\bibfnamefont {S.}~\bibnamefont {Xu}}, \bibinfo {author} {\bibfnamefont {J.}~\bibnamefont {Yang}}, \ and\ \bibinfo {author} {\bibfnamefont {F.}~\bibnamefont {Su}},\ }\bibfield  {title} {\enquote {\bibinfo {title} {{Probing photocarrier dynamics of pressurized graphene using time-resolved terahertz spectroscopy}},}\ }\href {\doibase 10.1088/1674-1056/acc2b1} {\bibfield  {journal} {\bibinfo  {journal} {Chinese Physics B}\ }\textbf {\bibinfo {volume} {32}},\ \bibinfo {pages} {067802} (\bibinfo {year} {2023})}\BibitemShut {NoStop}%
\bibitem [{\citenamefont {Guidi}\ \emph {et~al.}(2004)\citenamefont {Guidi}, \citenamefont {Nucara}, \citenamefont {Calvani}, \citenamefont {Postorino}, \citenamefont {Sacchetti}, \citenamefont {Congeduti}, \citenamefont {Piccinini}, \citenamefont {Marcelli},\ and\ \citenamefont {Burattini}}]{guidi-2004}%
  \BibitemOpen
  \bibfield  {author} {\bibinfo {author} {\bibfnamefont {M.~C.}\ \bibnamefont {Guidi}}, \bibinfo {author} {\bibfnamefont {A.}~\bibnamefont {Nucara}}, \bibinfo {author} {\bibfnamefont {P.}~\bibnamefont {Calvani}}, \bibinfo {author} {\bibfnamefont {P.}~\bibnamefont {Postorino}}, \bibinfo {author} {\bibfnamefont {A.}~\bibnamefont {Sacchetti}}, \bibinfo {author} {\bibfnamefont {A.}~\bibnamefont {Congeduti}}, \bibinfo {author} {\bibfnamefont {M.}~\bibnamefont {Piccinini}}, \bibinfo {author} {\bibfnamefont {A.}~\bibnamefont {Marcelli}}, \ and\ \bibinfo {author} {\bibfnamefont {E.}~\bibnamefont {Burattini}},\ }\bibfield  {title} {\enquote {\bibinfo {title} {{High-pressure far-infrared measurements at SINBAD}},}\ }\href {\doibase 10.1016/j.infrared.2004.01.011} {\bibfield  {journal} {\bibinfo  {journal} {Infrared Physics \& Technology}\ }\textbf {\bibinfo {volume} {45}},\ \bibinfo {pages} {365--368} (\bibinfo {year} {2004})}\BibitemShut {NoStop}%
\bibitem [{\citenamefont {Kimura}\ \emph {et~al.}(2010)\citenamefont {Kimura}, \citenamefont {Mizuno}, \citenamefont {Iizuka}, \citenamefont {Predoi-Cross},\ and\ \citenamefont {Billinghurst}}]{kimura-2010}%
  \BibitemOpen
  \bibfield  {author} {\bibinfo {author} {\bibfnamefont {S.-I.}\ \bibnamefont {Kimura}}, \bibinfo {author} {\bibfnamefont {T.}~\bibnamefont {Mizuno}}, \bibinfo {author} {\bibfnamefont {T.}~\bibnamefont {Iizuka}}, \bibinfo {author} {\bibfnamefont {A.}~\bibnamefont {Predoi-Cross}}, \ and\ \bibinfo {author} {\bibfnamefont {B.~E.}\ \bibnamefont {Billinghurst}},\ }\bibfield  {title} {\enquote {\bibinfo {title} {{Synchrotron Terahertz Spectroscopy of Solids under Extreme Conditions}},}\ }\href {\doibase 10.1063/1.3326353} {\bibfield  {journal} {\bibinfo  {journal} {AIP conference proceedings}\ ,\ \bibinfo {pages} {71--74}} (\bibinfo {year} {2010})}\BibitemShut {NoStop}%
\bibitem [{\citenamefont {Kimura}\ and\ \citenamefont {Okamura}(2013)}]{kimura-2013}%
  \BibitemOpen
  \bibfield  {author} {\bibinfo {author} {\bibfnamefont {S.-I.}\ \bibnamefont {Kimura}}\ and\ \bibinfo {author} {\bibfnamefont {H.}~\bibnamefont {Okamura}},\ }\bibfield  {title} {\enquote {\bibinfo {title} {{Infrared and Terahertz Spectroscopy of Strongly Correlated Electron Systems under Extreme Conditions}},}\ }\href {\doibase 10.7566/jpsj.82.021004} {\bibfield  {journal} {\bibinfo  {journal} {Journal of the Physical Society of Japan}\ }\textbf {\bibinfo {volume} {82}},\ \bibinfo {pages} {021004} (\bibinfo {year} {2013})}\BibitemShut {NoStop}%
\bibitem [{\citenamefont {Voute}\ \emph {et~al.}(2016)\citenamefont {Voute}, \citenamefont {Deutsch}, \citenamefont {Kalinko}, \citenamefont {Alabarse}, \citenamefont {Brubach}, \citenamefont {Capitani}, \citenamefont {Chapuis}, \citenamefont {Phuoc}, \citenamefont {Sopracase},\ and\ \citenamefont {Roy}}]{voute-2016}%
  \BibitemOpen
  \bibfield  {author} {\bibinfo {author} {\bibfnamefont {A.}~\bibnamefont {Voute}}, \bibinfo {author} {\bibfnamefont {M.}~\bibnamefont {Deutsch}}, \bibinfo {author} {\bibfnamefont {A.}~\bibnamefont {Kalinko}}, \bibinfo {author} {\bibfnamefont {F.}~\bibnamefont {Alabarse}}, \bibinfo {author} {\bibfnamefont {J.-b.}\ \bibnamefont {Brubach}}, \bibinfo {author} {\bibfnamefont {F.}~\bibnamefont {Capitani}}, \bibinfo {author} {\bibfnamefont {M.}~\bibnamefont {Chapuis}}, \bibinfo {author} {\bibfnamefont {V.~T.}\ \bibnamefont {Phuoc}}, \bibinfo {author} {\bibfnamefont {R.}~\bibnamefont {Sopracase}}, \ and\ \bibinfo {author} {\bibfnamefont {P.}~\bibnamefont {Roy}},\ }\bibfield  {title} {\enquote {\bibinfo {title} {{New high-pressure/low-temperature set-up available at the AILES beamline}},}\ }\href {\doibase 10.1016/j.vibspec.2016.05.007} {\bibfield  {journal} {\bibinfo  {journal} {Vibrational Spectroscopy}\ }\textbf {\bibinfo {volume} {86}},\ \bibinfo {pages} {17--23} (\bibinfo {year} {2016})}\BibitemShut {NoStop}%
\bibitem [{\citenamefont {Wenzel}\ \emph {et~al.}(2023)\citenamefont {Wenzel}, \citenamefont {Tsirlin}, \citenamefont {Capitani}, \citenamefont {Chan}, \citenamefont {Ortiz}, \citenamefont {Wilson}, \citenamefont {Dressel},\ and\ \citenamefont {Uykur}}]{wenzel-2023}%
  \BibitemOpen
  \bibfield  {author} {\bibinfo {author} {\bibfnamefont {M.}~\bibnamefont {Wenzel}}, \bibinfo {author} {\bibfnamefont {A.~A.}\ \bibnamefont {Tsirlin}}, \bibinfo {author} {\bibfnamefont {F.}~\bibnamefont {Capitani}}, \bibinfo {author} {\bibfnamefont {Y.~T.}\ \bibnamefont {Chan}}, \bibinfo {author} {\bibfnamefont {B.~R.}\ \bibnamefont {Ortiz}}, \bibinfo {author} {\bibfnamefont {S.~D.}\ \bibnamefont {Wilson}}, \bibinfo {author} {\bibfnamefont {M.}~\bibnamefont {Dressel}}, \ and\ \bibinfo {author} {\bibfnamefont {E.}~\bibnamefont {Uykur}},\ }\bibfield  {title} {\enquote {\bibinfo {title} {{Pressure evolution of electron dynamics in the superconducting kagome metal CsV3Sb5}},}\ }\href {\doibase 10.1038/s41535-023-00577-4} {\bibfield  {journal} {\bibinfo  {journal} {npj Quantum Materials}\ }\textbf {\bibinfo {volume} {8}} (\bibinfo {year} {2023}),\ 10.1038/s41535-023-00577-4}\BibitemShut {NoStop}%
\bibitem [{\citenamefont {Varma}\ \emph {et~al.}(2023)\citenamefont {Varma}, \citenamefont {Krottenmüller}, \citenamefont {Poswal},\ and\ \citenamefont {Kuntscher}}]{varma-2023}%
  \BibitemOpen
  \bibfield  {author} {\bibinfo {author} {\bibfnamefont {M.}~\bibnamefont {Varma}}, \bibinfo {author} {\bibfnamefont {M.}~\bibnamefont {Krottenmüller}}, \bibinfo {author} {\bibfnamefont {H.~K.}\ \bibnamefont {Poswal}}, \ and\ \bibinfo {author} {\bibfnamefont {C.~A.}\ \bibnamefont {Kuntscher}},\ }\bibfield  {title} {\enquote {\bibinfo {title} {{Pressure-Induced Structural Phase Transitions in the Chromium Spinel LiInCr4O8 with Breathing Pyrochlore Lattice}},}\ }\href {\doibase 10.3390/cryst13020170} {\bibfield  {journal} {\bibinfo  {journal} {Crystals}\ }\textbf {\bibinfo {volume} {13}},\ \bibinfo {pages} {170} (\bibinfo {year} {2023})}\BibitemShut {NoStop}%
\bibitem [{\citenamefont {Weir}\ \emph {et~al.}(1959)\citenamefont {Weir}, \citenamefont {Lippincott}, \citenamefont {Van~Valkenburg},\ and\ \citenamefont {Bunting}}]{weir-1959}%
  \BibitemOpen
  \bibfield  {author} {\bibinfo {author} {\bibfnamefont {C.}~\bibnamefont {Weir}}, \bibinfo {author} {\bibfnamefont {E.}~\bibnamefont {Lippincott}}, \bibinfo {author} {\bibfnamefont {A.}~\bibnamefont {Van~Valkenburg}}, \ and\ \bibinfo {author} {\bibfnamefont {E.}~\bibnamefont {Bunting}},\ }\bibfield  {title} {\enquote {\bibinfo {title} {{Infrared Studies In the 1- to I5-Micron Region to 30,000 Atmospheres}},}\ }\href {https://nvlpubs.nist.gov/nistpubs/jres/63A/jresv63An1p55_A1b.pdf} {\bibfield  {journal} {\bibinfo  {journal} {Journa l of Research of the National Bureau of Standards- A. Physics and Chemistry}\ }\textbf {\bibinfo {volume} {63A}} (\bibinfo {year} {1959})}\BibitemShut {NoStop}%
\bibitem [{\citenamefont {Jamieson}, \citenamefont {Lawson},\ and\ \citenamefont {Nachtrieb}(1959)}]{jamieson-1959}%
  \BibitemOpen
  \bibfield  {author} {\bibinfo {author} {\bibfnamefont {J.~C.}\ \bibnamefont {Jamieson}}, \bibinfo {author} {\bibfnamefont {A.~W.}\ \bibnamefont {Lawson}}, \ and\ \bibinfo {author} {\bibfnamefont {N.~D.}\ \bibnamefont {Nachtrieb}},\ }\bibfield  {title} {\enquote {\bibinfo {title} {{New Device for Obtaining X-Ray Diffraction Patterns from Substances Exposed to High Pressure}},}\ }\href {\doibase 10.1063/1.1716408} {\bibfield  {journal} {\bibinfo  {journal} {Review of Scientific Instruments}\ }\textbf {\bibinfo {volume} {30}},\ \bibinfo {pages} {1016--1019} (\bibinfo {year} {1959})}\BibitemShut {NoStop}%
\bibitem [{\citenamefont {Jayaraman}(1983)}]{jayaraman-1983}%
  \BibitemOpen
  \bibfield  {author} {\bibinfo {author} {\bibfnamefont {A.}~\bibnamefont {Jayaraman}},\ }\bibfield  {title} {\enquote {\bibinfo {title} {{Diamond anvil cell and high-pressure physical investigations}},}\ }\href {\doibase 10.1103/revmodphys.55.65} {\bibfield  {journal} {\bibinfo  {journal} {Reviews of Modern Physics}\ }\textbf {\bibinfo {volume} {55}},\ \bibinfo {pages} {65--108} (\bibinfo {year} {1983})}\BibitemShut {NoStop}%
\bibitem [{alm()}]{almax-easylab-2023}%
  \BibitemOpen
  \href@noop {} {\enquote {\bibinfo {title} {{Diacell® CryoDAC-Nitro - Almax EasyLab}},}\ }\bibinfo {howpublished} {\url{https://almax-easylab.com/product/diacell-cryodac-nitro/}},\ \bibinfo {note} {accessed: 2025-04-25}\BibitemShut {NoStop}%
\bibitem [{lak()}]{lakeshore-no-date}%
  \BibitemOpen
  \href {https://www.lakeshore.com/products/product-detail/janis/st-100-cryostats} {\enquote {\bibinfo {title} {{Lakeshore ST-100 Series cryostats}},}\ }\bibinfo {howpublished} {\url{https://www.lakeshore.com/products/product-detail/janis/st-100-cryostats}},\ \bibinfo {note} {accessed: 2024-04-25}\BibitemShut {NoStop}%
\bibitem [{\citenamefont {Dunstan}(1989)}]{dunstan-1989}%
  \BibitemOpen
  \bibfield  {author} {\bibinfo {author} {\bibfnamefont {D.~J.}\ \bibnamefont {Dunstan}},\ }\bibfield  {title} {\enquote {\bibinfo {title} {{Theory of the gasket in diamond anvil high-pressure cells}},}\ }\href {\doibase 10.1063/1.1140442} {\bibfield  {journal} {\bibinfo  {journal} {Review of Scientific Instruments}\ }\textbf {\bibinfo {volume} {60}},\ \bibinfo {pages} {3789--3795} (\bibinfo {year} {1989})}\BibitemShut {NoStop}%
\bibitem [{\citenamefont {Spain}\ and\ \citenamefont {Dunstan}(1989)}]{spain-1989}%
  \BibitemOpen
  \bibfield  {author} {\bibinfo {author} {\bibfnamefont {N.~I.~L.}\ \bibnamefont {Spain}}\ and\ \bibinfo {author} {\bibfnamefont {N.~D.~J.}\ \bibnamefont {Dunstan}},\ }\bibfield  {title} {\enquote {\bibinfo {title} {{The technology of diamond anvil high-pressure cells: II. Operation and use}},}\ }\href {\doibase 10.1088/0022-3735/22/11/005} {\bibfield  {journal} {\bibinfo  {journal} {Journal of Physics E Scientific Instruments}\ }\textbf {\bibinfo {volume} {22}},\ \bibinfo {pages} {923--933} (\bibinfo {year} {1989})}\BibitemShut {NoStop}%
\bibitem [{\citenamefont {Seifert}\ \emph {et~al.}(2016)\citenamefont {Seifert}, \citenamefont {Jaiswal}, \citenamefont {Martens}, \citenamefont {Hannegan}, \citenamefont {Braun}, \citenamefont {Maldonado}, \citenamefont {Freimuth}, \citenamefont {Kronenberg}, \citenamefont {Henrizi}, \citenamefont {Radu}, \citenamefont {Beaurepaire}, \citenamefont {Mokrousov}, \citenamefont {Oppeneer}, \citenamefont {Jourdan}, \citenamefont {Jakob}, \citenamefont {Turchinovich}, \citenamefont {Hayden}, \citenamefont {Wolf}, \citenamefont {Münzenberg}, \citenamefont {Kläui},\ and\ \citenamefont {Kampfrath}}]{seifert-2016}%
  \BibitemOpen
  \bibfield  {author} {\bibinfo {author} {\bibfnamefont {T.}~\bibnamefont {Seifert}}, \bibinfo {author} {\bibfnamefont {S.}~\bibnamefont {Jaiswal}}, \bibinfo {author} {\bibfnamefont {U.}~\bibnamefont {Martens}}, \bibinfo {author} {\bibfnamefont {J.}~\bibnamefont {Hannegan}}, \bibinfo {author} {\bibfnamefont {L.}~\bibnamefont {Braun}}, \bibinfo {author} {\bibfnamefont {P.}~\bibnamefont {Maldonado}}, \bibinfo {author} {\bibfnamefont {F.}~\bibnamefont {Freimuth}}, \bibinfo {author} {\bibfnamefont {A.}~\bibnamefont {Kronenberg}}, \bibinfo {author} {\bibfnamefont {J.}~\bibnamefont {Henrizi}}, \bibinfo {author} {\bibfnamefont {I.}~\bibnamefont {Radu}}, \bibinfo {author} {\bibfnamefont {E.}~\bibnamefont {Beaurepaire}}, \bibinfo {author} {\bibfnamefont {Y.}~\bibnamefont {Mokrousov}}, \bibinfo {author} {\bibfnamefont {P.~M.}\ \bibnamefont {Oppeneer}}, \bibinfo {author} {\bibfnamefont {M.}~\bibnamefont {Jourdan}}, \bibinfo {author} {\bibfnamefont {G.}~\bibnamefont {Jakob}}, \bibinfo {author} {\bibfnamefont
  {D.}~\bibnamefont {Turchinovich}}, \bibinfo {author} {\bibfnamefont {L.~M.}\ \bibnamefont {Hayden}}, \bibinfo {author} {\bibfnamefont {M.}~\bibnamefont {Wolf}}, \bibinfo {author} {\bibfnamefont {M.}~\bibnamefont {Münzenberg}}, \bibinfo {author} {\bibfnamefont {M.}~\bibnamefont {Kläui}}, \ and\ \bibinfo {author} {\bibfnamefont {T.}~\bibnamefont {Kampfrath}},\ }\bibfield  {title} {\enquote {\bibinfo {title} {{Efficient metallic spintronic emitters of ultrabroadband terahertz radiation}},}\ }\href {\doibase 10.1038/nphoton.2016.91} {\bibfield  {journal} {\bibinfo  {journal} {Nature Photonics}\ }\textbf {\bibinfo {volume} {10}},\ \bibinfo {pages} {483--488} (\bibinfo {year} {2016})}\BibitemShut {NoStop}%
\bibitem [{las()}]{laserglow-technologies-no-date}%
  \BibitemOpen
  \href@noop {} {\enquote {\bibinfo {title} {{Laswerglow Technologies, 532 nm Low-Cost DPSS Laser System | 15 - 50 mW Output Power}},}\ }\bibinfo {howpublished} {https://www.laserglow.com/product/C53-C-532-nm-Low-Cost-DPSS-Laser-System},\ \bibinfo {note} {accessed: 2025-04-25}\BibitemShut {NoStop}%
\bibitem [{oce()}]{ocean-optics-no-date}%
  \BibitemOpen
  \href@noop {} {\enquote {\bibinfo {title} {{Ocean Optics, High-Sensitivity UV spectrometers Maya}},}\ }\bibinfo {howpublished} {https://www.gmp.ch/spectroscopy/spectrometer/high-sensitivity-uv-spectrometers-maya},\ \bibinfo {note} {accessed: 2025-04-25}\BibitemShut {NoStop}%
\bibitem [{\citenamefont {Shen}\ \emph {et~al.}(2020)\citenamefont {Shen}, \citenamefont {Wang}, \citenamefont {Dewaele}, \citenamefont {Wu}, \citenamefont {Fratanduono}, \citenamefont {Eggert}, \citenamefont {Klotz}, \citenamefont {Dziubek}, \citenamefont {Loubeyre}, \citenamefont {Fat’yanov}, \citenamefont {Asimow}, \citenamefont {Mashimo},\ and\ \citenamefont {Wentzcovitch}}]{shen-2020}%
  \BibitemOpen
  \bibfield  {author} {\bibinfo {author} {\bibfnamefont {G.}~\bibnamefont {Shen}}, \bibinfo {author} {\bibfnamefont {Y.}~\bibnamefont {Wang}}, \bibinfo {author} {\bibfnamefont {A.}~\bibnamefont {Dewaele}}, \bibinfo {author} {\bibfnamefont {C.}~\bibnamefont {Wu}}, \bibinfo {author} {\bibfnamefont {D.~E.}\ \bibnamefont {Fratanduono}}, \bibinfo {author} {\bibfnamefont {J.}~\bibnamefont {Eggert}}, \bibinfo {author} {\bibfnamefont {S.}~\bibnamefont {Klotz}}, \bibinfo {author} {\bibfnamefont {K.~F.}\ \bibnamefont {Dziubek}}, \bibinfo {author} {\bibfnamefont {P.}~\bibnamefont {Loubeyre}}, \bibinfo {author} {\bibfnamefont {O.~V.}\ \bibnamefont {Fat’yanov}}, \bibinfo {author} {\bibfnamefont {P.~D.}\ \bibnamefont {Asimow}}, \bibinfo {author} {\bibfnamefont {T.}~\bibnamefont {Mashimo}}, \ and\ \bibinfo {author} {\bibfnamefont {R.~M.~M.}\ \bibnamefont {Wentzcovitch}},\ }\bibfield  {title} {\enquote {\bibinfo {title} {{Toward an international practical pressure scale: A proposal for an IPPS ruby gauge
  (IPPS-Ruby2020)}},}\ }\href {\doibase 10.1080/08957959.2020.1791107} {\bibfield  {journal} {\bibinfo  {journal} {High Pressure Research}\ }\textbf {\bibinfo {volume} {40}},\ \bibinfo {pages} {299--314} (\bibinfo {year} {2020})}\BibitemShut {NoStop}%
\bibitem [{\citenamefont {Kubarev}(2009)}]{kubarev-2009}%
  \BibitemOpen
  \bibfield  {author} {\bibinfo {author} {\bibfnamefont {V.~V.}\ \bibnamefont {Kubarev}},\ }\bibfield  {title} {\enquote {\bibinfo {title} {{Optical properties of CVD-diamond in terahertz and infrared ranges}},}\ }\href {\doibase 10.1016/j.nima.2008.12.121} {\bibfield  {journal} {\bibinfo  {journal} {Nuclear Instruments and Methods in Physics Research Section A: Accelerators, Spectrometers, Detectors and Associated Equipment}\ }\textbf {\bibinfo {volume} {603}},\ \bibinfo {pages} {22--24} (\bibinfo {year} {2009})}\BibitemShut {NoStop}%
\bibitem [{\citenamefont {Saleh}\ and\ \citenamefont {Teich}(2007)}]{saleh-2007}%
  \BibitemOpen
  \bibfield  {author} {\bibinfo {author} {\bibfnamefont {B.~E.~A.}\ \bibnamefont {Saleh}}\ and\ \bibinfo {author} {\bibfnamefont {M.~C.}\ \bibnamefont {Teich}},\ }\href@noop {} {\emph {\bibinfo {title} {{Fundamentals of photonics}}}}\ (\bibinfo  {publisher} {Wiley-Interscience},\ \bibinfo {year} {2007})\BibitemShut {NoStop}%
\bibitem [{\citenamefont {O’Bannon}\ \emph {et~al.}(2018)\citenamefont {O’Bannon}, \citenamefont {Jenei}, \citenamefont {Cynn}, \citenamefont {Lipp},\ and\ \citenamefont {Jeffries}}]{obannon-2018}%
  \BibitemOpen
  \bibfield  {author} {\bibinfo {author} {\bibfnamefont {E.~F.}\ \bibnamefont {O’Bannon}}, \bibinfo {author} {\bibfnamefont {Z.}~\bibnamefont {Jenei}}, \bibinfo {author} {\bibfnamefont {H.}~\bibnamefont {Cynn}}, \bibinfo {author} {\bibfnamefont {M.~J.}\ \bibnamefont {Lipp}}, \ and\ \bibinfo {author} {\bibfnamefont {J.~R.}\ \bibnamefont {Jeffries}},\ }\bibfield  {title} {\enquote {\bibinfo {title} {{Contributed Review: Culet diameter and the achievable pressure of a diamond anvil cell: Implications for the upper pressure limit of a diamond anvil cell}},}\ }\href {\doibase 10.1063/1.5049720} {\bibfield  {journal} {\bibinfo  {journal} {Review of Scientific Instruments}\ }\textbf {\bibinfo {volume} {89}} (\bibinfo {year} {2018}),\ 10.1063/1.5049720}\BibitemShut {NoStop}%
\bibitem [{\citenamefont {Verseils}\ \emph {et~al.}(2023)\citenamefont {Verseils}, \citenamefont {Hemme}, \citenamefont {Bounoua}, \citenamefont {Cervasio}, \citenamefont {Brubach}, \citenamefont {Houver}, \citenamefont {Gallais}, \citenamefont {Sacuto}, \citenamefont {Colson}, \citenamefont {Iijima}, \citenamefont {Mochizuki}, \citenamefont {Roy},\ and\ \citenamefont {Cazayous}}]{verseils-2023}%
  \BibitemOpen
  \bibfield  {author} {\bibinfo {author} {\bibfnamefont {M.}~\bibnamefont {Verseils}}, \bibinfo {author} {\bibfnamefont {P.}~\bibnamefont {Hemme}}, \bibinfo {author} {\bibfnamefont {D.}~\bibnamefont {Bounoua}}, \bibinfo {author} {\bibfnamefont {R.}~\bibnamefont {Cervasio}}, \bibinfo {author} {\bibfnamefont {J.-b.}\ \bibnamefont {Brubach}}, \bibinfo {author} {\bibfnamefont {S.}~\bibnamefont {Houver}}, \bibinfo {author} {\bibfnamefont {Y.}~\bibnamefont {Gallais}}, \bibinfo {author} {\bibfnamefont {A.}~\bibnamefont {Sacuto}}, \bibinfo {author} {\bibfnamefont {D.}~\bibnamefont {Colson}}, \bibinfo {author} {\bibfnamefont {T.}~\bibnamefont {Iijima}}, \bibinfo {author} {\bibfnamefont {M.}~\bibnamefont {Mochizuki}}, \bibinfo {author} {\bibfnamefont {P.}~\bibnamefont {Roy}}, \ and\ \bibinfo {author} {\bibfnamefont {M.}~\bibnamefont {Cazayous}},\ }\bibfield  {title} {\enquote {\bibinfo {title} {{Stabilizing electromagnons in CuO under pressure}},}\ }\href {\doibase 10.1038/s41535-023-00542-1} {\bibfield  {journal}
  {\bibinfo  {journal} {npj Quantum Materials}\ }\textbf {\bibinfo {volume} {8}} (\bibinfo {year} {2023}),\ 10.1038/s41535-023-00542-1}\BibitemShut {NoStop}%
\bibitem [{\citenamefont {Lagarias}\ \emph {et~al.}(1998)\citenamefont {Lagarias}, \citenamefont {Reeds}, \citenamefont {Wright},\ and\ \citenamefont {Wright}}]{lagarias-1998}%
  \BibitemOpen
  \bibfield  {author} {\bibinfo {author} {\bibfnamefont {J.~C.}\ \bibnamefont {Lagarias}}, \bibinfo {author} {\bibfnamefont {J.~A.}\ \bibnamefont {Reeds}}, \bibinfo {author} {\bibfnamefont {M.~H.}\ \bibnamefont {Wright}}, \ and\ \bibinfo {author} {\bibfnamefont {P.~E.}\ \bibnamefont {Wright}},\ }\bibfield  {title} {\enquote {\bibinfo {title} {{Convergence properties of the Nelder--Mead Simplex method in low dimensions}},}\ }\href {\doibase 10.1137/s1052623496303470} {\bibfield  {journal} {\bibinfo  {journal} {SIAM journal on optimization}\ }\textbf {\bibinfo {volume} {9}},\ \bibinfo {pages} {112--147} (\bibinfo {year} {1998})}\BibitemShut {NoStop}%
\bibitem [{\citenamefont {Kennedy}\ and\ \citenamefont {Eberhart}(1995)}]{kennedy-1995}%
  \BibitemOpen
  \bibfield  {author} {\bibinfo {author} {\bibfnamefont {J.}~\bibnamefont {Kennedy}}\ and\ \bibinfo {author} {\bibfnamefont {R.}~\bibnamefont {Eberhart}},\ }\bibfield  {title} {\enquote {\bibinfo {title} {{Particle swarm optimization}},}\ \ }(\bibinfo  {publisher} {Proceedings of ICNN'95 - International Conference on Neural Networks},\ \bibinfo {year} {1995})\BibitemShut {NoStop}%
\bibitem [{\citenamefont {Wang}, \citenamefont {Tan},\ and\ \citenamefont {Liu}(2017)}]{wang-2017}%
  \BibitemOpen
  \bibfield  {author} {\bibinfo {author} {\bibfnamefont {D.}~\bibnamefont {Wang}}, \bibinfo {author} {\bibfnamefont {D.}~\bibnamefont {Tan}}, \ and\ \bibinfo {author} {\bibfnamefont {L.}~\bibnamefont {Liu}},\ }\bibfield  {title} {\enquote {\bibinfo {title} {{Particle swarm optimization algorithm: an overview}},}\ }\href {\doibase 10.1007/s00500-016-2474-6} {\bibfield  {journal} {\bibinfo  {journal} {Soft computing}\ }\textbf {\bibinfo {volume} {22}},\ \bibinfo {pages} {387--408} (\bibinfo {year} {2017})}\BibitemShut {NoStop}%
\bibitem [{\citenamefont {Kaplan}(1991)}]{kaplan-1991}%
  \BibitemOpen
  \bibfield  {author} {\bibinfo {author} {\bibfnamefont {W.}~\bibnamefont {Kaplan}},\ }\href@noop {} {\emph {\bibinfo {title} {{Advanced Calculus}}}}\ (\bibinfo  {publisher} {Addison Wesley Publishing Company},\ \bibinfo {year} {1991})\BibitemShut {NoStop}%
\bibitem [{sil()}]{silicone_oil}%
  \BibitemOpen
  \href@noop {} {\enquote {\bibinfo {title} {{Silicone Oil Xiameter PMX-200 Fluid, Sigma Aldrich 181838}},}\ }\bibinfo {howpublished} {\url{https://www.sigmaaldrich.com/CH/en/product/sial/181838}},\ \bibinfo {note} {accessed: 2025-04-25}\BibitemShut {NoStop}%
\bibitem [{\citenamefont {Bell}\ and\ \citenamefont {Mao}(1981)}]{bell-1981}%
  \BibitemOpen
  \bibfield  {author} {\bibinfo {author} {\bibfnamefont {P.}~\bibnamefont {Bell}}\ and\ \bibinfo {author} {\bibfnamefont {H.}~\bibnamefont {Mao}},\ }\href@noop {} {\emph {\bibinfo {title} {Carnegie Institution of Washington Yearbook, 80}}}\ (\bibinfo {year} {1981})\ pp.\ \bibinfo {pages} {404--406}\BibitemShut {NoStop}%
\bibitem [{\citenamefont {Syassen}(2008)}]{syassen-2008}%
  \BibitemOpen
  \bibfield  {author} {\bibinfo {author} {\bibfnamefont {K.}~\bibnamefont {Syassen}},\ }\bibfield  {title} {\enquote {\bibinfo {title} {{Ruby under pressure}},}\ }\href {\doibase 10.1080/08957950802235640} {\bibfield  {journal} {\bibinfo  {journal} {High Pressure Research}\ }\textbf {\bibinfo {volume} {28}},\ \bibinfo {pages} {75--126} (\bibinfo {year} {2008})}\BibitemShut {NoStop}%
\bibitem [{\citenamefont {Celeste}, \citenamefont {Borondics},\ and\ \citenamefont {Capitani}(2019)}]{celeste-2019}%
  \BibitemOpen
  \bibfield  {author} {\bibinfo {author} {\bibfnamefont {A.}~\bibnamefont {Celeste}}, \bibinfo {author} {\bibfnamefont {F.}~\bibnamefont {Borondics}}, \ and\ \bibinfo {author} {\bibfnamefont {F.}~\bibnamefont {Capitani}},\ }\bibfield  {title} {\enquote {\bibinfo {title} {{Hydrostaticity of pressure-transmitting media for high pressure infrared spectroscopy}},}\ }\href {\doibase 10.1080/08957959.2019.1666844} {\bibfield  {journal} {\bibinfo  {journal} {High Pressure Research}\ }\textbf {\bibinfo {volume} {39}},\ \bibinfo {pages} {608--618} (\bibinfo {year} {2019})}\BibitemShut {NoStop}%
\bibitem [{\citenamefont {Staško}\ \emph {et~al.}(2020)\citenamefont {Staško}, \citenamefont {Prchal}, \citenamefont {Klicpera}, \citenamefont {Aoki},\ and\ \citenamefont {Murata}}]{stasko-2020}%
  \BibitemOpen
  \bibfield  {author} {\bibinfo {author} {\bibfnamefont {D.}~\bibnamefont {Staško}}, \bibinfo {author} {\bibfnamefont {J.}~\bibnamefont {Prchal}}, \bibinfo {author} {\bibfnamefont {M.}~\bibnamefont {Klicpera}}, \bibinfo {author} {\bibfnamefont {S.}~\bibnamefont {Aoki}}, \ and\ \bibinfo {author} {\bibfnamefont {K.}~\bibnamefont {Murata}},\ }\bibfield  {title} {\enquote {\bibinfo {title} {{Pressure media for high pressure experiments, Daphne Oil 7000 series}},}\ }\href {\doibase 10.1080/08957959.2020.1825706} {\bibfield  {journal} {\bibinfo  {journal} {High Pressure Research}\ }\textbf {\bibinfo {volume} {40}},\ \bibinfo {pages} {525--536} (\bibinfo {year} {2020})}\BibitemShut {NoStop}%
\bibitem [{\citenamefont {Klotz}\ \emph {et~al.}(2009)\citenamefont {Klotz}, \citenamefont {Chervin}, \citenamefont {Munsch},\ and\ \citenamefont {Marchand}}]{klotz-2009}%
  \BibitemOpen
  \bibfield  {author} {\bibinfo {author} {\bibfnamefont {S.}~\bibnamefont {Klotz}}, \bibinfo {author} {\bibfnamefont {J.~C.}\ \bibnamefont {Chervin}}, \bibinfo {author} {\bibfnamefont {P.}~\bibnamefont {Munsch}}, \ and\ \bibinfo {author} {\bibfnamefont {G.}~\bibnamefont {Marchand}},\ }\bibfield  {title} {\enquote {\bibinfo {title} {{Hydrostatic limits of 11 pressure transmitting media}},}\ }\href {\doibase 10.1088/0022-3727/42/7/075413} {\bibfield  {journal} {\bibinfo  {journal} {Journal of Physics D: Applied Physics}\ }\textbf {\bibinfo {volume} {42}},\ \bibinfo {pages} {075413} (\bibinfo {year} {2009})}\BibitemShut {NoStop}%
\bibitem [{\citenamefont {Lowndes}(1974)}]{lowndes-1974}%
  \BibitemOpen
  \bibfield  {author} {\bibinfo {author} {\bibfnamefont {R.}~\bibnamefont {Lowndes}},\ }\bibfield  {title} {\enquote {\bibinfo {title} {{High pressure far infrared spectroscopy of ionic solids}},}\ }\href {\doibase 10.1109/tmtt.1974.1128433} {\bibfield  {journal} {\bibinfo  {journal} {IEEE transactions on microwave theory and techniques}\ }\textbf {\bibinfo {volume} {22}},\ \bibinfo {pages} {1076--1080} (\bibinfo {year} {1974})}\BibitemShut {NoStop}%
\bibitem [{\citenamefont {Postmus}, \citenamefont {Ferraro},\ and\ \citenamefont {Mitra}(1968)}]{postmus-1968}%
  \BibitemOpen
  \bibfield  {author} {\bibinfo {author} {\bibfnamefont {C.}~\bibnamefont {Postmus}}, \bibinfo {author} {\bibfnamefont {J.~R.}\ \bibnamefont {Ferraro}}, \ and\ \bibinfo {author} {\bibfnamefont {S.~S.}\ \bibnamefont {Mitra}},\ }\bibfield  {title} {\enquote {\bibinfo {title} {{Pressure dependence of infrared eigenfrequencies of KCL and KBR}},}\ }\href {\doibase 10.1103/physrev.174.983} {\bibfield  {journal} {\bibinfo  {journal} {Physical review}\ }\textbf {\bibinfo {volume} {174}},\ \bibinfo {pages} {983--987} (\bibinfo {year} {1968})}\BibitemShut {NoStop}%
\bibitem [{\citenamefont {Yomogida}\ \emph {et~al.}(2010{\natexlab{a}})\citenamefont {Yomogida}, \citenamefont {Sato}, \citenamefont {Nozaki}, \citenamefont {Mishina},\ and\ \citenamefont {Nakahara}}]{yomogida-2010A}%
  \BibitemOpen
  \bibfield  {author} {\bibinfo {author} {\bibfnamefont {Y.}~\bibnamefont {Yomogida}}, \bibinfo {author} {\bibfnamefont {Y.}~\bibnamefont {Sato}}, \bibinfo {author} {\bibfnamefont {R.}~\bibnamefont {Nozaki}}, \bibinfo {author} {\bibfnamefont {T.}~\bibnamefont {Mishina}}, \ and\ \bibinfo {author} {\bibfnamefont {J.}~\bibnamefont {Nakahara}},\ }\bibfield  {title} {\enquote {\bibinfo {title} {{Dielectric study of normal alcohols with THz time-domain spectroscopy}},}\ }\href {\doibase 10.1016/j.molliq.2010.03.007} {\bibfield  {journal} {\bibinfo  {journal} {Journal of molecular liquids}\ }\textbf {\bibinfo {volume} {154}},\ \bibinfo {pages} {31--35} (\bibinfo {year} {2010}{\natexlab{a}})}\BibitemShut {NoStop}%
\bibitem [{\citenamefont {Yomogida}\ \emph {et~al.}(2010{\natexlab{b}})\citenamefont {Yomogida}, \citenamefont {Sato}, \citenamefont {Nozaki}, \citenamefont {Mishina},\ and\ \citenamefont {Nakahara}}]{yomogida-2010B}%
  \BibitemOpen
  \bibfield  {author} {\bibinfo {author} {\bibfnamefont {Y.}~\bibnamefont {Yomogida}}, \bibinfo {author} {\bibfnamefont {Y.}~\bibnamefont {Sato}}, \bibinfo {author} {\bibfnamefont {R.}~\bibnamefont {Nozaki}}, \bibinfo {author} {\bibfnamefont {T.}~\bibnamefont {Mishina}}, \ and\ \bibinfo {author} {\bibfnamefont {J.}~\bibnamefont {Nakahara}},\ }\bibfield  {title} {\enquote {\bibinfo {title} {{Comparative dielectric study of monohydric alcohols with terahertz time-domain spectroscopy}},}\ }\href {\doibase 10.1016/j.molstruc.2010.08.002} {\bibfield  {journal} {\bibinfo  {journal} {Journal of molecular structure}\ }\textbf {\bibinfo {volume} {981}},\ \bibinfo {pages} {173--178} (\bibinfo {year} {2010}{\natexlab{b}})}\BibitemShut {NoStop}%
\bibitem [{\citenamefont {Zhang}\ \emph {et~al.}(2020)\citenamefont {Zhang}, \citenamefont {Zhang}, \citenamefont {Zhao}, \citenamefont {Cao}, \citenamefont {Yu}, \citenamefont {Li}, \citenamefont {Li}, \citenamefont {Chen},\ and\ \citenamefont {Ren}}]{zhang-2020}%
  \BibitemOpen
  \bibfield  {author} {\bibinfo {author} {\bibfnamefont {T.}~\bibnamefont {Zhang}}, \bibinfo {author} {\bibfnamefont {Z.}~\bibnamefont {Zhang}}, \bibinfo {author} {\bibfnamefont {X.}~\bibnamefont {Zhao}}, \bibinfo {author} {\bibfnamefont {C.}~\bibnamefont {Cao}}, \bibinfo {author} {\bibfnamefont {Y.}~\bibnamefont {Yu}}, \bibinfo {author} {\bibfnamefont {X.}~\bibnamefont {Li}}, \bibinfo {author} {\bibfnamefont {Y.}~\bibnamefont {Li}}, \bibinfo {author} {\bibfnamefont {Y.}~\bibnamefont {Chen}}, \ and\ \bibinfo {author} {\bibfnamefont {Q.}~\bibnamefont {Ren}},\ }\bibfield  {title} {\enquote {\bibinfo {title} {{Molecular polarizability investigation of polar solvents: water, ethanol, and acetone at terahertz frequencies using terahertz time-domain spectroscopy}},}\ }\href {\doibase 10.1364/ao.392780} {\bibfield  {journal} {\bibinfo  {journal} {Applied optics}\ }\textbf {\bibinfo {volume} {59}},\ \bibinfo {pages} {4775} (\bibinfo {year} {2020})}\BibitemShut {NoStop}%
\bibitem [{\citenamefont {Franta}\ \emph {et~al.}(2017)\citenamefont {Franta}, \citenamefont {Dubroka}, \citenamefont {Wang}, \citenamefont {Giglia}, \citenamefont {Vohánka}, \citenamefont {Franta},\ and\ \citenamefont {Ohlídal}}]{franta-2017}%
  \BibitemOpen
  \bibfield  {author} {\bibinfo {author} {\bibfnamefont {D.}~\bibnamefont {Franta}}, \bibinfo {author} {\bibfnamefont {A.}~\bibnamefont {Dubroka}}, \bibinfo {author} {\bibfnamefont {C.}~\bibnamefont {Wang}}, \bibinfo {author} {\bibfnamefont {A.}~\bibnamefont {Giglia}}, \bibinfo {author} {\bibfnamefont {J.}~\bibnamefont {Vohánka}}, \bibinfo {author} {\bibfnamefont {P.}~\bibnamefont {Franta}}, \ and\ \bibinfo {author} {\bibfnamefont {I.}~\bibnamefont {Ohlídal}},\ }\bibfield  {title} {\enquote {\bibinfo {title} {{Temperature-dependent dispersion model of float zone crystalline silicon}},}\ }\href {\doibase 10.1016/j.apsusc.2017.02.021} {\bibfield  {journal} {\bibinfo  {journal} {Applied Surface Science}\ }\textbf {\bibinfo {volume} {421}},\ \bibinfo {pages} {405--419} (\bibinfo {year} {2017})}\BibitemShut {NoStop}%
\bibitem [{\citenamefont {Martienssen}\ and\ \citenamefont {Warlimont}(2005)}]{martienssen-2005}%
  \BibitemOpen
  \bibfield  {author} {\bibinfo {author} {\bibfnamefont {W.}~\bibnamefont {Martienssen}}\ and\ \bibinfo {author} {\bibfnamefont {H.}~\bibnamefont {Warlimont}},\ }\href {\doibase 10.1007/3-540-30437-1} {\emph {\bibinfo {title} {{Springer Handbook of Condensed Matter and Materials Data}}}}\ (\bibinfo {year} {2005})\BibitemShut {NoStop}%
\bibitem [{\citenamefont {Li}\ \emph {et~al.}(2012)\citenamefont {Li}, \citenamefont {Li}, \citenamefont {Jin},\ and\ \citenamefont {Ma}}]{li-2012}%
  \BibitemOpen
  \bibfield  {author} {\bibinfo {author} {\bibfnamefont {G.}~\bibnamefont {Li}}, \bibinfo {author} {\bibfnamefont {D.}~\bibnamefont {Li}}, \bibinfo {author} {\bibfnamefont {Z.}~\bibnamefont {Jin}}, \ and\ \bibinfo {author} {\bibfnamefont {G.}~\bibnamefont {Ma}},\ }\bibfield  {title} {\enquote {\bibinfo {title} {{Photocarriers dynamics in silicon wafer studied with optical-pump terahertz-probe spectroscopy}},}\ }\href {\doibase 10.1016/j.optcom.2012.05.053} {\bibfield  {journal} {\bibinfo  {journal} {Optics Communications}\ }\textbf {\bibinfo {volume} {285}},\ \bibinfo {pages} {4102--4106} (\bibinfo {year} {2012})}\BibitemShut {NoStop}%
\bibitem [{\citenamefont {Brunner}, \citenamefont {Schneider},\ and\ \citenamefont {Günter}(2009)}]{brunner-2009}%
  \BibitemOpen
  \bibfield  {author} {\bibinfo {author} {\bibfnamefont {F.~D.~J.}\ \bibnamefont {Brunner}}, \bibinfo {author} {\bibfnamefont {A.}~\bibnamefont {Schneider}}, \ and\ \bibinfo {author} {\bibfnamefont {P.}~\bibnamefont {Günter}},\ }\bibfield  {title} {\enquote {\bibinfo {title} {{A terahertz time-domain spectrometer for simultaneous transmission and reflection measurements at normal incidence}},}\ }\href {\doibase 10.1364/oe.17.020684} {\bibfield  {journal} {\bibinfo  {journal} {Optics Express}\ }\textbf {\bibinfo {volume} {17}},\ \bibinfo {pages} {20684} (\bibinfo {year} {2009})}\BibitemShut {NoStop}%
\bibitem [{\citenamefont {Singh}\ \emph {et~al.}(2000)\citenamefont {Singh}, \citenamefont {Singh}, \citenamefont {Singh}, \citenamefont {Singhal},\ and\ \citenamefont {Chopra}}]{singh-2000}%
  \BibitemOpen
  \bibfield  {author} {\bibinfo {author} {\bibfnamefont {S.}~\bibnamefont {Singh}}, \bibinfo {author} {\bibfnamefont {R.}~\bibnamefont {Singh}}, \bibinfo {author} {\bibfnamefont {B.}~\bibnamefont {Singh}}, \bibinfo {author} {\bibfnamefont {S.}~\bibnamefont {Singhal}}, \ and\ \bibinfo {author} {\bibfnamefont {R.}~\bibnamefont {Chopra}},\ }\bibfield  {title} {\enquote {\bibinfo {title} {{Structural properties of KBR: elastic behaviour and pressure effects}},}\ }\href {\doibase 10.1002/1521-396x(200008)180:2} {\bibfield  {journal} {\bibinfo  {journal} {Physica status solidi. A, Applied research}\ }\textbf {\bibinfo {volume} {180}},\ \bibinfo {pages} {459--466} (\bibinfo {year} {2000})}\BibitemShut {NoStop}%
\bibitem [{\citenamefont {Zhao}\ and\ \citenamefont {Ross}(2015)}]{zhao-2015}%
  \BibitemOpen
  \bibfield  {author} {\bibinfo {author} {\bibfnamefont {J.}~\bibnamefont {Zhao}}\ and\ \bibinfo {author} {\bibfnamefont {N.~L.}\ \bibnamefont {Ross}},\ }\bibfield  {title} {\enquote {\bibinfo {title} {{Non-hydrostatic behavior of KBr as a pressure medium in diamond anvil cells up to 5.63 GPa}},}\ }\href {\doibase 10.1088/0953-8984/27/18/185402} {\bibfield  {journal} {\bibinfo  {journal} {Journal of physics. Condensed matter}\ }\textbf {\bibinfo {volume} {27}},\ \bibinfo {pages} {185402} (\bibinfo {year} {2015})}\BibitemShut {NoStop}%
\bibitem [{\citenamefont {Dewaele}\ \emph {et~al.}(2012)\citenamefont {Dewaele}, \citenamefont {Belonoshko}, \citenamefont {Garbarino}, \citenamefont {Occelli}, \citenamefont {Bouvier}, \citenamefont {Hanfland},\ and\ \citenamefont {Mezouar}}]{dewaele-2012}%
  \BibitemOpen
  \bibfield  {author} {\bibinfo {author} {\bibfnamefont {A.}~\bibnamefont {Dewaele}}, \bibinfo {author} {\bibfnamefont {A.~B.}\ \bibnamefont {Belonoshko}}, \bibinfo {author} {\bibfnamefont {G.}~\bibnamefont {Garbarino}}, \bibinfo {author} {\bibfnamefont {F.}~\bibnamefont {Occelli}}, \bibinfo {author} {\bibfnamefont {P.}~\bibnamefont {Bouvier}}, \bibinfo {author} {\bibfnamefont {M.}~\bibnamefont {Hanfland}}, \ and\ \bibinfo {author} {\bibfnamefont {M.}~\bibnamefont {Mezouar}},\ }\bibfield  {title} {\enquote {\bibinfo {title} {{High-pressure–high-temperature equation of state of KCl and KBr}},}\ }\href {\doibase 10.1103/physrevb.85.214105} {\bibfield  {journal} {\bibinfo  {journal} {Physical review. B, Condensed matter and materials physics}\ }\textbf {\bibinfo {volume} {85}} (\bibinfo {year} {2012}),\ 10.1103/physrevb.85.214105}\BibitemShut {NoStop}%
\bibitem [{\citenamefont {D’Angelo}\ \emph {et~al.}(2014)\citenamefont {D’Angelo}, \citenamefont {Mics}, \citenamefont {Bonn},\ and\ \citenamefont {Turchinovich}}]{dangelo-2014}%
  \BibitemOpen
  \bibfield  {author} {\bibinfo {author} {\bibfnamefont {F.}~\bibnamefont {D’Angelo}}, \bibinfo {author} {\bibfnamefont {Z.}~\bibnamefont {Mics}}, \bibinfo {author} {\bibfnamefont {M.}~\bibnamefont {Bonn}}, \ and\ \bibinfo {author} {\bibfnamefont {D.}~\bibnamefont {Turchinovich}},\ }\bibfield  {title} {\enquote {\bibinfo {title} {{Ultra-broadband THz time-domain spectroscopy of common polymers using THz air photonics}},}\ }\href {\doibase 10.1364/oe.22.012475} {\bibfield  {journal} {\bibinfo  {journal} {Optics express}\ }\textbf {\bibinfo {volume} {22}},\ \bibinfo {pages} {12475} (\bibinfo {year} {2014})}\BibitemShut {NoStop}%
\bibitem [{\citenamefont {Kirby}(1956)}]{kirby-1956}%
  \BibitemOpen
  \bibfield  {author} {\bibinfo {author} {\bibfnamefont {R.~K.}\ \bibnamefont {Kirby}},\ }\bibfield  {title} {\enquote {\bibinfo {title} {{Thermal expansion of polytetrafluoroethylene (Teflon) from -190 to +300 C}},}\ }\href {\doibase 10.6028/jres.057.010} {\bibfield  {journal} {\bibinfo  {journal} {Journal of research of the National Bureau of Standards}\ }\textbf {\bibinfo {volume} {57}},\ \bibinfo {pages} {91} (\bibinfo {year} {1956})}\BibitemShut {NoStop}%
\bibitem [{\citenamefont {Wu}, \citenamefont {Hewitt},\ and\ \citenamefont {Zhang}(1996)}]{wu-1996}%
  \BibitemOpen
  \bibfield  {author} {\bibinfo {author} {\bibfnamefont {Q.}~\bibnamefont {Wu}}, \bibinfo {author} {\bibfnamefont {T.~D.}\ \bibnamefont {Hewitt}}, \ and\ \bibinfo {author} {\bibfnamefont {X.-c.}\ \bibnamefont {Zhang}},\ }\bibfield  {title} {\enquote {\bibinfo {title} {{Two-dimensional electro-optic imaging of THz beams}},}\ }\href {\doibase 10.1063/1.116920} {\bibfield  {journal} {\bibinfo  {journal} {Applied Physics Letters}\ }\textbf {\bibinfo {volume} {69}},\ \bibinfo {pages} {1026--1028} (\bibinfo {year} {1996})}\BibitemShut {NoStop}%
\bibitem [{\citenamefont {Usami}\ \emph {et~al.}(2005)\citenamefont {Usami}, \citenamefont {Yamashita}, \citenamefont {Fukushima}, \citenamefont {Otani},\ and\ \citenamefont {Kawase}}]{usami-2005}%
  \BibitemOpen
  \bibfield  {author} {\bibinfo {author} {\bibfnamefont {M.}~\bibnamefont {Usami}}, \bibinfo {author} {\bibfnamefont {M.}~\bibnamefont {Yamashita}}, \bibinfo {author} {\bibfnamefont {K.}~\bibnamefont {Fukushima}}, \bibinfo {author} {\bibfnamefont {C.}~\bibnamefont {Otani}}, \ and\ \bibinfo {author} {\bibfnamefont {K.}~\bibnamefont {Kawase}},\ }\bibfield  {title} {\enquote {\bibinfo {title} {{Terahertz wideband spectroscopic imaging based on two-dimensional electro-optic sampling technique}},}\ }\href {\doibase 10.1063/1.1899259} {\bibfield  {journal} {\bibinfo  {journal} {Applied Physics Letters}\ }\textbf {\bibinfo {volume} {86}} (\bibinfo {year} {2005}),\ 10.1063/1.1899259}\BibitemShut {NoStop}%
\bibitem [{\citenamefont {Blanchard}, \citenamefont {Arikawa},\ and\ \citenamefont {Tanaka}(2022)}]{blanchard-2022}%
  \BibitemOpen
  \bibfield  {author} {\bibinfo {author} {\bibfnamefont {F.}~\bibnamefont {Blanchard}}, \bibinfo {author} {\bibfnamefont {T.}~\bibnamefont {Arikawa}}, \ and\ \bibinfo {author} {\bibfnamefont {K.}~\bibnamefont {Tanaka}},\ }\bibfield  {title} {\enquote {\bibinfo {title} {{Real-Time Megapixel Electro-Optical Imaging of THz Beams with Probe Power Normalization}},}\ }\href {\doibase 10.3390/s22124482} {\bibfield  {journal} {\bibinfo  {journal} {Sensors}\ }\textbf {\bibinfo {volume} {22}},\ \bibinfo {pages} {4482} (\bibinfo {year} {2022})}\BibitemShut {NoStop}%
\end{thebibliography}%

\end{document}